\documentclass[article,12pt]{article}

\usepackage{amscd,amsmath,amssymb,amsfonts,latexsym,mathrsfs,amsthm}
\usepackage{graphicx} 
\usepackage{multirow}
\usepackage{setspace} 
\usepackage[font=small,labelfont=bf]{caption}

\usepackage{graphics,epsfig}
\usepackage{sectsty}
\usepackage{natbib}
\usepackage[english]{babel}
\usepackage{amsmath}
\usepackage{amsfonts}
\usepackage{amssymb,amsthm}
\usepackage{bm}
\usepackage{mathrsfs}

\usepackage{natbib}

\usepackage[FIGTOPCAP]{subfigure}

\usepackage{setspace}

\usepackage{hyperref}
\usepackage[usenames]{color}
\usepackage{rotating}

\usepackage{colortbl}

\definecolor{MYCOLOR0}{rgb}{0.92,0.92,0.92}
\definecolor{MYCOLOR}{rgb}{1,1,0}
\definecolor{MYCOLOR2}{rgb}{0.5,1,0.5}

\definecolor{MYCOLOR3}{rgb}{0.88,1,1}

\newtheorem{theo}{Theorem}
\newtheorem{defi}{Definition}
\newtheorem*{PR}{Proof}

\newtheorem{rem}{{\bf Remark}}

\def\x{{\mathbf x}}

\def\v{{\mathbf v}}
\def\y{{\mathbf y}}

\newcounter{alg}

\title{Group Importance Sampling for Particle Filtering and MCMC} 

\author{Luca Martino$^\star$, V\'ictor Elvira$^\dag$, Gustau Camps-Valls$^\top$ \\
{$^\star$ \small Universidad Carlos III de Madrid (Spain). }\\
{$^\star,^\top$ \small Image Processing Laboratory, Universitat de Val\`encia (Spain). }\\
{$^\dag$ \small IMT Lille Douai CRISTAL (UMR 9189), Villeneuve d'Ascq (France).}
}

\date{}
 
\begin{document}

\maketitle

\thispagestyle{empty}

\begin{abstract}
 Bayesian methods and their implementations by means of sophisticated Monte Carlo techniques have become very popular in signal processing over the last years. Importance Sampling (IS) is a well-known Monte Carlo technique that approximates integrals involving a posterior distribution by means of weighted samples. In this work, we study the assignation of a single weighted sample which compresses the information contained in a population of weighted samples. Part of the theory that we present as Group Importance Sampling (GIS) has been employed implicitly in different works in the literature. The provided analysis yields several theoretical and practical consequences. For instance, we discuss the application of GIS into the Sequential Importance Resampling framework and show that Independent Multiple Try Metropolis schemes can be interpreted as a standard Metropolis-Hastings algorithm, following the  GIS approach. We also introduce two novel Markov Chain Monte Carlo (MCMC) techniques based on GIS. The first one, named Group Metropolis Sampling method, produces a Markov chain of sets of weighted samples. All these sets are then employed for obtaining a unique global estimator. The second one is the Distributed Particle Metropolis-Hastings technique, where different parallel particle filters are jointly used to drive an MCMC algorithm. Different resampled trajectories are compared and then tested with a proper acceptance probability. The novel schemes are tested in different numerical experiments such as learning the hyperparameters of Gaussian Processes, two localization problems in a wireless sensor network  (with synthetic and real data) and the tracking of vegetation parameters given satellite observations, where they are compared with several benchmark Monte Carlo techniques. Three illustrative Matlab demos are also provided.
\end{abstract}

\section{Introduction}

Bayesian signal processing, which has become very popular over the last years in statistical signal processing, requires the study of complicated distributions of variables of interested conditioned on observed data \citep{Liu04b,Martino09,IA2RMS14,Robert04}. Unfortunately, the computation of statistical features related to these posterior distributions (such as moments or credible intervals)  is analytically impossible in many real-world applications. Monte Carlo methods are state-of-the-art tools for approximating complicated integrals involving sophisticated multidimensional densities \citep{Liang10,Liu04b,Robert04}. The most popular classes of MC methods are the Importance Sampling (IS) techniques and the Markov chain Monte Carlo (MCMC) algorithms~\citep{Liang10,Robert04}. IS schemes produce a random discrete approximation of the posterior distribution by a population of weighted samples \citep{Bugallo15,LAIS17,EBEB14,Liu04b,Robert04}. MCMC techniques generate a Markov chain (i.e., a sequence of correlated samples) with a pre-established target probability density function (pdf) as invariant density~\citep{Liang10,Liu04b}.  Both families are widely used in the signal processing community. Several exhaustive overviews regarding the application of Monte Carlo methods in statistical signal processing, communications and machine learning can be found in the literature: some of them specifically focused on MCMC algorithms \citep{Andrieu:MCMachineLearning2003,DSRL06,Fitzgerald01,MARTINO2018134}, others specifically focused on IS techniques (and related methods) \citep{7974876,Bugallo15,Djuric03,ElviraMIS15} or with a broader view \citep{Candy09,Wang:MCWirelessComm2002,Doucet:MCSignalProcessing2005,Pereyra16,Ruanaidh12}.

In this work, we introduce theory and practice of a novel approach, called Group Importance Sampling (GIS), where the information contained in different sets of weighted samples is compressed by using only one, yet properly selected, particle, and one suitable weight.\footnote{A preliminary version of this work has been published in \citep{GMS17}. With respect to that paper, here we provide a complete theoretical support of the Group Importance Sampling (GIS) approach (and of the derived methods), given in the main body of the text (Sections \ref{GISsect} and \ref{AppGIS}) and in five additional appendices.  Moreover, we provide  an additional method based on GIS in Section \ref{P-PF-PMH} and a discussion regarding particle Metropolis schemes and the standard Metropolis-Hastings method in Section \ref{MTM_StMH}. We also provide several additional numerical studies, one considering real data. Related Matlab software is also given at  \url{https://github.com/lukafree/GIS.git}.} This general idea supports the validity of different Monte Carlo algorithms in the literature: interacting parallel particle filters \citep{Bolic05,Miguez16,Read2014}, particle island schemes and related techniques \citep{Pisland15, Pisland15_2,Whiteley16}, particle filters for model selection \citep{Drovandi14,Martino15PF,PetarCopy}, nested Sequential Monte Carlo (SMC) methods \citep{NSMC,NosCitan,RafaelThesis} are some examples. We point out some consequences of the application of GIS in Sequential Importance Resampling (SIR) schemes, allowing partial resampling procedures and the use of different marginal likelihood estimators. Then, we show that the Independent Multiple Try Metropolis (I-MTM) techniques and the Particle Metropolis-Hastings (PMH) algorithm can be interpreted as a classical Independent Metropolis-Hastings method by the application of GIS. 

Furthermore, we present two novel techniques based on GIS. The first one is the {\it Group Metropolis Sampling} (GMS) algorithm that generates a Markov chain of sets of weighted samples. All these resulting sets of samples are jointly exploited to obtain a unique particle approximation of the target distribution. 
On the one hand, GMS can be considered an MCMC method since it produces a Markov chain of sets of samples.
On the other hand, the GMS can be also considered as an iterated importance sampler where different estimators are finally combined in order to build a unique IS estimator. This combination is obtained {\it dynamically} through random repetitions given by MCMC-type acceptance tests. GMS is closely related to Multiple Try Metropolis (MTM) techniques and Particle Metropolis-Hastings (PMH) algorithms \citep{PMCMC,Bedard12, Casarin13, Craiu07, LucaJesse2,MTMissue17}, as we discuss below. The GMS algorithm can be also seen as an extension of the method  in \citep{Casella96},  for recycling auxiliary samples in a MCMC method. 

The second novel algorithm based on GIS is the Distributed PMH (DPMH) technique where the outputs of several parallel particle filters are compared by an MH-type acceptance function. The proper design of DPMH is a direct application of GIS.  The benefit of DPMH is twofold: different type of particle filters (for instance, with different proposal densities) can be jointly employed, and the computational effort can be distributed in several machines speeding up the resulting algorithm.  As the standard PMH method, DPMH is useful for filtering and smoothing the estimation of the trajectory of a variable of interest in a state-space model. Furthermore, the marginal version of DPMH can be used for the joint estimation of dynamic and static parameters. When the approximation of only one specific moment of the posterior is required, like GMS, the DPMH output can be expressed as a chain of IS estimators.  
The novel schemes are tested in different numerical experiments: hyperparameter tuning for Gaussian Processes, two localization problems in a wireless sensor network (one with real data), and finally a filtering problem of Leaf Area Index (LAI), which is a parameter widely used to monitor vegetation from satellite observations. The comparisons with other benchmark Monte Carlo methods show the benefits of the proposed algorithms.\footnote{Three illustrative Matlab demos are also provided at \url{https://github.com/lukafree/GIS.git}.}

The remainder of the paper has the following structure. Section \ref{BackSect} recalls some background material. The basis of the GIS theory is introduced in Section \ref{GISsect}. The applications of GIS in particle filtering and Multiple Try Metropolis algorithms are discussed in Section \ref{AppGIS}. In Section \ref{NoTech}, we introduce the novel techniques based on GIS. Section \ref{GPexample} provides the numerical results and in Section \ref{Conclsect} we discuss some conclusions.

\section{Problem statement and background}
 \label{BackSect}
 
 In many applications,  the goal is to infer a variable of interest,  
 ${\bf x}=x_{1:D}=[x_1,x_2,\ldots,x_D]\in \mathcal{X}\subseteq\mathbb{R}^{D\times \xi}$,
where $x_d\in \mathbb{R}^{\xi}$ for all $d=1,\ldots,D$, given a set of related observations or measurements, ${\bf y}\in \mathbb{R}^{d_Y}$. In the Bayesian framework all the statistical information is summarized by the posterior probability density function (pdf), i.e.,
\begin{equation}
	\bar{\pi}({\bf x})= p({\bf x}| {\bf y})= \frac{\ell({\bf y}|{\bf x}) g({\bf x})}{Z({\bf y})},
\label{eq:posterior}
\end{equation}
where $\ell({\bf y}|{\bf x})$ is the likelihood function, $g({\bf x})$ is the prior pdf and $Z({\bf y})$ is the marginal likelihood (a.k.a., Bayesian evidence).
In general, $Z\equiv Z({\bf y})$ is unknown and difficult to estimate in general, so we assume to be able to evaluate the unnormalized target function,
\begin{equation}
\pi({\bf x})=\ell({\bf y}|{\bf x}) g({\bf x}).
\label{eq:target}
\end{equation}
The computation of integrals involving $\bar{\pi}({\bf x})=\frac{1}{Z} \pi({\bf x})$ is often intractable. We consider the Monte Carlo approximation of complicated integrals involving the target $ \bar{\pi}({\bf x})$ and an integrable function $h({\bf x})$ with respect to $\bar{\pi}$, i.e.,  
 \begin{equation}
 \label{EqI}
I=E_{\bar \pi}[h({\bf X})]=\int_{\mathcal{X}} h({\bf x}) \bar{\pi}({\bf x}) d{\bf x},
\end{equation}
where we denote ${\bf X}\sim \bar{\pi}({\bf x})$.  The basic Monte Carlo (MC) procedure consists in drawing $N$ independent samples from the  target pdf, i.e., ${\bf x}_{1},\ldots,{\bf x}_{N} \sim \bar{\pi}({\bf x})$, so that ${\widehat I}_N=\frac{1}{N} \sum_{n=1}^N h({\bf x}_n)$ is an unbiased estimator of $I$ \citep{Liu04b,Robert04}. However, in general, direct methods for drawing samples from $\bar{\pi}({\bf x})$ do not exist so that alternative procedures are required. Below, we describe the most popular approaches. Table \ref{tab:notation} summarizes the main notation of the work. Note that the words {\it sample} and {\it particle} are used as synonyms along this work. Moreover, Table \ref{tab:acro} shows the main used acronyms.

\paragraph{Marginal Likelihood.} As shown above, we consider a target function $\bar{\pi}({\bf x})=\frac{1}{Z} \pi({\bf x})$ that is posterior density, i.e., $\bar{\pi}({\bf x})= p({\bf x}| {\bf y})= \frac{\ell({\bf y}|{\bf x}) g({\bf x})}{Z({\bf y})}$ and $\pi({\bf x})=\ell({\bf y}|{\bf x}) g({\bf x})$. In this case, $Z=Z({\bf y})=\int_{\mathcal{X}} \ell({\bf y}|{\bf x}) g({\bf x}) d{\bf x}$ represents the marginal probability of ${\bf y}$, i.e.,   $Z({\bf y})=p({\bf y})$ that is usually called marginal likelihood (or Bayesian evidence). This quantity is important for model selection purpose.  More generally, the considerations in the rest of the work are valid also for generic target densities $\bar{\pi}({\bf x})=\frac{1}{Z} \pi({\bf x})$ where $\pi({\bf x}) \geq 0$ and $Z=\int_{\mathcal{X}} \pi({\bf x}) d{\bf x}$. In this scenario, $Z$ represents a normalizing constant and could not have any other statistical meanings. However, we often refer to $Z$ as marginal likelihood, without loss of generality.

\subsection{Markov Chain Monte Carlo (MCMC) algorithms}
\label{MCMCback}

 An MCMC method generates an ergodic Markov chain with invariant (a.k.a., stationary) density given by the posterior pdf $ \bar{\pi}({\bf x})$ \citep{Liang10,Robert04,DANI06}. Specifically, given a starting state ${\bf x}_0$, a sequence of {\it correlated} samples is generated, 
$\{\x_t\}_{t=1}^{T}$. Even if the samples are now correlated, the estimator ${\widehat I}_T=\frac{1}{T} \sum_{t=1}^T f({\bf x}_t)$
is consistent, regardless the starting vector ${\bf x}_0$ \citep{DANI06,Robert04}. 
The Metropolis-Hastings (MH) method is one of the most popular MCMC algorithm \citep{Liang10,Liu04b,Robert04}. Given a simpler proposal density $ q(\x|\x_{t-1})$ depending on the previous state of the chain, the MH method is outlined below: 
  \begin{enumerate}
\item Choose an initial state $\x_0$.
\item For $t=1,\ldots,T:$
\begin{enumerate}
\item  Draw a sample $\v'\sim q(\x|\x_{t-1})$.
\item Accept the new state, $\x_t={\bf v}'$, with probability 
\begin{equation}
\label{AlfaMH_back}
\alpha(\x_{t-1},{\bf v}') = \min\left[1, \frac{\pi({\bf v}')  q(\x_{t-1}|{\bf v}')}{\pi({\bf x}_{t-1})  q({\bf v}'|\x_{t-1})}\right],
\end{equation}
Otherwise, with probability $1-\alpha(\x_{t-1},{\bf v}')$, set $\x_t=\x_{t-1}$.
\end{enumerate}
\item Return $\{\x_t\}_{t=1}^T$.
   \end{enumerate} 
Due to the correlation the chain requires a burn-in period  before converging to the invariant distribution. Therefore a certain number of initial samples should be discarded, i.e., not included in the resulting estimator. However, the length of the burn-in period is in general unknown. Several studies in order to estimate the length of the burn-in period   can be found in the literature \citep{Brooks98,Gelman92,Propp96}.

\begin{table}[!t]
\centering
\caption{Main notation of the work. }
\vspace{0.1cm}
	\begin{tabular}{|c|l||c|l|}
	 \hline
      \cellcolor{MYCOLOR0} ${\bf x}=[x_1,\ldots,x_D]$ & \multicolumn{3}{l|}{Variable of interest, $\x \in \mathcal{X}\subseteq\mathbb{R}^{D\times\xi}$, with $x_d \in \mathbb{R}^{\xi}$  for all $d$ } \\
  \cellcolor{MYCOLOR0} $\bar{\pi}({\bf x})$ & \multicolumn{3}{l|}{Normalized posterior pdf, $\bar{\pi}({\bf x})= p({\bf x}| {\bf y})$} \\
  \cellcolor{MYCOLOR0} $\pi({\bf x})$ & \multicolumn{3}{l|}{Unnormalized posterior function,  $\pi({\bf x}) \propto \bar{\pi}({\bf x})$} \\
   \cellcolor{MYCOLOR0} $\widehat{\pi}(\x|\x_{1:N})$ &  \multicolumn{3}{l|}{Particle approximation of $\bar{\pi}({\bf x})$ using the set of samples $\x_{1:N}=\{\x_{n}\}_{n=1}^N$} \\
   \cellcolor{MYCOLOR0} ${\widetilde\x}$  &   \multicolumn{3}{l|}{Resampled particle, ${\widetilde\x} \sim \widehat{\pi}(\x|\x_{1:N})$ (note that ${\widetilde\x}\in\{\x_{1},\dots,\x_{N}\}$)}  \\
    \hline
     \cellcolor{MYCOLOR0} $w_n=w(\x_n)$ &  \multicolumn{3}{l|}{Unnormalized standard IS weight of the particle $\x_n$} \\
       \cellcolor{MYCOLOR0} $\bar{w}_n=\bar{w}(\x_n)$ &  \multicolumn{3}{l|}{Normalized weight associated to $\x_n$} \\
       \cellcolor{MYCOLOR0} $\widetilde{w}_m=\widetilde{w}({\widetilde\x}_m)$ &  \multicolumn{3}{l|}{Unnormalized proper weight associated to the resampled particle  ${\widetilde\x}_m$} \\  
       \cellcolor{MYCOLOR0} $W_m$ &  \multicolumn{3}{l|}{Summary weight of $m$-th set $\mathcal{S}_m$} \\  
       \cellcolor{MYCOLOR0} $\overline{I}_N$ &  \multicolumn{3}{l|}{Standard self-normalized IS estimator using $N$ samples} \\     
       \cellcolor{MYCOLOR0} $\widetilde{I}_N$ &  \multicolumn{3}{l|}{Self-normalized estimator using $N$ samples and based on GIS theory} \\            
     \hline
       \cellcolor{MYCOLOR0}  $Z$ &  \multicolumn{3}{l|}{Marginal likelihood; normalizing constant of $\pi(\x)$} \\ 
              \cellcolor{MYCOLOR0}  $\widehat{Z}$, $\overline{Z}$ &  \multicolumn{3}{l|}{Estimators of the marginal likelihood $Z$} \\ 
     \hline
\end{tabular}
\label{tab:notation}
\end{table}

\begin{table}[!h]
\centering
\caption{Main acronyms in the work. }
\vspace{0.1cm}
	\begin{tabular}{|c|c|}
	 \hline
	   \cellcolor{MYCOLOR0} IS &  Importance Sampling\\
	      \cellcolor{MYCOLOR0} SIS & Sequential Importance Sampling\\
	            \cellcolor{MYCOLOR0} SIR & Sequential Importance Resampling\\
	               \cellcolor{MYCOLOR0} PF &  Particle Filter\\
	                \cellcolor{MYCOLOR0} SMC &  Sequential Monte Carlo\\
	        \hline   
	     \cellcolor{MYCOLOR0} MCMC &  Markov  Chain Monte Carlo\\
  \cellcolor{MYCOLOR0} MH & Metropolis-Hastings \\
    \cellcolor{MYCOLOR0} IMH & Independent Metropolis-Hastings \\
    \cellcolor{MYCOLOR0} MTM & Multiple Try Metropolis \\
      \cellcolor{MYCOLOR0} I-MTM & Independent Multiple Try Metropolis\\
        \cellcolor{MYCOLOR0} I-MTM2 & Independent Multiple Try Metropolis (version 2)\\
       \cellcolor{MYCOLOR0} PMH & Particle Metropolis-Hastings \\ 
          \cellcolor{MYCOLOR0} PMMH & Particle Marginal Metropolis-Hastings \\ 
    \hline   
    \cellcolor{MYCOLOR0} GIS &  Group Importance Sampling\\
    \cellcolor{MYCOLOR0} GSM & Group Metropolis Sampling \\
      \cellcolor{MYCOLOR0} PGSM & Particle Group Metropolis Sampling \\
    \cellcolor{MYCOLOR0} DPMH & Distributed Particle Metropolis-Hastings \\ 
    \cellcolor{MYCOLOR0} DPMMH & Distributed Particle Marginal Metropolis-Hastings \\
     \hline
\end{tabular}
\label{tab:acro}
\end{table}

\subsection{Importance Sampling}
\label{ISsect}

Let us consider again the use of  a simpler proposal pdf, $q({\bf x})$, and rewrite the integral $I$ in Eq. \eqref{EqI} as 
 \begin{eqnarray}
I=E_{\bar \pi}[h({\bf X})]=E_q[h({\bf X})w({\bf X})]=\frac{1}{Z}\int_{\mathcal{X}} h({\bf x}) \frac{\pi({\bf x})}{q({\bf x})} q({\bf x}) d{\bf x},
\end{eqnarray}
where $w({\bf x})=\frac{\pi({\bf x})}{q({\bf x})}: \mathcal{X}\rightarrow \mathbb{R}$. This suggests an alternative procedure. Indeed, we can draw  $N$  samples ${\bf x}_1,\ldots,{\bf x}_N$ from $q({\bf x})$,\footnote{We assume that $q({\bf x})>0$ for all ${\bf x}$ where $\bar{\pi}({\bf x})\neq 0$, and $q({\bf x})$ has heavier tails than $\bar{\pi}({\bf x})$.} and then assign to each sample the unnormalized weights
 \begin{equation} 
	w_n=w({\bf x}_n)= \frac{\pi({\bf x}_n)}{{q({\bf x}_n)}}, \quad n=1,\ldots,N.
\label{is_weights_static}
\end{equation} 
If $Z$ is known, an unbiased IS estimator \citep{Liu04b,Robert04} is defined as
${\widehat I}_N=\frac{1}{ZN}\sum_{n=1}^N  w_n h({\bf x}_n)$, 
where ${\bf x}_n \sim q({\bf x})$. If $Z$ is unknown, defining the normalized weights,
${\bar w}_n=\frac{ w_n }{\sum_{i=1}^N w_i}$ with $n=1,\ldots,N$, an alternative consistent IS estimator of $I$ in Eq. \eqref{EqI} (i.e., still asymptotically unbiased) is given by \citep{Liu04b,Robert04}
 \begin{equation}
 \label{EstNorm}
{\overline I}_N=\sum_{n=1}^N  {\bar w}_n h({\bf x}_n). 
\end{equation}
Moreover, an unbiased estimator of marginal likelihood, $Z=\int_{\mathcal{X}} \pi({\bf x}) d{\bf x}$, is given by ${\widehat Z}= \frac{1}{N} \sum_{i=1}^N w_i$. More generally, the pairs $\{\x_i,w_i\}_{i=1}^N$ can be used to build a particle approximation of the posterior measure,
\begin{eqnarray}
\label{ApprIS}
\widehat{\pi}(\x|\x_{1:N})=\frac{1}{N{\widehat Z}}\sum_{n=1}^N w_n \delta(\x-\x_n)=\sum_{n=1}^N {\bar w}_n \delta(\x-\x_n),
\end{eqnarray}
where $\delta(\x)$ denotes the Dirac delta function.  Given a specific integrand function $h(\x)$ in Eq. \eqref{EqI}, it is possible to show that the optimal proposal pdf, which minimizes the variance of the corresponding IS estimator, is given by $q_{\texttt{opt}}(\x)\propto |h(\x)| {\bar \pi}(\x)$.


\subsection{Concept of proper weighting}
 \label{LiuSect}

In this section, we discuss a generalization of the classical importance sampling (IS) technique.
 The standard IS weights in Eq. \eqref{is_weights_static} are broadly used in the literature. However, the definition of {\it properly weighted sample} can be extended as suggested in \cite[Section 14.2]{Robert04}, \cite[Section 2.5.4]{Liu04b} and in \citep{ElviraMIS15}.
  More specifically, given a set of samples, they are properly weighted with respect to the target ${\bar \pi}$ if, for any integrable function $h$, 
\begin{equation}
E_q[w({\bf x}_n) h({\bf x}_n)]=c E_{\bar \pi}[h({\bf x}_n)], \quad \forall n\in\{1,\dots,N\},
\label{eq_liu_1}
\end{equation} 
where $c>0$ is a constant value, also independent from the index $n$, and the expectation of the left hand side is performed, in general, w.r.t. to the joint pdf of $w({\bf x})$ and ${\bf x}$, i.e., $q(w,{\bf x})$. Namely, the weight  $w({\bf x})$, conditioned to a given value of ${\bf x}$, could even be considered a random variable. Thus, in order to obtain consistent estimators, one can design any joint $q(w,{\bf x})$ as long as the restriction of Eq. \eqref{eq_liu_1} is fulfilled. Based on this idea, dynamic weighting algorithms that mix MCMC and IS approaches have been proposed \citep{DW97}. When different proposal pdfs $q_1(\x),q_2(\x),\ldots, q_N(\x)$ are jointly used in an IS scheme, the class of proper weighting schemes is even broader, as shown in \citep{ElviraMIS15,LetterVictor,7544571}.


\section{Group Importance Sampling: weighting a set of samples} \label{GISsect}
 In this section, we use  the general definition in Eq.  \eqref{eq_liu_1} for designing proper weights and summary samples assigned to different sets of samples. Let us consider $M$ sets of weighted samples,
$\mathcal{S}_1=\{\x_{1,n},w_{1,n}\}_{n=1}^{N_1}$, $\mathcal{S}_2=\{\x_{2,n},w_{2,n}\}_{n=1}^{N_2}$, ...., $\mathcal{S}_M=\{\x_{M,n},w_{M,n}\}_{n=1}^{N_M}$,
where $\x_{m,n} \sim q_m(\x)$, i.e., a different proposal pdf for each set $\mathcal{S}_m$ and in general $N_i\neq N_j$, for all $i \neq j$, $i,j\in\{1,...,M\}$.  
In some applications and different Monte Carlo schemes, it is  convenient (and often required) to compress the statistical information contained in each set using a pair of summary sample, $\widetilde{\x}_m$, and summary weight, $W_m$, $m=1,\ldots,M$, in such a way that the following expression
\begin{eqnarray}
\label{GroupIS}
\widetilde{I}_M&=&\frac{1}{\sum_{j=1}^M  W_j} \sum_{m=1}^M  W_m h(\widetilde{\x}_m),Ê
\end{eqnarray}
is still a consistent estimator of $I$, for a generic integrable function $h(\x)$. Thus, although the compression is lossy, we  
still have a suitable particle approximation ${\widehat \pi}$ of the target ${\bar \pi}$ as shown below.  
%
%
In the following, we denote the importance weight of the $n$-th sample in the $m$-th group as  $w_{m,n}=w(\x_{m,n})=\frac{\pi(\x_{m,n})}{q_m(\x_{m,n})}$, the $m$-th marginal likelihood estimator as
 \begin{equation}
 \label{ParaMGroups}
   \widehat{Z}_m=\frac{1}{N_m}\sum_{n=1}^{N_m}w_{m,n}, 
\end{equation}
and the normalized weights within a set, $\bar{w}_{m,n}=\frac{w_{m,n}}{\sum_{j=1}^{N_m} w_{m,j}}=\frac{w_{m,n}}{N_m \widehat{Z}_m}$, for $n=1,\ldots,N_m$ and $m=1,\ldots,M$.

{\defi \label{Def3}  A resampled particle, i.e., 
  \begin{equation}
 \label{Def2}
  \widetilde{\x}_m \sim \widehat{\pi}_m(\x) =\widehat{\pi}(\x|\x_{m,1:N_m})=\sum_{n=1}^{N_m} \bar{w}_{m,n} \delta(\x-\x_{m,n}),
\end{equation}  
is a summary particle $\widetilde{\x}_m$ for the $m$-group. Note that $\widetilde{\x}_m$ is selected within $\{\x_{m,1},\ldots,\x_{m,N_m}\}$ according to the probability mass function (pmf) defined by $\bar{w}_{m,n}$,  $n=1,\ldots,N_m$. 
 }
 \newline
\newline 
 It is possible to use the Liu's definition in order to assign a proper importance weight to a resampled particle \citep{GISssp16}, as stated in the following theorem.  

\begin{figure}[htbp]
\centering
\includegraphics[width=15cm]{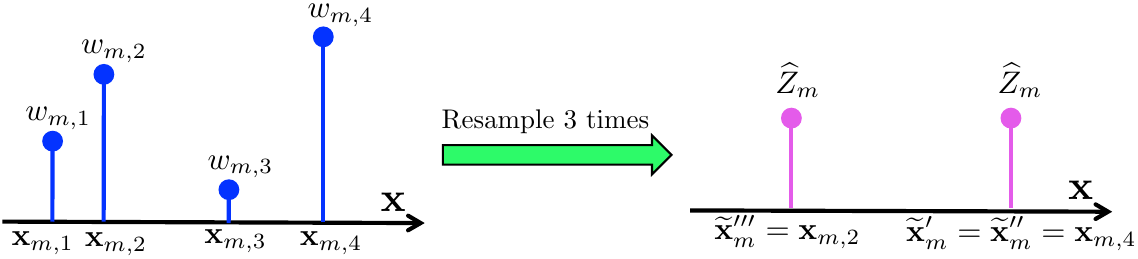} 
\caption{Example of generation (one run) and proper weighting of $3$ resampled particles (with replacement), $\widetilde{\x}_m'$, $\widetilde{\x}_m''$ and  $\widetilde{\x}_m'''$, from the $m$-th group, where $N_m=4$ and $\widehat{Z}_m=\frac{1}{4}\sum_{n=1}^4 w_{m,n}$.  }
\label{Fig0}
\end{figure}

 {\theo  \label{Theo1} Let us consider a resampled particle $\widetilde{\x}_mÊ\sim \widehat{\pi}_m(\x)=\widehat{\pi}(\x|\x_{m,1:N_m})$. A proper unnormalized weight following the Liu's definition in Eq. \eqref{eq_liu_1} for this resampled particle is $\widetilde{w}_m= \widehat{Z}_m$, defined in Eq. \eqref{ParaMGroups}.}
\newline
\newline
The proof is given in  \ref{WeightingAResPar}. Note that two (or more) particles, $\widetilde{\x}_m'$, $\widetilde{\x}_m''$,  resampled with replacement from the same set and hence from the same approximation,  $\widetilde{\x}_m', \widetilde{\x}_m'' \sim\widehat{\pi}_m(\x)$, have the same weight $\widetilde{w}(\widetilde{\x}_m')=\widetilde{w}(\widetilde{\x}_m'')= \widehat{Z}_m$, as depicted in Figure \ref{Fig0}.  Note that the classical importance weight cannot be computed for a resampled particle, as explained in  \ref{WeightingAResPar} and pointed out in \citep{Lamberti16,GISssp16,NSMC}, \cite[App. C1]{LAIS17}. 

  {\defi  The summary weight for the $m$-th group of samples is 
 $W_m= N_m \widetilde{w}_m= N_m\widehat{Z}_m$.} 

\begin{figure}[htbp]
\centering
\includegraphics[width=10cm]{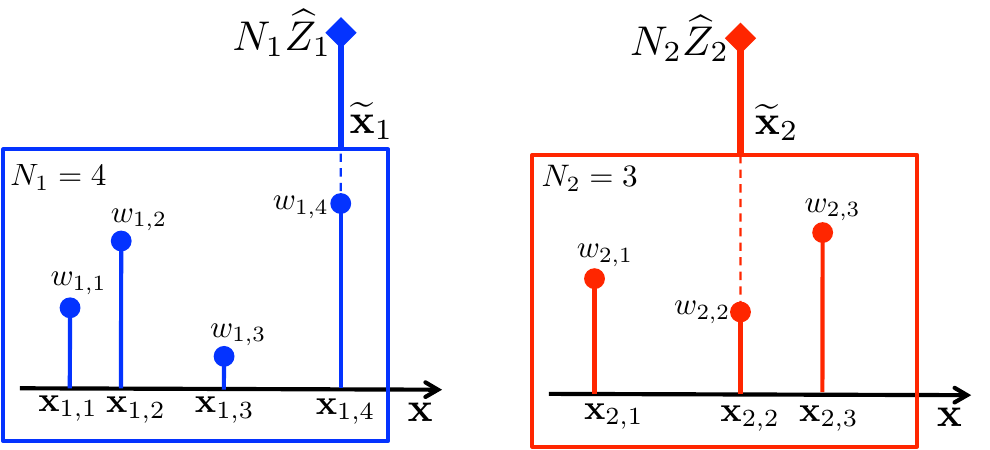} 
\caption{ Graphical representation of GIS. In this case, $M=2$ groups of $N_1=4$ and $N_2=3$ weighted samples are summarized with a resampled particle and one summary weight $\widetilde{w}_m=N_m\widehat{Z}_m$, $m=1,2$. 
 }
\label{Fig1}
\end{figure}

\paragraph{Particle approximation.}  Figure \ref{Fig1} represents graphically an example of GIS with $M=2$ and $N_1=4$, $N_2=3$. Given the $M$ summary pairs $\{\widetilde{\x}_m,\widetilde{w}_{m}\}_{m=1}^M$ in a common computational node, we can obtain the following particle approximation of $\bar{\pi}(\x)$, i.e., 
\begin{eqnarray}
\label{PiGroup}
\widehat{\pi}(\x|\widetilde{\x}_{1:M})=\frac{1}{\sum_{j=1}^M N_j\widehat{Z}_j}  \sum_{m=1}^M N_m\widehat{Z}_m \delta(\x-\widetilde{\x}_m),
\end{eqnarray} 
involving $M$ weighted samples in this case (see \ref{ImportantApp}). For a given function $h(\x)$, the corresponding specific GIS estimator in Eq. \eqref{GroupIS} is
\begin{eqnarray}
\label{GroupIS2}
\widetilde{I}_M&=&\frac{1}{\sum_{j=1}^M N_j\widehat{Z}_j} \sum_{m=1}^M  N_m\widehat{Z}_m h(\widetilde{\x}_m).
\end{eqnarray}
It is a consistent estimator of $I$, as we show in  \ref{ImportantApp}. The expression in Eq. \eqref{GroupIS2} can be interpreted as a standard IS estimator where $\widetilde{w}(\widetilde{\x}_m)= \widehat{Z}_m$ is a proper weight of a resampled particle \citep{GISssp16}, and we give more importance to the resampled particles belonging to a set with higher cardinality.  See DEMO-2 at \url{https://github.com/lukafree/GIS.git}.

\paragraph{Combination of estimators.} If we are only interested in computing the integral $I$  for a specific function $h(\x)$, we can summarize the statistical information by the pairs $\{{\overline I}_{N_m}^{(m)}, \widetilde{w}_{m}\}$ where  
\begin{equation}
{\overline I}_{N_m}^{(m)}= \sum_{n=1}^{N_m}  \bar{w}_{m,n} h(\x_{m,n}),
\end{equation}
 is the $m$-th partial IS estimator obtained by using $N_m$ samples in $\mathcal{S}_m$. Given all the $S=\sum_{j=1}^M N_j$ weighted samples in the $M$ sets, the complete estimator ${\overline I}_{S}$ in Eq. \eqref{EstNorm} can be written as a convex combination of the $M$ partial IS estimators, ${\overline I}_{N_m}^{(m)}$, i.e.,
 \begin{eqnarray}
 \label{EqAllSETS}
  {\overline I}_{S}&=&\frac{1}{\sum_{j=1}^M N_j\widehat{Z}_j}\sum_{m=1}^M\sum_{n=1}^{N_m}w_{m,n} h(\x_{m,n}),  \\
&=&\frac{1}{\sum_{j=1}^M N_j\widehat{Z}_j}\sum_{m=1}^M N_m\widehat{Z}_m \sum_{n=1}^{N_m}  \bar{w}_{m,n} h(\x_{m,n}), \\
 \label{EqAllSETS3}
&=&\frac{1}{\sum_{j=1}^MW_{m}} \sum_{m=1}^M  W_{m} {\overline I}_{N_m}^{(m)}.
\end{eqnarray}
The equation above shows that the summary weight $W_{m}$ measures the importance of the $m$-th estimator ${\overline I}_{N_m}^{(m)}$. This confirms that $W_{m}$  is a proper weight the group of samples $\mathcal{S}_m$, and also suggests another valid compression scheme.

\begin{figure}[htbp]
\centering
\subfigure[]{\includegraphics[width=7cm]{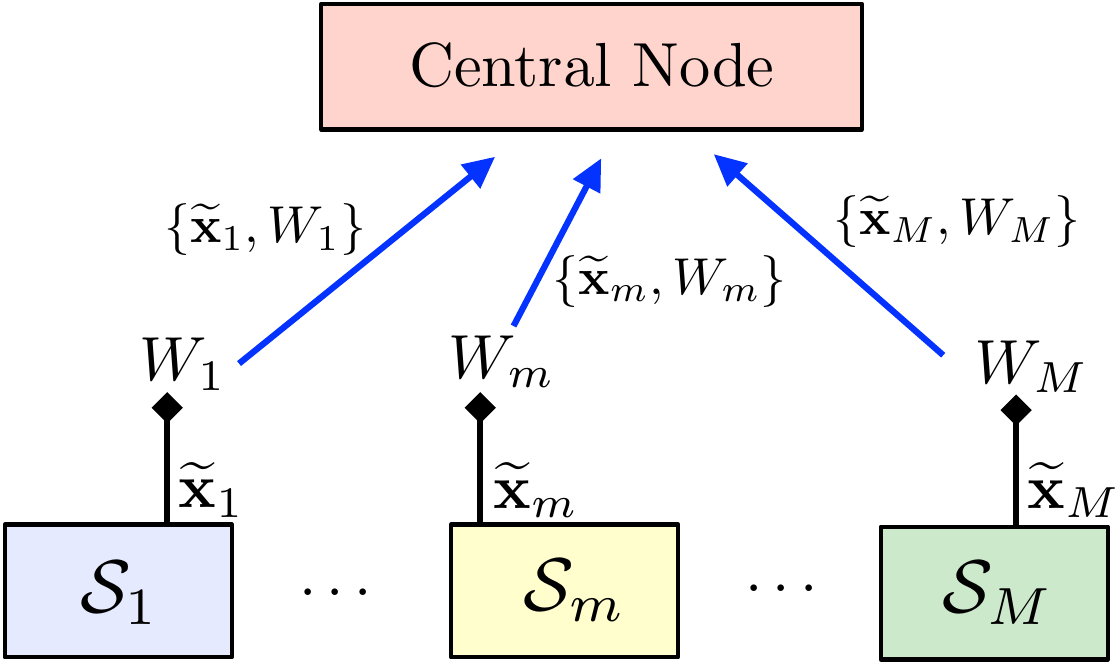}}
\hspace{1cm}
\subfigure[]{\includegraphics[width=7cm]{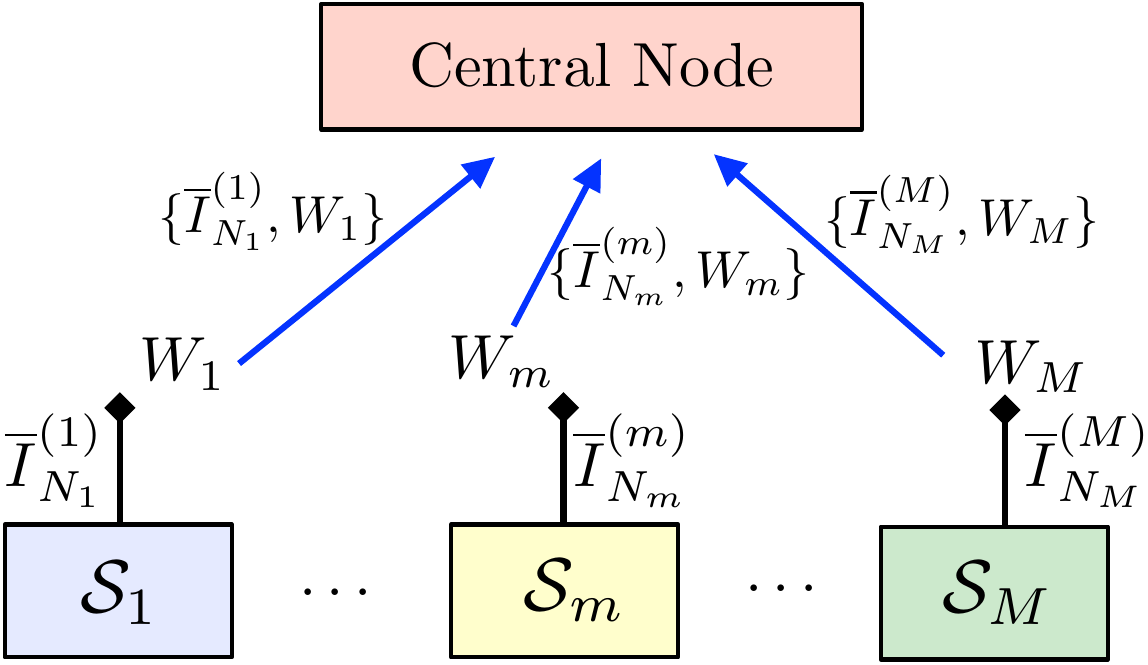}}
\caption{Graphical overview of GIS in a parallel/distributed framework. {\bf (a)} The central node obtains all the pairs $\{\widetilde{\x}_m,W_m\}_{m=1}^M$, and provides $\widehat{\pi}(\x|\widetilde{\x}_{1:M})$ or ${\overline I}_M$. Note that only $M$ particles, $\widetilde{\x}_m\in\mathbb{R}^D$, and $M$ scalar weights, $W_m\in\mathbb{R}$, are transmitted, instead of $S$ samples and $S$ weights, with $S=\sum_{m=1}^{M} N_m$. {\bf (b)} Alternatively, if we are interested only in a specific moment of the target, we can transmit the pairs $\{{\overline I}_{N_m}^{(m)},W_m\}_{m=1}^M$ and then combine them as in Eq. \eqref{EqAllSETS3}.}
\label{Fig2}
\end{figure}

\begin{figure}[htbp]
\centering
\includegraphics[width=15cm]{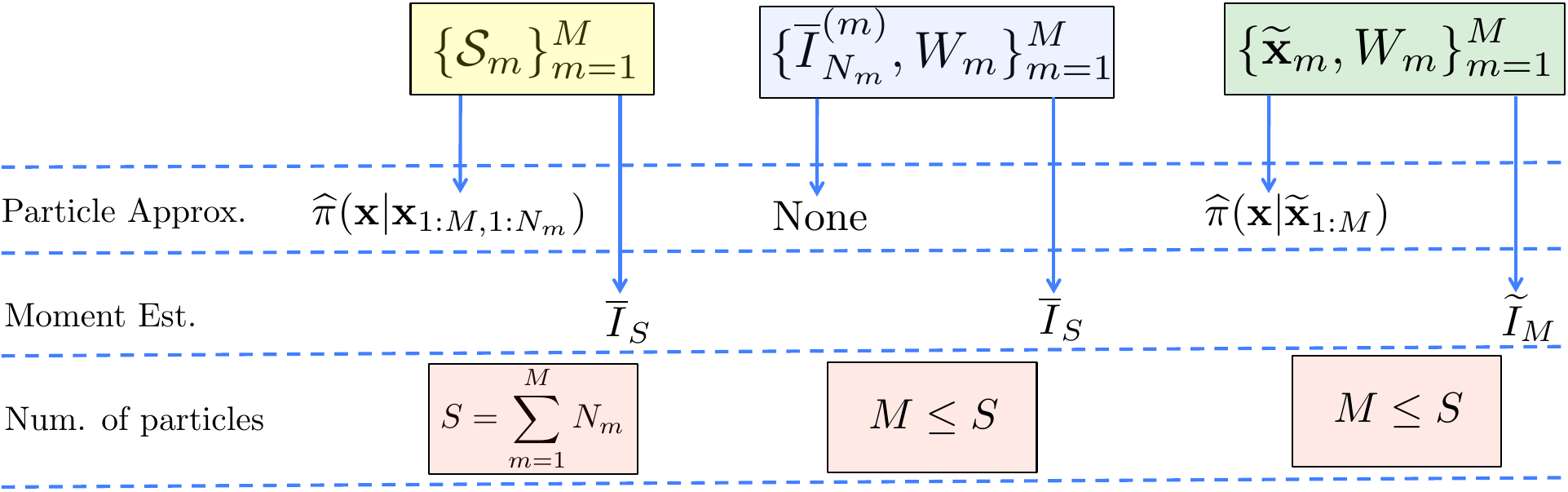}
\caption{Possible outputs of different GIS compression schemes. On the left, $\{\mathcal{S}_m\}_{m=1}^M$, no compression is applied.  In the center, $\{{\overline I}_{N_m}^{(m)},W_m\}_{m=1}^M$, we can perfectly reconstruct the estimator ${\overline I}_{S}$ in Eq. \eqref{EqAllSETS} where $S=\sum_{m=1}^M N_m$, but we cannot approximate other moments. Using $\{{\widetilde{\x}}_{N_m}^{(m)},W_m\}_{m=1}^M$, we always obtain a lossy compression, but any moments of ${\bar \pi}(\x)$ can be approximated, as shown in Eqs. \eqref{PiGroup}-\eqref{GroupIS2}. 
}
\label{FigSummary}
\end{figure}

{\rem  In order to approximate only one specific moment $I$ of ${\bar \pi}(\x)$, we can summarize the $m$-group with the pair $\{{\overline I}_{N_m}^{(m)},W_m\}_{m=1}^M$, thus all the $M$ partial estimators can be combined following Eq. \eqref{EqAllSETS3}.
}
\newline
\newline
In this case, there is no loss of information w.r.t. storing all the weighted samples. However, the approximation of other moments of  ${\bar \pi}(\x)$ is not possible. Figures \ref{Fig2}-\ref{FigSummary} depict the graphical representations of the two possible approaches for GIS. 

\section{GIS in other Monte Carlo schemes}
\label{AppGIS}

 

\subsection{Application in particle filtering}
\label{GIS_SIR}

In Section \ref{ISsect}, we have described the IS procedure in a batch way, i.e., generating directly a $D$-dimensional vector ${\bf x}' \sim q(\x)$ and then compute the weight $\frac{\pi(\x')}{q(\x')}$. This procedure can be performed sequentially if the target density is factorized. In this case, the method is known as Sequential Importance Sampling (SIS). It is the basis of particle filtering, along with the use of the resampling procedure. Below, we describe the SIS method.

\subsubsection{Sequential Importance Sampling (SIS) }
Let us that recall ${\bf x}=x_{1:D}=[x_1,x_2,\ldots,x_D]\in \mathcal{X}\subseteq\mathbb{R}^{D\times\xi}$ where $x_d\in \mathbb{R}^{\xi}$ for all $d=1,\ldots,D$, and let us consider a target pdf $\bar{\pi}({\bf x})$ factorized as 
\begin{eqnarray}
\label{DecTarget}
\bar{\pi}({\bf x})= \frac{1}{Z} \pi({\bf x})&=&\frac{1}{Z} \gamma_1(x_1) \prod_{d=2}^D\gamma_d(x_d|x_{1:d-1}), 
\end{eqnarray}
where $\gamma_1(x_1) $ is a marginal pdf and  $\gamma_d(x_d|x_{1:d-1})$ are conditional pdfs. 
We can also consider a proposal pdf decomposed in the same way,
$q({\bf x})=q_1(x_1) \prod_{d=2}^{D}q_d(x_d|x_{d-1})$.
In a batch IS scheme, given the $n$-th sample ${\bf x}_n=x_{1:D}^{(n)}\sim q({\bf x})$, we assign the importance weight  
\begin{eqnarray}
w({\bf x}_n)&=&\frac{\pi({\bf x}_n)}{q({\bf x}_n)}=\frac{ \gamma_1(x_1^{(n)}) \gamma_2(x_2^{(n)}|x_1^{(n)}) \cdots  \gamma_D(x_D^{(n)}|x_{1:D-1}^{(n)})}{q_1(x_1^{(n)}) q_2(x_2^{(n)}|x_1^{(n)}) \cdots  q_D(x_D^{(n)}|x_{1:D-1}^{(n)})}
= \prod_{d=1}^{D} \beta_d, \label{SISweight}
\end{eqnarray}
where $\beta_1^{(n)}=\frac{\pi(x_1^{(n)})}{q(x_1^{(n)})}$ and $\beta_d^{(n)}=\frac{\gamma_d(x_d^{(n)}|x_{1:d-1}^{(n)})}{q_d(x_d^{(n)}|x_{1:d-1}^{(n)})}$, with $d=2,\ldots, D$. Let us also denote the joint probability of $[x_1,\ldots, x_d]$ as
\begin{equation}
\bar{\pi}_d(x_{1:d})=\frac{1}{Z_d} \pi_d(x_{1:d})=\frac{1}{Z_d}\gamma_1(x_1) \prod_{j=2}^d\gamma_j(x_j|x_{1:j-1}),
\end{equation}
where $Z_d=\int_{\mathbb{R}^{d\times \xi}} \pi_d(x_{1:d}) d x_{1:d}$.
Note that  $\bar{\pi}_D(x_{1:D})\equiv\bar{\pi}({\bf x}) $ and $Z_D \equiv Z$. Thus, we can draw samples 
generating sequentially each component $x_d^{(n)}\sim q_d(x_d|x_{1:d-1}^{(n)})$, $d=1,\ldots, D$, so that $\x_n=x_{1:D}^{(n)} \sim q({\bf x})=q_1(x_1) \prod_{d=2}^{D}q_d(x_d|x_{d-1})$, and compute recursively the corresponding IS weight in Eq. \eqref{SISweight}.  
Indeed, considering the definition 
\begin{eqnarray}
w_d^{(n)}=\frac{\pi_d(x_{1:d}^{(n)})}{q_1(x_1^{(n)})\prod_{j=2}^{d}q_j(x_j^{(n)}|x_{1:j-1}^{(n)})}
\end{eqnarray}
we have the recursion $w_d^{(n)}=w_{d-1}^{(n)} \beta_d^{(n)} =\prod_{j=1}^{d} \beta_j^{(n)}$, with $w_0^{(n)}=1$, and  we recall that $\beta_d^{(n)}=\frac{\gamma_d(x_d^{(n)}|x_{1:d-1}^{(n)})}{q_d(x_d^{(n)}|x_{1:d-1}^{(n)})}$. 
The SIS technique is also given in Table \ref{alg:SIRpartialRes} by setting $\eta=0$. Note also that $\widehat{Z}_d=\frac{1}{N} \sum_{n=1}^N w_d^{(n)}$ is an unbiased estimator of $Z_d$. Defining the normalized weights ${\bar w}_{d}^{(n)}=\frac{w_d^{(n)}}{\sum_{i=1}^N w_d^{(i)}}$, in SIS we have another equivalent formulation of the same estimator as shown below. 

{\rem In SIS, there are two possible formulations of the estimator of the normalizing constant $Z_d =\int_{\mathbb{R}^{d\times \xi}} \pi_d(x_{1:d}) d x_{1:d}$, 
\begin{eqnarray}
\widehat{Z}_d&=&\frac{1}{N} \sum_{n=1}^N w_d^{(n)}= \frac{1}{N} \sum_{n=1}^N w_{d-1}^{(n)}  \beta_d^{(n)},  \label{Est1}\\
\overline{Z}_d&=& \prod_{j=1}^d\left[\sum_{n=1}^N{\bar w}_{j-1}^{(n)}\beta_j^{(n)}\right]. \label{Est2}
\end{eqnarray}
In SIS, both estimators are equivalent $\overline{Z}_d\equiv \widehat{Z}_d$. See  \ref{SuperAppMargLikeEstimator} for further details.
}

\subsubsection{Sequential Importance Resampling (SIR) }


The expression in Eq. \eqref{SISweight} suggests a recursive procedure for generating the samples and computing the importance weights, as shown in Steps \ref{PropStep} and \ref{WStep} of Table \ref{alg:SIRpartialRes}. In Sequential Importance Resampling (SIR), a.k.a., standard particle filtering, resampling steps are incorporated during the recursion as in step \ref{StepResampling2} of Table \ref{alg:SIRpartialRes} \citep{Djuric03,Doucet01b}. In general, the resampling steps are applied only in certain iterations in order to avoid the path degeneration, taking into account an approximation $\widehat{ESS}$ of the Effective Sampling Size (ESS) \citep{Huggins15,ESSarxiv16}. If $\widehat{ESS}$ is smaller than a pre-established threshold, the particles are resampled. Two examples of ESS approximation are $\widehat{ESS}=\frac{1}{\sum_{n=1}^N (\bar{w}_d^{(n)})^2}$ and $\widehat{ESS}=\frac{1}{\max \bar{w}_d^{(n)}}$ where $\bar{w}_d^{(n)}=\frac{w_d^{(n)}}{\sum_{i=1}^N w_d^{(i)}}$. Note that, in both cases, $1\leq \widehat{ESS}\leq N$.
Hence, the condition for the adaptive resampling can be expressed as $\widehat{ESS} < \eta N$ where $\eta\in [0,1]$. SIS is given when $\eta=0$ and SIR  for $\eta\in(0,1]$. When $\eta=1$, the resampling is applied at each iteration and in this case SIR is often called {\it bootstrap particle filter} \citep{Djuric03,Doucet01b,Doucet08tut}. If $\eta= 0$, no resampling is applied, we only apply Steps \ref{PropStep} and \ref{WStep}, and we have the SIS method described above, that after $D$ iterations is completely equivalent to the batch IS approach,  since $w_n=w({\bf x}_n)\equiv w_D^{(n)}$ where $\x_n=x_{1:D}$.

\paragraph{Partial resampling.} In Table \ref{alg:SIRpartialRes}, we have considered that only a {\it subset} of $R\leq N$ particles are resampled. In this case, step \ref{StepProperGIS} including the GIS weighting is strictly required in order to provide final proper weighted samples and hence consistent estimators. The partial resampling procedure is an alternative approach to prevent the loss of particle diversity \citep{GISssp16}.
 In the classical description of SIR \citep{Rubin88}, we have $R=N$ (i.e., all the particles are resampled) and
the weight recursion follows setting the unnormalized weights of the resampled particles to any equal value. Since {\it all} the $N$ particles have been resampled, the selection of this value has no impact in the weight recursion and in the estimation of $I$.

\paragraph{Marginal likelihood estimators.} Even in the case $R=N$,  i.e., all the particle are resampled as in the standard SIR method, without using the GIS weighting only the formulation $\overline{Z}_d$ in Eq. \eqref{Est2} provides a consistent estimator of $Z_d$, since it involves the normalized weights ${\bar w}_{d-1}^{(n)}$, instead of the unnormalized ones, $w_{d-1}^{(n)}$. 

{\rem 
If the GIS weighting is applied in SIR, both formulations $\widehat{Z}_d$  and $\overline{Z}_d$ in Eqs. \eqref{Est1}-\eqref{Est2} provide consistent estimator of $Z_d$ and they are equivalent, $\widehat{Z}_d\equiv \overline{Z}_d$ (as in SIS). See an exhaustive discussion in  \ref{SuperAppMargLikeEstimator}.
}
\newline
See DEMO-1 at \url{https://github.com/lukafree/GIS.git}.


\begin{table}[!h]
	\centering
	\caption{{\bf SIR with partial resampling}}
	    \begin{tabular}{|p{0.95\columnwidth}|}
		\hline
  \begin{enumerate}
\item Choose $N$ the number of particles, $R\leq N$ the number of particles to be resampled, the initial particles $x_0^{(n)}$,  with $w_0^{(n)}=1$, $n=1,\ldots,N$, an ESS approximation $\widehat{ESS}$ \citep{ESSarxiv16}  and a constant value $\eta\in[0,1]$.
    \item For $d = 1, \ldots, D$:
    	\begin{enumerate}
        \item {\bf Propagation:}\label{PropStep} Draw  $x_d^{(n)} \sim q_d(x_d|x_{1:d-1}^{(n)})$, for $n=1,\ldots,N$.
        \item  {\bf Weighting:}\label{WStep} Compute the weights       
                 \begin{equation}
                 \label{RecWeights}
                   w_d^{(n)}=w_{d-1}^{(n)} \beta_d^{(n)} =\prod_{j=1}^{d} \beta_j^{(n)},   \quad \quad n=1,\ldots,N,   
                \end{equation}       
                   where  $\beta_d^{(n)}=\frac{\gamma_d(x_d^{(n)}|x_{1:d-1}^{(n)})}{q_d(x_d^{(n)}|x_{1:d-1}^{(n)})}$. 
             \item\label{StepResampling}  if $\widehat{ESS} < \eta N$ then: 
             \begin{enumerate}
  \item Select randomly, without repetition, a set of particles $\mathcal{S}=\{x_{1:d}^{(j_r)}\}_{r=1}^{R}$  where $R\leq N$,  $j_r\in\{1,\ldots,N\}$ for all $r$, and  $j_r\neq j_k$ for $r\neq k$. 
\item {\bf Resampling:}\label{StepResampling2} Resample $R$ times within the set $\mathcal{S}$ according to the probabilities $\bar{w}_d^{(j_r)}=\frac{w_d^{(j_r)}}{\sum_{k=1}^R w_d^{(j_k)}}$, obtaining $\{\bar{x}_{1:d}^{(j_r)}\}_{r=1}^R$. Then, set $x_{1:d}^{(j_r)}=\bar{x}_{1:d}^{(j_r)}$, for $r=1,\ldots,R$.
\item\label{StepProperGIS}  {\bf GIS weighting:} Compute $\widehat{Z}_{\mathcal{S}}=\frac{1}{R} \sum\limits_{r=1}^Rw_{d}^{(j_r)}$ and set $w_{d}^{(j_r)}=\widehat{Z}_{\mathcal{S}}$ for all $r=1,\ldots,R$.
\end{enumerate}             
         \end{enumerate}
         \item Return $\{\x_n=x_{1:D}^{(n)},  w_n=w_D^{(n)}\}_{n=1}^N$.
   \end{enumerate} 	
\\
	\hline
	\end{tabular}		
	\label{alg:SIRpartialRes}
\end{table}


 \paragraph{GIS in Sequential Monte Carlo (SMC).}
The idea of summary sample and summary weight have been implicitly used in different SMC schemes proposed in literature, for instance, for the communication among parallel particle filters \citep{Bolic05,Miguez16,Read2014}, and in the particle island methods \citep{Pisland15,Pisland15_2,Whiteley16}. GIS also appears indirectly in particle filtering for model selection \citep{Drovandi14,Martino15PF,PetarCopy} and in the so-called Nested Sequential Monte Carlo techniques  \citep{NSMC,NosCitan,RafaelThesis}.

\subsection{Multiple Try Metropolis schemes as a Standard Metropolis-Hastings method}
 \label{MTM_StMH}
 
 The Metropolis-Hastings (MH) method, described in Section \ref{MCMCback}, is a simple and popular MCMC algorithm  \citep{Liang10,Liu04b,Robert04}. It  generates a Markov chain $\{\x_t\}_{t=1}^{\infty}$ where $\bar{\pi}(\x)$ is the invariant density. Considering a proposal pdf $q(\x)$ independent from the previous state $\x_{t-1}$, the corresponding Independent MH (IMH) scheme is formed by the steps in Table \ref{alg:MH} \citep{Robert04}.

 \begin{table}[!h]
	\centering
	\caption{{\bf The Independent Metropolis-Hastings (IMH) algorithm}}
	    \begin{tabular}{|p{0.95\columnwidth}|}
		\hline
  \begin{enumerate}
\item Choose an initial state $\x_0$.
\item For $t=1,\ldots,T:$
\begin{enumerate}
\item  Draw a sample $\v'\sim q(\x)$.
\item Accept the new state, $\x_t={\bf v}'$, with probability 
\begin{equation}
\label{AlfaMH}
\alpha(\x_{t-1},{\bf v}') = \min\left[1, \frac{\pi({\bf v}')  q(\x_{t-1})}{\pi({\bf x}_{t-1})  q({\bf v}')}\right]= \min\left[1, \frac{w({\bf v}')  }{w({\bf x}_{t-1})}\right],
\end{equation}
where $w(\x)=\frac{\pi(\x)}{q(\x)}$ (standard importance weight). Otherwise, set $\x_t=\x_{t-1}$.
\end{enumerate}
\item Return $\{\x_t\}_{t=1}^T$.
   \end{enumerate} 	
\\
	\hline
	\end{tabular}		
	\label{alg:MH}
\end{table}

Observe that  $\alpha(\x_{t-1},{\bf v}')=\min\left[1, \frac{w({\bf v}')  }{w({\bf x}_{t-1})}\right]$ in Eq. \eqref{AlfaMH} involves the ratio between the importance weight of the proposed sample $\v'$ at the $t$-th iteration, and the importance weight of the previous state $\x_{t-1}$. Furthermore, note that at each iteration only one new sample $\v'$ is generated and compared with the previous state $\x_{t-1}$ by the acceptance probability $\alpha(\x_{t-1},{\bf v}')$ (in order to obtain the next state $\x_t$). 
The Particle Metropolis-Hastings (PMH) method\footnote{PMH is used for filtering and smoothing a variable of interest in state-space models (see, for instance, Figure \ref{Fig2_Ex3}). The {\it Particle Marginal MH} (PMMH) algorithm \citep{PMCMC} is an extension of PMH employed in order to infer both dynamic and static variables. PMMH is described in  \ref{MarginalPMH}. } \citep{PMCMC} and the alternative version of the Independent Multiply Try Metropolis technique \citep{MTM_PMH14} (denoted as I-MTM2) are jointly described in Table \ref{alg:IMTM2}.
 They are two MCMC algorithms where at each iteration several candidates $\{\v_1,\ldots,\v_N\}$ are generated. After computing the IS weights $w(\v_n)$, one candidate is selected $\v_j$ within the $N$ possible values, i.e., $j\in\{1,\ldots,N\}$, applying a resampling step according to the probability mass $\bar{w}_n=\frac{w(\v_n)}{\sum_{i=1}^N w(\v_i)}=\frac{w(\v_n)}{N\widehat{Z}'}$, $n=1,\ldots,N$. Then the selected sample $\v_j$ is tested with a proper probability $\alpha(\x_{t-1},\v_j)$ in Eq. \eqref{AlfaPMH}.  


 \begin{table}[!h]
	\centering
	\caption{{\bf PMH and I-MTM2 techniques}}
	    \begin{tabular}{|p{0.95\columnwidth}|}
		\hline
  \begin{enumerate}
\item Choose an initial state $\x_0$ and $\widehat{Z}_{0}$.
\item For $t=1,\ldots,T:$
\begin{enumerate}
\item  Draw $N$ particles $\v_1,\ldots,\v_N$ from $q(\x)$ and weight them with the proper importance weight $w(\v_n)$, $n=1,\ldots,N$, using a sequential approach (PMH), or a batch approach (I-MTM2). Thus, denoting $\widehat{Z}'=\frac{1}{N}\sum_{n=1}^N w(\v_n)$, we obtain the particle approximation
$\widehat{\pi}(\x|\v_{1:N})=\frac{1}{N \widehat{Z}'}\sum_{n=1}^N w(\v_n) \delta(\x-\v_n)$.
\item Draw $\v_j\sim \widehat{\pi}(\x|\v_{1:N})$.
\item Set $\x_t={\bf v}_j$ and $\widehat{Z}_t=\widehat{Z}'$, with probability 
\begin{equation}
\label{AlfaPMH}
\alpha(\x_{t-1},{\bf v}_j) = \min\left[1, \frac{\widehat{Z}'}{\widehat{Z}_{t-1}}\right].
\end{equation}
Otherwise, set $\x_t=\x_{t-1}$ and $\widehat{Z}_t=\widehat{Z}_{t-1}$.
\end{enumerate}
\item Return $\{\x_t\}_{t=1}^T$.
   \end{enumerate} 	
\\
	\hline
	\end{tabular}		
	\label{alg:IMTM2}
\end{table}

The difference between PMH and I-MTM2 is the procedure employed for the generation of the $N$ candidates and for the construction of the weights.  PMH employs a  sequential approach, whereas I-MTM2 uses a standard batch approach \citep{MTM_PMH14}. Namely, PMH generates sequentially the components $v_{j,k}$ of the candidates, $\v_j=[v_{j,1},\ldots,v_{j,D}]^{\top}$, and compute recursively the weights as shown in Section \ref{GIS_SIR}. Since resampling steps are often used, then the resulting candidates $\v_1,\ldots,\v_N$ are correlated, whereas in I-MTM2 they are independent.  I-MTM2 coincides with PMH if the candidates are generated sequentially but without applying resampling steps, so that I-MTM2 can be considered a special case of PMH.

Note that $\widetilde{w}({\bf v}_j)=\widehat{Z}'$ and $\widetilde{w}({\x}_{t-1})=\widehat{Z}_{t-1}$ are the GIS  weights of the resampled particles ${\bf v}_j$ and $\x_{t-1}$ respectively, as stated in Definition \ref{Def3} and Theorem \ref{Theo1}.\footnote{Note that the number  of candidates per iteration is constant ($N$), so that $\frac{W_t}{W_{t-1}}=\frac{N\widetilde{w}({\bf v}_j)}{N\widetilde{w}({\x}_{t-1})}=\frac{\widetilde{w}({\bf v}_j)}{\widetilde{w}({\x}_{t-1})}$.} Hence, considering the GIS theory, we can write
\begin{equation}
\label{AlfaPMH2}
\alpha(\x_{t-1},{\bf v}_j) =  \min\left[1, \frac{\widehat{Z}'}{\widehat{Z}_{t-1}}\right]=\min\left[1, \frac{\widetilde{w}({\bf v}_j)}{\widetilde{w}({\x}_{t-1})}\right],
\end{equation}
which has the form of the acceptance function of the classical IMH method in Table \ref{alg:MH}. Therefore, PMH and I-MTM2 algorithms can be also summarized as in Table \ref{alg:PMH_IMTM2}.


 \begin{table}[!h]
	\centering
	\caption{{\bf Alternative description of PMH and I-MTM2}}
	    \begin{tabular}{|p{0.95\columnwidth}|}
		\hline
  \begin{enumerate}
\item Choose an initial state $\x_0$.
\item For $t=1,\ldots,T:$
\begin{enumerate}
\item  Draw $\widetilde{\x}' \sim \widetilde{q}({\bf x})$, where
 \begin{equation}
 \label{EqProp}
\widetilde{q}({\bf x})=\int_{\mathcal{X}^N} \left[\prod_{i=1}^N q({\bf v}_i)\right] \widehat{\pi}({\bf x}|{\bf v}_{1:N}) d{\bf v}_{1:N},
\end{equation} 
is the equivalent proposal pdf associated to a resampled particle \citep{Lamberti16,GISssp16}.
\item Set $\x_t=\widetilde{\x}'$, with probability 
\begin{equation}
\label{AlfaPMH_alternative}
\alpha(\x_{t-1},\widetilde{\x}') = \min\left[1, \frac{\widetilde{w}(\widetilde{\x}')}{\widetilde{w}(\x_{t-1})}\right].
\end{equation}
Otherwise, set $\x_t=\x_{t-1}$.
\end{enumerate}
\item Return $\{\x_t\}_{t=1}^T$.
   \end{enumerate} 	
\\
	\hline
	\end{tabular}		
	\label{alg:PMH_IMTM2}
\end{table}

\begin{figure}[htbp]
\centering
\centerline{
\includegraphics[width=15cm]{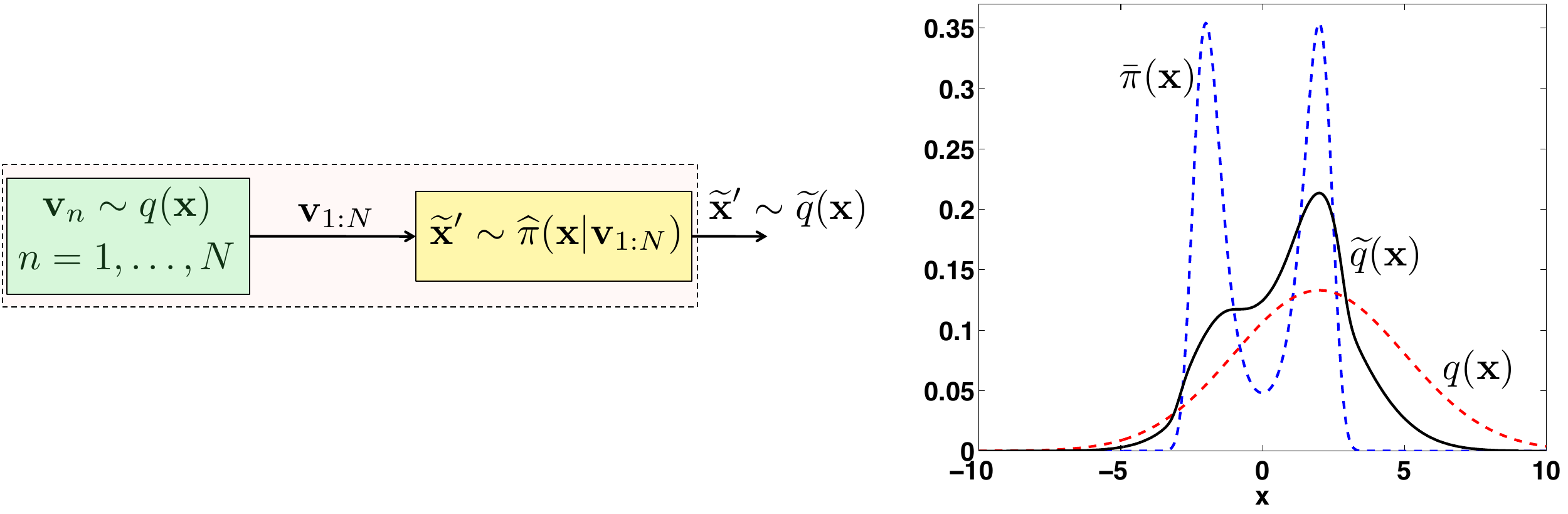} 
}
\caption{{\bf (Left)} Graphical representation of the generation of one sample ${\bf x}'$ from the equivalent proposal pdf $\widetilde{q}({\bf x})$ in Eq. \eqref{EqProp}.{\bf (Right)} Example of the equivalent density $\widetilde{q}({\bf x})$ (solid line) with $N=2$. The target, $\bar{\pi}(\x)$, and proposal, $q(\x)$, pdfs are shown with dashed lines. See  DEMO-3 at \url{https://github.com/lukafree/GIS.git}. 
 }
\label{FigEqProb}
\end{figure}

{\rem The PMH and I-MTM2 algorithms  
take the form of the classical IMH method employing the equivalent proposal pdf $\widetilde{q}({\bf x})$ in Eq. \eqref{EqProp} (depicted in Figure \ref{FigEqProb}; see also  \ref{WeightingAResPar}), and using the GIS weight $\widetilde{w}(\widetilde{\x}')$ of a resampled particle $\widetilde{\x}' \sim \widetilde{q}({\bf x})$, within the acceptance function $\alpha(\x_t,\widetilde{\x}')$.
}

\section{Novel MCMC techniques based on GIS}
 \label{NoTech}

 In this section, 
 we provide two examples of novel MCMC algorithms based on GIS. First of all, we introduce a Metropolis-type method producing a chain of sets of weighted samples. Secondly, we present a PMH technique driven by $M$ parallel particle filters. In the first scheme, we exploit the concept of summary weight and all the weighted samples are stored. In the second one,  both concepts of summary weight and summary particle are used. The consistency of the resulting estimators and the ergodicity of both schemes is ensured and discussed. 

\subsection{Group Metropolis Sampling}
 \label{GMS} 
\begin{figure}[htbp]
\centering
\centerline{
\subfigure[]{\includegraphics[width=10cm]{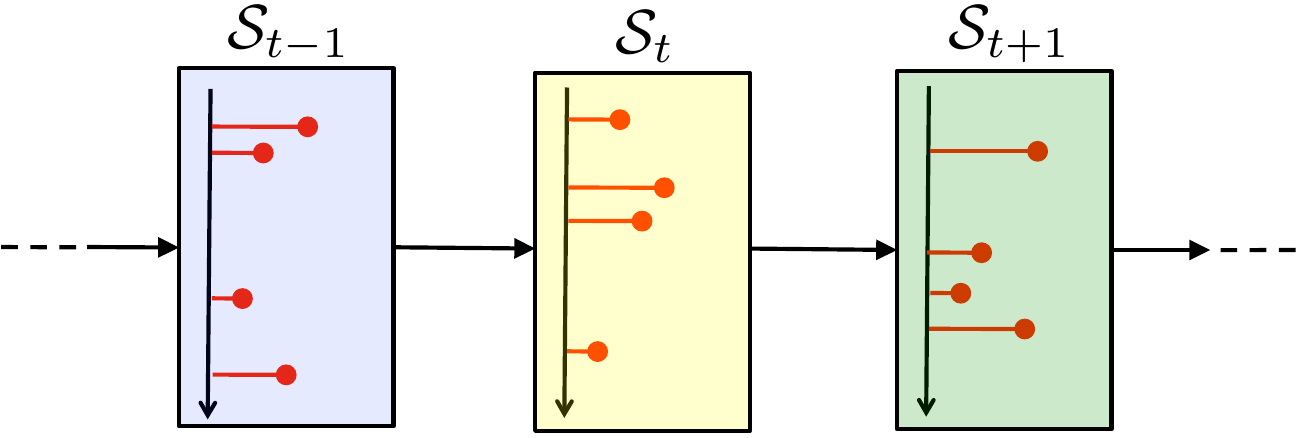} }
\hspace{1cm}
\subfigure[]{\includegraphics[width=6cm]{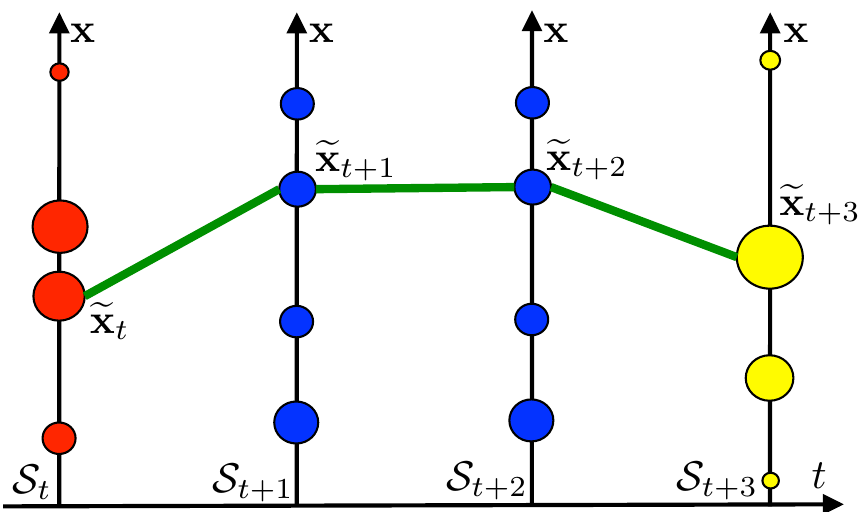} }}
\caption{{\bf (a)} Chain of sets $\mathcal{S}_t=\{\x_{n,t},\rho_{n,t}\}_{n=1}^N$ generated by the GMS method (graphical representation with $N=4$). {\bf (b)} Graphical examples of GMS outputs, $\mathcal{S}_t$, $\mathcal{S}_{t+1}$, $\mathcal{S}_{t+2}$ and $\mathcal{S}_{t+3}$, where $\mathcal{S}_{t+2}=\mathcal{S}_{t+1}$. The weights of the samples are denoted by the size of the circles.  A possible recovered MTM chain is also depicted with solid line, where the states are ${\bf \widetilde{x}}_\tau$ with $\tau=t,t+1,t+2, t+3$ and ${\bf \widetilde{x}}_{t+2}={\bf \widetilde{x}}_{t+1}$.
 }
\label{FigGMS}
\end{figure}

Here, we describe an MCMC procedure that yields a sequence of sets of weighted samples. All the samples are then employed for a joint particle approximation of the target distribution. The Group Metropolis Sampling (GMS) is outlined in Table \ref{alg:GMS}. Figures \ref{FigGMS}(a)-(b) give two graphical representations of GMS outputs  $\mathcal{S}_t=\{\x_{n,t}=\v_n,\rho_{n,t}=w_n\}_{n=1}^N$ (with $N=4$ in both cases).
Note that the GMS algorithm uses the idea of summary weight for comparing sets. Let us denote as $\rho_{n,t}$ the importance weights assigned to the samples $\x_{n,t}$ contained in the current set $\mathcal{S}_{t}$.  Given the generated sets $\mathcal{S}_{t}=\{\x_{n,t},\rho_{n,t}\}_{n=1}^N$, for $t=1,\ldots,T$, GMS provides the global particle approximation
\begin{eqnarray}
\widehat{\pi}(\x|\x_{1:N,1:T})&=&\frac{1}{T}\sum_{t=1}^T \sum_{n=1}^N\frac{\rho_{n,t}}{ \sum_{i=1}^N \rho_{i,t}}  \delta(\x-\x_{n,t}), \\
&=&\frac{1}{T}\sum_{t=1}^T \sum_{n=1}^N \bar{\rho}_{n,t}  \delta(\x-\x_{n,t}),
\end{eqnarray}
where $\bar{\rho}_{n,t}=\frac{\rho_{n,t}}{ \sum_{i=1}^N \rho_{i,t}}$. Thus, the estimator of a specific moment of the target is 
\begin{eqnarray}
\label{Echecazzo}
\widetilde{I}_{NT}=\frac{1}{T}\sum_{t=1}^T \sum_{n=1}^N \bar{\rho}_{n,t} h(\x_{n,t})= \sum_{n=1}^N \bar{\rho}_{n,t} \widetilde{I}_T^{(n)}=\frac{1}{T}\sum_{t=1}^T \widetilde{I}_N^{(t)},  
\end{eqnarray}
 where we have denoted 
 \begin{eqnarray}
\label{Echecazzo2}
 \widetilde{I}_T^{(n)}=\frac{1}{T}\sum_{t=1}^T h(\x_{n,t}), \quad  \widetilde{I}_N^{(t)}=\sum_{n=1}^N \bar{\rho}_{n,t} h(\x_{n,t}).
\end{eqnarray}
See also \ref{ConsistencyGMS_2} for further details.
If the $N$ candidates at step \ref{GMSstep1}, $\v_1,\ldots,\v_N$, and the associated weights, $w_1,\ldots, w_N$, are built sequentially by a particle filtering method, we have a Particle GMS (PGMS) algorithm (see Section \ref{LAI}) and marginal versions can be also considered (see  \ref{MarginalPMH}).

\begin{table}[!h]
	\centering
	\caption{{\bf Group Metropolis Sampling}}
	    \begin{tabular}{|p{0.95\columnwidth}|}
		\hline
  \begin{enumerate}
\item Build an initial set $\mathcal{S}_{0}=\{\x_{n,0},\rho_{n,0}\}_{n=1}^N$ and $\widehat{Z}_{0}=\frac{1}{N}\sum_{n=1}^N \rho_{n,0}$.
\item For $t=1,\ldots,T:$
\begin{enumerate}
\item\label{GMSstep1} Draw $N$ samples, $\v_1,\ldots,\v_N \sim q({\bf x})$ following a sequential or a batch procedure.
\item  Compute the weights, $w_n=\frac{\pi(\v_n)}{q(\v_n)}$, $n=1,\dots,N$; define $\mathcal{S}'=\{\v_n,w_n\}_{n=1}^N$; and compute $\widehat{Z}'=\frac{1}{N}\sum_{n=1}^N w_n$. 
\item Set $\mathcal{S}_t=\{\x_{n,t}=\v_n,\rho_{n,t}=w_n\}_{n=1}^N$, i.e., $\mathcal{S}_{t}=\mathcal{S}'$, and $\widehat{Z}_{t}=\widehat{Z}'$, with probability 
\begin{equation}
\label{AlfaGMS}
\alpha(\mathcal{S}_{t-1},\mathcal{S}') = \min\left[1, \frac{\widehat{Z}'}{\widehat{Z}_{t-1}}\right].
\end{equation}
Otherwise, set $\mathcal{S}_t=\mathcal{S}_{t-1}$ and $\widehat{Z}_{t}=\widehat{Z}_{t-1}$.
\end{enumerate}
\item Return $\{\mathcal{S}_t\}_{t=1}^T$, or $\{\widetilde{I}_N^{(t)}\}_{t=1}^T$ where $\widetilde{I}_N^{(t)}= \sum\limits_{n=1}^N\frac{\rho_{n,t}}{\sum_{i=1}^N \rho_{i,t}}h(\x_{n,t})$. 
   \end{enumerate} 	
\\
	\hline
	\end{tabular}		
	\label{alg:GMS}
\end{table}

\paragraph{Relationship with IMH.} The acceptance probability $\alpha$ in Eq. \eqref{AlfaGMS} is the extension of the acceptance probability of IMH in Eq. \eqref{AlfaMH}, by considering the proper GIS weighting of a set of weighted samples. Note that, in this version of GMS, all the sets contain the same number of samples. 

\paragraph{Relationship with MTM methods.} GMS is strictly related to Multiple Try Metropolis (MTM) schemes \citep{Casarin13,LucaJesse1,LucaJesse2,MTM_PMH14} and Particle Metropolis Hastings (PMH) techniques \citep{PMCMC,MTM_PMH14}. The main difference is that GMS  uses no resampling steps at each iteration for generating summary samples, indeed GMS uses the entire set. 
 However, considering a sequential of a batch procedure for generating the $N$ tries at each iteration, we can recover an  MTM (or the PMH) chain by the GMS output applying one resampling step when $\mathcal{S}_t\neq  \mathcal{S}_{t-1}$,       
\begin{gather}
\widetilde{\x}_{t}= \left\{
\begin{split}
\label{RecChain}
&\widetilde{\v}_{t} \sim \sum_{n=1}^N \bar{\rho}_{n,t}  \delta(\x-\x_{n,t}),   \quad\quad \mbox{ if }  \quad \mathcal{S}_t\neq  \mathcal{S}_{t-1}, \\
&\widetilde{\x}_{t-1}, \quad\quad\quad\quad\quad\quad\quad\quad\quad\mbox{ } \mbox{ if } \quad \mathcal{S}_t=  \mathcal{S}_{t-1},
\end{split}
\right. 
\end{gather}
for $t=1,\ldots,T$. Namely, $\{\widetilde{\x}_t\}_{t=1}^T$ is the chain obtained by one run of the MTM (or PMH) technique. Figure \ref{FigGMS}(b) provides a graphical representation of a MTM chain recovered by GMS outputs.

\paragraph{Ergodicity.} As also discussed above, (a) the sample generation, (b) the acceptance probability function and hence (c) the dynamics of GMS exactly coincides with the corresponding steps of PMH or MTM (with a sequential or batch particle generation, respectively). Hence, the ergodicity of the chain is ensured  \citep{Casarin13,LucaJesse2,PMCMC,MTM_PMH14}. Indeed, we can recover the MTM (or PMH) chain as shown in Eq. \eqref{RecChain}. 

\paragraph{Recycling samples.} The GMS algorithm can be seen as a method of recycling auxiliary weighted samples in MTM schemes (or PMH schemes, if the candidates are generated by SIR). However, GMS does not recycle all the samples generated at the step \ref{GMSstep1} of Table \ref{alg:GMS}. Indeed, when a set is rejected, GMS discards these samples and repeats the previous set. Therefore, GMS also decides which samples will be either recycled or not.  In \citep{Casella96}, the authors show how recycling and including the samples rejected in one run of a standard MH method into a unique consistent estimator. GMS can be considered an extension of this technique where $N\geq 1$ candidates are drawn at each iteration. 

\paragraph{Iterated IS.} GMS can be also interpreted as an iterative importance sampling scheme where an IS approximation of $N$ samples is built at each iteration and compared with the previous IS approximation. This procedure is iterated $T$ times and all the accepted IS estimators $\widetilde{I}_N^{(t)}$ are finally combined to provide a unique global approximation of $NT$ samples. Note that the temporal combination of the IS estimators is obtained dynamically by the random repetitions due to the rejections in the MH test. Hence, the complete procedure for weighting the samples generated by GMS can be interpreted as the composition of two weighting schemes: (a) by an importance sampling approach building $\{\rho_{n,t}\}_{n=1}^N$ and (b) by the possible random repetitions due to the rejections in the MH-type test.  

\paragraph{Connection with dynamic weighting schemes.} For its hybrid nature between an IS method and a MCMC technique,  GMS could recall the dynamic weighting schemes proposed in \citep{DW97}.  The authors in  \citep{DW97}  have proposed different kinds of {\it moves} considering weighted samples, that are  suitable according to the so-called  {\it ``invariance with respect to the importance weights''} condition. However, these moves are completely different from the GMS scheme. The dynamic of GMS is totally based on standard ergodic theory, indeed, we can recover a standard MCMC chain from the GMS output, as shown in Eq. \eqref{RecChain}.  


\paragraph{Consistency of the GMS estimator.}  Recovering the MTM chain $\{\widetilde{\x}_{t}\}_{t=1}^T$ as in Eq. \eqref{RecChain}, the estimator obtained by the recovered chain, $\widetilde{I}_{T}=\frac{1}{T}\sum_{t=1}^T h(\widetilde{\x}_{t})$, is consistent. Namely, $\widetilde{I}_{T}$ converges almost-surely to $I=E_{\bar{\pi}}[h(\x)]$ as $T\rightarrow \infty$, since $\{\widetilde{\x}_{t}\}_{t=1}^T$ is an ergodic chain \citep{Robert04}.\footnote{The estimator $\widetilde{I}_{T}=\frac{1}{T}\sum_{t=1}^T h(\widetilde{\x}_{t})$ considers only the samples $\{\widetilde{\x}_{t}\}_{t=1}^T$ obtained after applying several resampling steps and thus recovering a MTM chain, following Eq. \eqref{RecChain}. Observe also that, unlike the previous estimator, $\widetilde{I}_{T}^{(n)}$ in Eq. \eqref{Echecazzo2} considers all the GMS output samples $\x_{n,t}$'s, and then each of $\widetilde{I}_{T}^{(n)}$ is weighted according to the corresponding weight $ \bar{\rho}_{n,t}$, in order to provide the final complete estimator $\widetilde{I}_{NT}$, as shown in Eq. \eqref{Echecazzo}. }
For $\mathcal{S}_t\neq\mathcal{S}_{t-1}$, note that
$E_{\widehat{\pi}}[h(\widetilde{\x}_{t})|\mathcal{S}_{t}]= \sum_{n=1}^N \bar{\rho}_{n,t} h(\x_{n,t})=\widetilde{I}_N^{(t)}$ in Eq. \eqref{Echecazzo2}, 
where $\widehat{\pi}(\x|\x_{1:N,t})=\sum_{n=1}^N \bar{\rho}_{n,t}  \delta(\x-\x_{n,t})$. If $\mathcal{S}_t=\mathcal{S}_{t-1}$, then $E_{\widehat{\pi}}[h(\widetilde{\x}_{t})|\mathcal{S}_{t}]=E_{\widehat{\pi}}[h(\widetilde{\x}_{t-1})|\mathcal{S}_{t-1}]=\widetilde{I}_N^{(t-1)}$ and, since $\widetilde{I}_N^{(t)}=\widetilde{I}_N^{(t-1)}$, we have again $E_{\widehat{\pi}}[h(\widetilde{\x}_{t})|\mathcal{S}_{t}]=\widetilde{I}_N^{(t)}$. Therefore, we have
  \begin{eqnarray}
  E[\widetilde{I}_{T}|\mathcal{S}_{1:T}]&=&\frac{1}{T}\sum_{t=1}^T E_{\widehat{\pi}}[h(\widetilde{\x}_{t})|\mathcal{S}_{t}] \\
  &=&\frac{1}{T}\sum_{t=1}^T \widetilde{I}_N^{(t)}=\widetilde{I}_{NT},
\end{eqnarray}
whee the last equality comes from Eq. \eqref{Echecazzo}. Thus, the GSM estimator $\widetilde{I}_{NT}$ in Eq. \eqref{Echecazzo} can be expressed as
$\widetilde{I}_{NT}=E[\widetilde{I}_{T}\big|\mathcal{S}_{1:T}]$, where $\mathcal{S}_{1:T}$ represents all the weighted samples obtained by GMS and  $\widetilde{I}_{T}=\frac{1}{T}\sum_{t=1}^T h(\widetilde{\x}_{t})$ is the estimator obtained by a given MTM chain recovered by using Eq. \eqref{RecChain}. 
Hence, $\widetilde{I}_{NT}$ is consistent for $T\rightarrow \infty$ since $\widetilde{I}_{T}$ is consistent, owing to the MTM chain is ergodic. Furthermore, fixing $T$, the GMS estimator $\widetilde{I}_{NT}$ in Eq. \eqref{Echecazzo} is also consistent when $N\rightarrow \infty$, due to the standard IS arguments \citep{Liu04b}. The consistency can be also shown considering GMS as the limit case of an in finite number of recovered parallel IMTM2 chains, as in Eq. \eqref{RecChain}, as shown in  \ref{ConsistencyGMS_2}. 



\subsection{Distributed Particle Metropolis-Hastings algorithm}
 \label{P-PF-PMH}
 
 The PMH algorithm is an MCMC technique particularly designed for filtering and smoothing a dynamic variable in a state-space model \citep{PMCMC,MTM_PMH14} (see for instance Figure \ref{Fig2_Ex3}). 
In PMH, different trajectories obtained by different runs of a particle filter (see Section \ref{GIS_SIR}) are compared according to suitable MH-type acceptance probabilities, as shown in Table \ref{alg:IMTM2}. 
In this section, we show how several parallel particle filters (for instance, each one consider a different proposal pdf) can drive a PMH-type technique.

\begin{figure}[htbp]
\centering
\centerline{
\subfigure[]{\includegraphics[width=9cm]{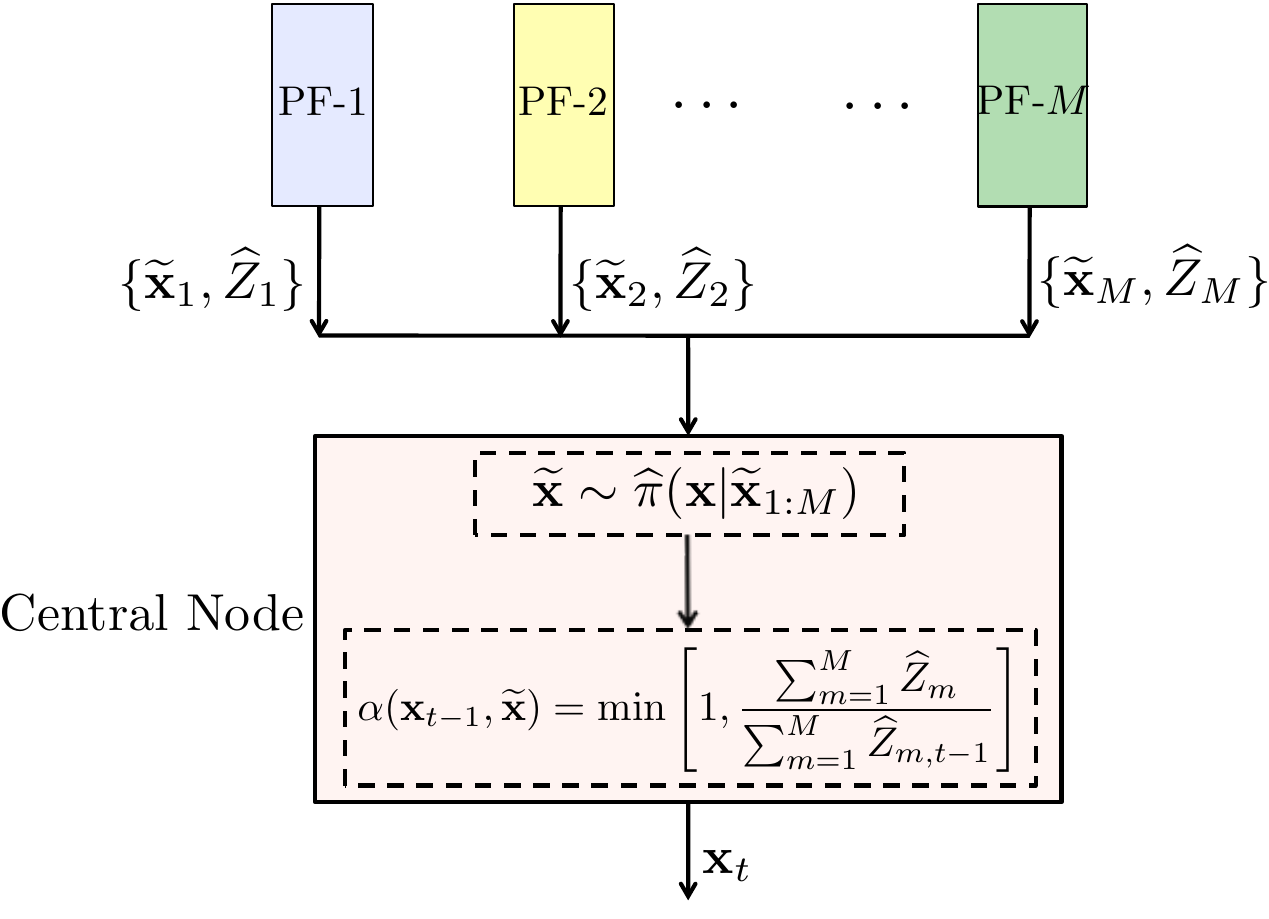} }
\subfigure[]{\includegraphics[width=9.1cm]{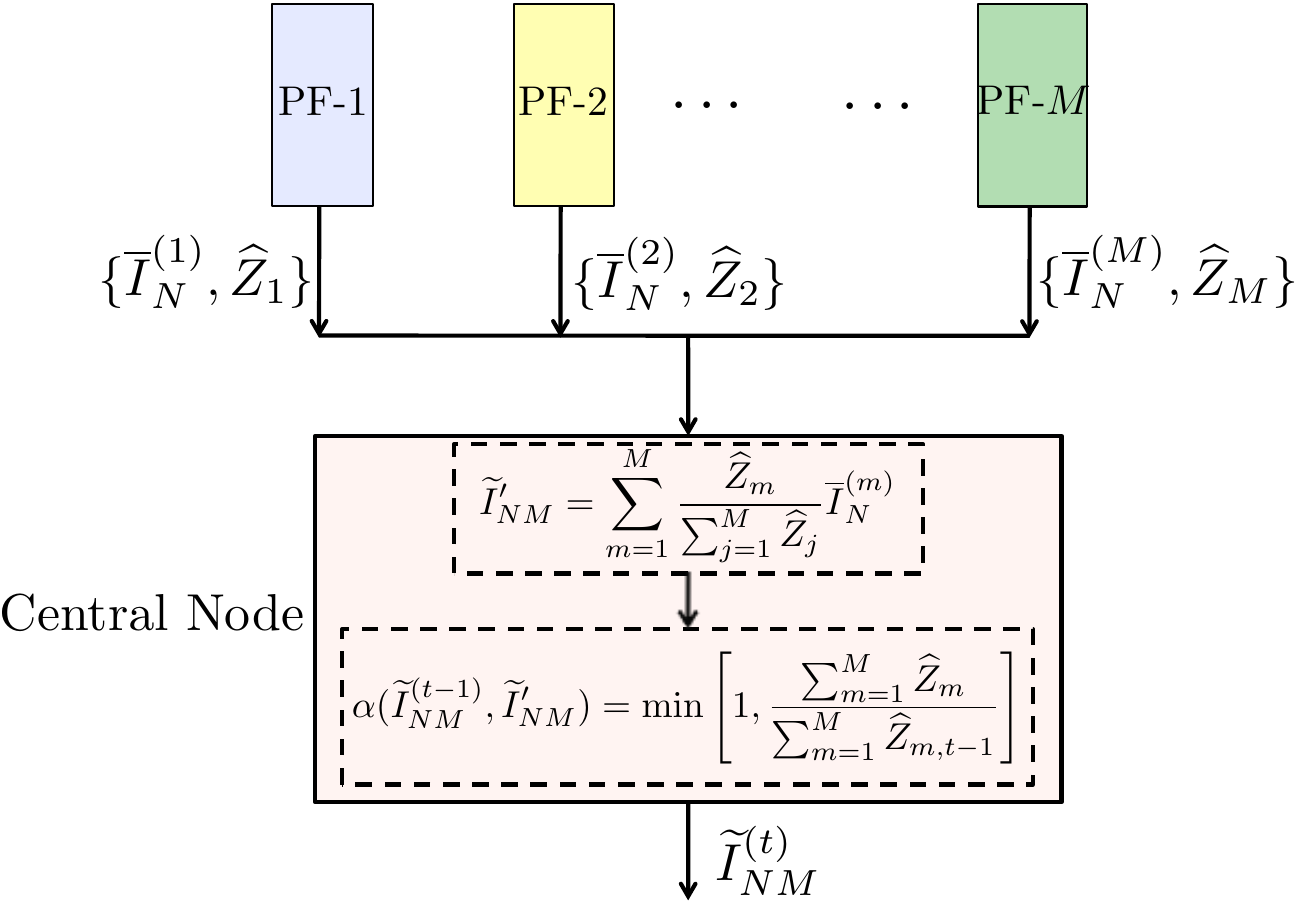} }
}
\caption{Graphical representation of Distributed Particle Metropolis-Hastings (DPMH) method, {\bf(a)} for estimating a generic moment, or {\bf (b)} for estimating of a specific moment of the target.
 }
\label{FigParallelPMH}
\end{figure}

\noindent
The classical PMH method uses a single proposal pdf $q(\x)=q_{1}(x_1)\prod_{d=2}^D q_{d}(x_d|x_{1:d-1})$, employed in single SIR method in order to generate new candidates before of the MH-type test (see Table \ref{alg:IMTM2}). Let us consider the problem of tracking a variable of interest $\x=[x_1,\ldots,x_D]^{\top}\in \mathbb{R}^{D\times \xi}$ with target pdf $\pi(\x)=\pi_1(x_1)\prod_{d=2}^D \pi_d(x_d|x_{1:d-1})$. We assume that $M$ independent processing units are available jointly with a central node as shown Fig. \ref{FigParallelPMH}. We use $M$ parallel particle filters, each one with a different proposal pdf,   $q_m(\x)=q_{m,1}(x_1)\prod_{d=2}^D q_{m,d}(x_d|x_{1:d-1})$, one per each processor. Then, after one run of the parallel particle filters, we obtain $M$ particle approximations $\widehat{\pi}_m(\x)$. Since, we aim to reduce the communication cost to the central node (see Figs. \ref{Fig2} and \ref{FigParallelPMH}), we consider that each machine only transmits the pair $\{\widehat{Z}_m, \widetilde{\x}_m\}$, where $\widetilde{\x}_m\sim \widehat{\pi}_m(\x)$ (we set $N_1=\ldots=N_M$, for simplicity). 
 Applying the GIS theory, then it is straightforward to outline the method, called  {\it Distributed Particle Metropolis-Hastings} (DPMH) technique, shown in Table \ref{alg:M_PMH}.

\begin{table}[!h]
	\centering
	\caption{{\bf Distributed Particle Metropolis-Hastings algorithm}}
	    \begin{tabular}{|p{0.95\columnwidth}|}
		\hline
  \begin{enumerate}
\item Choose an initial state $\x_0$ and $\widehat{Z}_{m,0}$ for $m=1,\ldots,M$ (e.g., both obtained with a first run of a particle filter).
\item For $t=1,\ldots,T:$
\begin{enumerate}
\item\label{stepONE} {\bf(Parallel Processors)} Draw $N$ particles $\v_{m,1},\ldots,\v_{m,N}$ from $q_m(\x)$ and weight them with IS weights $w(\v_{m,n})$, $n=1,\ldots,N$, using a particle filter (or a batch approach), for each $m=1,\ldots,M$. Thus, denoting $\widehat{Z}_m=\frac{1}{N}\sum_{n=1}^N w(\v_{m,n})$, we obtain the $M$ particle approximations
$\widehat{\pi}_m(\x)=\widehat{\pi}(\x|\v_{m,1:N})=\frac{1}{N \widehat{Z}_m}\sum_{n=1}^N w(\v_{m,n}) \delta(\x-\v_{m,n})$.
\item\label{stepMRes} {\bf(Parallel Processors)} Draw $\widetilde{\x}_m\sim \widehat{\pi}(\x|\v_{m,1:N})$, for $m=1,\ldots,M$.
\item\label{stepRestot}  {\bf(Central Node)} Resample $\widetilde{\x}\in \{\widetilde{\x}_1,\ldots,\widetilde{\x}_M\}$ according to the pmf $\frac{\widehat{Z}_m}{\sum_{j=1}^M \widehat{Z}_j}$, $m=1,\ldots,M$, i.e., $\widetilde{\x}\sim \widehat{\pi}({\bf x}| \widetilde{\x}_1,\ldots, \widetilde{\x}_M)$.
\item  {\bf(Central Node)} Set $\x_t=\widetilde{\x}$ and $\widehat{Z}_{m,t}=\widehat{Z}_m$, for $m=1,\ldots,M$, with probability 
\begin{equation}
\label{AlfaPMH3}
\alpha(\x_{t-1},\widetilde{\x}) = \min\left[1, \frac{\sum_{m=1}^M\widehat{Z}_m}{\sum_{m=1}^M\widehat{Z}_{m,t-1}}\right].
\end{equation}
Otherwise, set $\x_t=\x_{t-1}$ and $\widehat{Z}_{m,t}=\widehat{Z}_{m,t-1}$, for $m=1,\ldots,M$.
\end{enumerate}
   \end{enumerate} 	
\\
	\hline
	\end{tabular}		
	\label{alg:M_PMH}
\end{table}

The method in Table \ref{alg:M_PMH} has the structure of a Multiple Try Metropolis (MTM) algorithm using different proposal pdfs \citep{Casarin13,LucaJesse2}. More generally, in step \ref{stepONE}, the scheme described above can even employ different kinds of particle filtering algorithms.
In step \ref{stepMRes}, $M$ total resampling steps are performed,one per processor. Then, one resampling step is performed in the central node (step \ref{stepRestot}). Finally, the resampled particle is accepted as new state with probability $\alpha$ in Eq. \eqref{AlfaPMH3}.

\paragraph{Ergodicity.} The ergodicity of DPMH is ensured since it can be interpreted as a standard PMH method considering a single particle approximation\footnote{This particle approximation can be interpreted as being obtained by a single particle filter splitting the particles in $M$ disjoint sets and then applying the partial resampling described in Section \ref{GIS_SIR}, i.e., performing resampling steps within the sets. See also Eq. \eqref{EqAppPartApprox}.} 
\begin{equation}
\widehat{\pi}(\x|\v_{1:M,1:N})=\sum_{m=1}^M\frac{\widehat{Z}_m}{\sum_{j=1}^M \widehat{Z}_j} \widehat{\pi}(\x|\v_{m,1:N})=\sum_{m=1}^M \overline{W}_m \widehat{\pi}(\x|\v_{m,1:N}),
\end{equation}
 and then we resample once, i.e.,  draw $\widetilde{\x}\sim \widehat{\pi}(\x|\v_{1:M,1:N})$. Then, the proper weight of this resampled particle is $\widehat{Z}=\frac{1}{M}\sum_{m=1}^M \widehat{Z}_m$, so that the acceptance function of the equivalent classical PMH method is $\alpha(\x_{t-1},\widetilde{\x})=\min\left[1,\frac{\widehat{Z}}{\widehat{Z}_{t-1}}\right]=\min\left[1, \frac{\frac{1}{M}\sum_{m=1}^M\widehat{Z}_m}{\frac{1}{M}\sum_{m=1}^M\widehat{Z}_{m,t-1}}\right]$, where $\widehat{Z}_{t-1}=\frac{1}{M}\sum_{m=1}^M\widehat{Z}_{m,t-1}$ (see Table \ref{alg:IMTM2}).
  \paragraph{Using partial IS estimators.} If we are interested in approximating only one moment of the target pdf, as shown in Figures  \ref{Fig2}-\ref{FigSummary}, at each iteration we can transmit the $M$ partial estimators $\overline{I}_{N}^{(m)}$ and  combine them  in the central node as in Eq. \eqref{EqAllSETS3}, obtaining $\widetilde{I}_{NM}'=\frac{1}{\sum_{j=1}^M\widehat{Z}_{j}}\sum_{m=1}^M \widehat{Z}_{m}  {\overline I}_{N}^{(m)}$. Then, a sequence of estimators, $\widetilde{I}_{NM}^{(t)}$, is created according to the acceptance probability $\alpha$ in Eq. \eqref{AlfaPMH3}. Finally, we obtain the global estimator
 \begin{equation}
\widetilde{I}_{NMT}=\frac{1}{T}\sum_{t=1}^T \widetilde{I}_{NM}^{(t)}.
\end{equation}
 This scheme is depicted in Figure \ref{FigParallelPMH}(b).

 \paragraph{Benefits.} The main advantage of the DPMH scheme is that the generation of samples can be parallelized (i.e., fixing the computational cost, DPMH allows the use of $M$  processors in parallel) and the communication to the central node requires the transfer of only $M$ particles, $\widetilde{\x}_m'$, and $M$ weights, $\widehat{Z}_m'$, instead of $NM$ particles and $NM$ weights.  Figure \ref{FigParallelPMH} provides a general sketch of DPMH.  Its marginal version is described in  \ref{MarginalPMH}. Another benefit of DPMH is that different types of particle filters can be jointly employed, for instance, different proposal pdfs can be used. 
 
 \paragraph{Special cases and extensions.} The classical PMH method is  as a special case of the proposed algorithm of Table \ref{alg:M_PMH} when $M=1$. If the partial estimators are transmitted to the central node, as shown in Figure \ref{FigParallelPMH}(b), DPMH coincides with PGMS when $M=1$.  Adaptive versions of DPMH can be designed in order select automatically the best proposal pdf among the $M$ densities, based of the weights $\overline{W}_m =\frac{\widehat{Z}_m}{\sum_{j=1}^M \widehat{Z}_j}$, $m=1,\ldots,M$. For instance, Figure \ref{Fig1_Ex3}(b) shows that DPMH is able to detect the best scale parameters within the $M$ used values. 



\section{Numerical Experiments}
\label{Simu}

In this section, we test the novel techniques considering several experimental scenarios and three different applications: hyperparameters estimation for Gaussian Processes  ($D=2$), two different localization problems,\footnote{The first problem also provides the automatic tuning of the sensor network ($D=8$), whereas the second one is a bidimensional positioning problem considering real data ($D=2$).} and the online filtering of a remote sensing variable called  Leaf Area Index  (LAI; $D=365$). We compare the novel algorithms with different benchmark methods such adaptive MH algorithm, MTM and PMH techniques, parallel MH chains with random walk proposal pdfs, IS schemes, and the Adaptive Multiple Importance Sampling (AMIS) method. 

\subsection{Hyperparameter tuning for Gaussian Process (GP) regression models}
\label{GPexample}

We test the proposed GMS approach for the estimation of hyperparameters of a Gaussian process (GP) regression model~\citep{Bishop}, \citep{rasmussen2006gaussian}. 
Let us assume observed data pairs $\{y_j,{\bf z}_j\}_{j=1}^{P}$, with $y_j\in \mathbb{R}$ and ${\bf z}_j\in \mathbb{R}^{L}$. We also denote the corresponding $P\times 1$ output vector as ${\bf  y}=[y_1,\ldots,y_P]^{\top}$ and the $L\times P$ input matrix as ${\bf  Z}=[{\bf z}_1,\ldots,{\bf z}_P]$. We address the regression problem of inferring the unknown function $f$ which links the variable $y$ and ${\bf z}$. Thus, the assumed model is $y=f({\bf z})+e$,
where $e\sim N(e;0,\sigma^2)$, and that $f({\bf z})$ is a realization of a GP~\citep{rasmussen2006gaussian}. Hence $f({\bf z}) \sim \mathcal{GP}(\mu({\bf z}),\kappa({\bf z},{\bf r}))$ where $\mu({\bf z})=0$, ${\bf z},{\bf r} \in \mathbb{R}^{L}$, and we consider the kernel function 
\begin{equation}
\label{EqKernel}
\kappa({\bf z},{\bf r})=\exp\left(-\sum_{\ell=1}^{L}\frac{(z_\ell-r_\ell)^2}{2\delta^2}\right).
\end{equation}
Given these assumptions, the vector ${\bf  f}=[f({\bf z}_1),\ldots, f({\bf z}_P)]^\top$ is distributed as 
$p({\bf  f}|{\bf  Z},\delta, \kappa)=\mathcal{N}({\bf  f};{\bf  0},{\bf  K})$,
where ${\bf  0}$ is a $P\times 1$ null vector, and ${\bf  K}_{ij}:=\kappa({\bf z}_i,{\bf z}_j)$, for all $i,j=1,\ldots,P$, is a $P\times P$ matrix.
 Therefore, the vector containing all the hyperparameters of the model is
$\x=[\delta, \sigma]$,
i.e., all the parameters of the kernel function in Eq.~\eqref{EqKernel} and standard deviation $\sigma$ of the observation noise. In this experiment, we focus on the marginal posterior density of the hyperparameters,
${\bar \pi}(\x|{\bf  y}, {\bf  Z}, \kappa)\propto \pi(\x|{\bf  y}, {\bf  Z}, \kappa)=p({\bf  y}|\x, {\bf  Z}, \kappa) p(\x)$,
which can be evaluated analytically, but we cannot compute integrals involving it \citep{rasmussen2006gaussian}. Considering a uniform prior within $[0,20]^2$, $p(\x)$ and since $p({\bf  y}|\x, {\bf  Z}, \kappa)=\mathcal{N}({\bf  y};{\bf  0},{\bf K}+\sigma^2 {\bf  I})$, we have 
\begin{gather}
\begin{split}
\log \left[\pi(\x|{\bf  y}, {\bf  Z}, \kappa)\right] = -\frac{1}{2} {\bf  y}^{\top} ({\bf  K}+\sigma^2 {\bf  I})^{-1} {\bf  y} -\frac{1}{2} \log\left[\mbox{det} \left({\bf  K} +\sigma^2 {\bf  I}\right)\right],  
\end{split}
\end{gather}
 where clearly ${\bf  K}$ depends on $\delta$~\citep{rasmussen2006gaussian}. The moments of this marginal posterior cannot be computed analytically. Then, in order to compute the Minimum Mean Square Error (MMSE) estimator $\widehat{\x}=[\widehat{\delta},\widehat{\sigma}]$, i.e., the expected value $E[{\bf X}]$ with 
${\bf X} \sim {\bar \pi}(\x|{\bf  y}, {\bf  Z}, \kappa)$, we approximate $E[{\bf X}]$ via Monte Carlo quadrature. More specifically, we apply a the novel GMS technique and compare with an MTM sampler, a MH scheme with a longer chain and a static IS method. For all these methodologies, we consider the same number of target evaluations, denoted as $E$, in order to provide a fair comparison. 

\begin{figure*}[htbp]
\begin{center}
\centerline{
\subfigure[]{\includegraphics[width=0.45\textwidth]{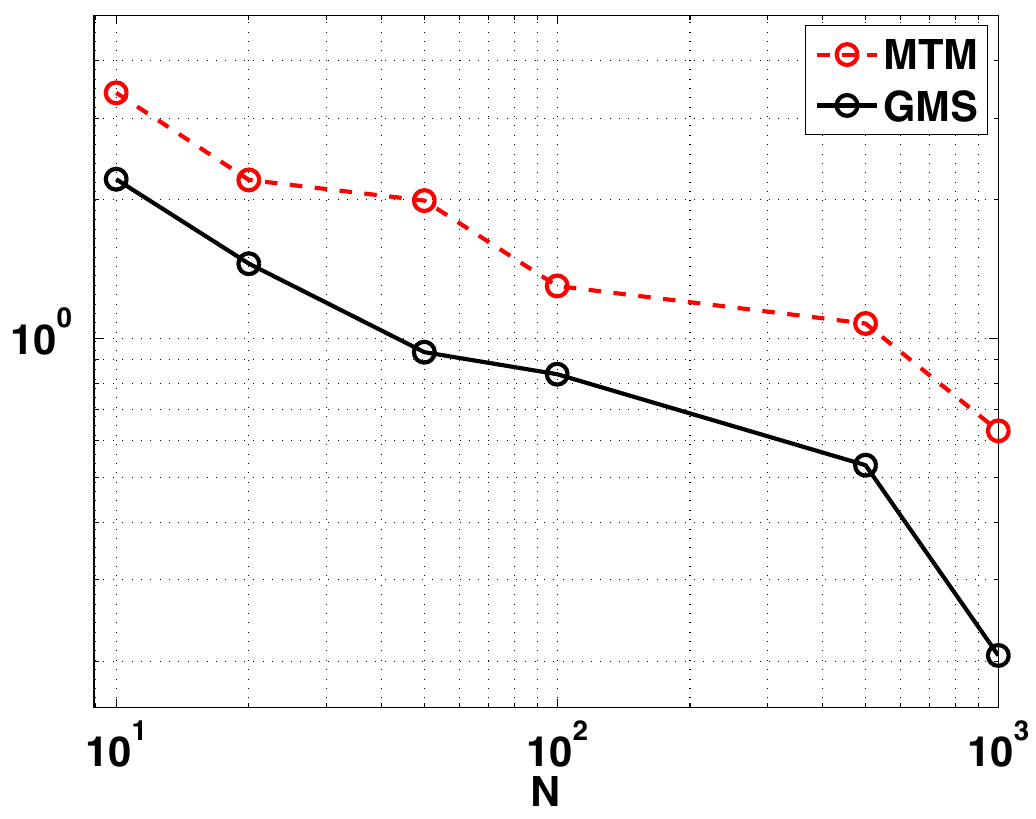}}
\subfigure[]{\includegraphics[width=0.45\textwidth]{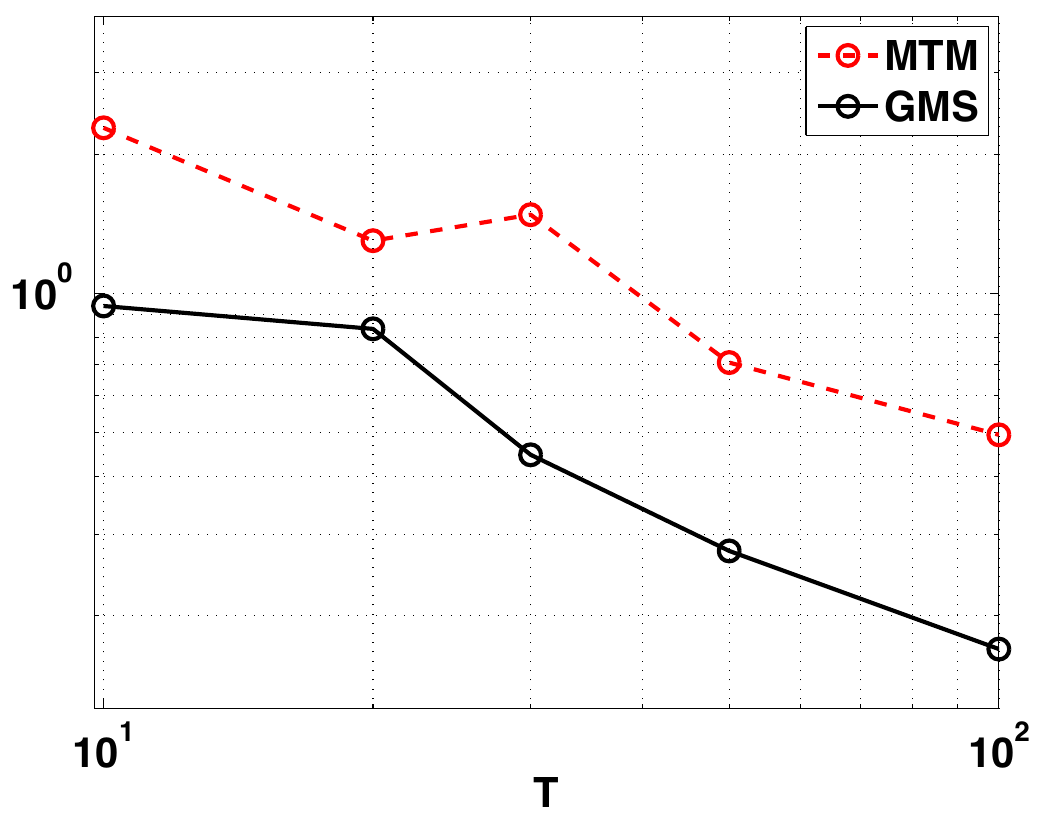}}
}
\vspace{-0.25cm}
\caption{MSE (loglog-scale; averaged over $10^3$ independent runs) obtained with the MTM and GMS algorithms (using the same proposal pdf and the same values of $N$ and $T$) {\bf (a)} as function of $N$ with $T=20$ and {\bf (b)} as function of $T$ with $N=100$. 
}
\label{FigSIMU}
\end{center}
\end{figure*}

\begin{figure*}[htbp]
\begin{center}
\centerline{
\subfigure[]{\includegraphics[width=0.45\textwidth]{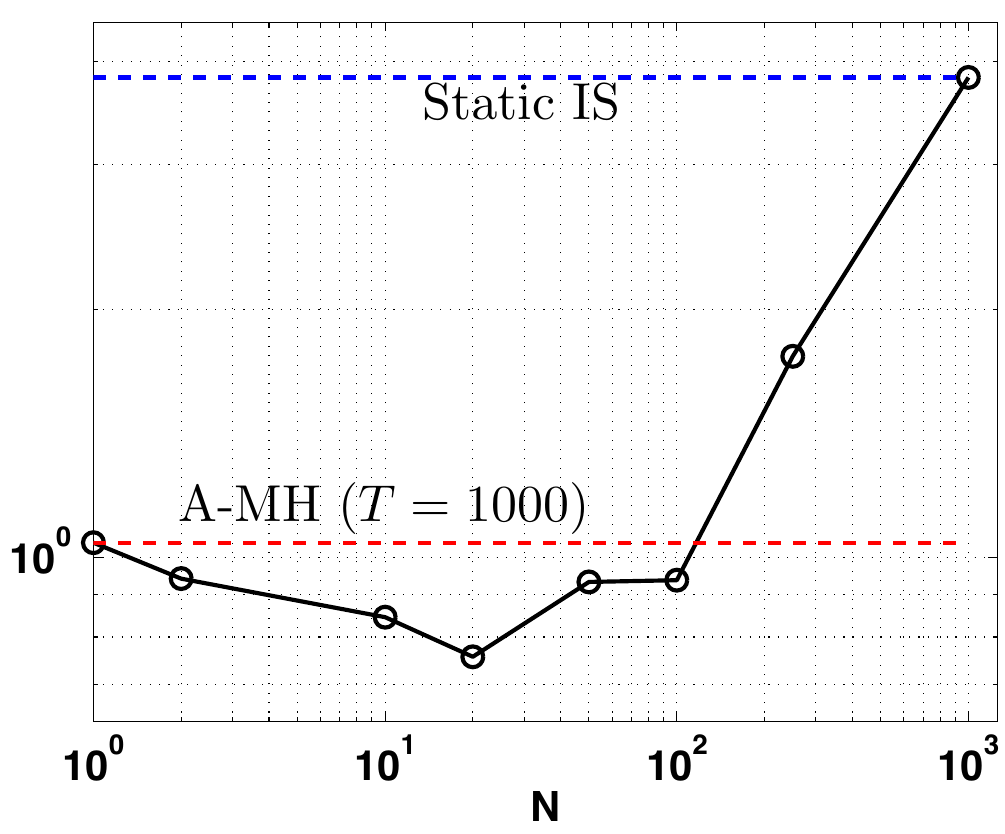}}
\hspace{0.3cm}
\subfigure[]{\includegraphics[width=0.47\textwidth]{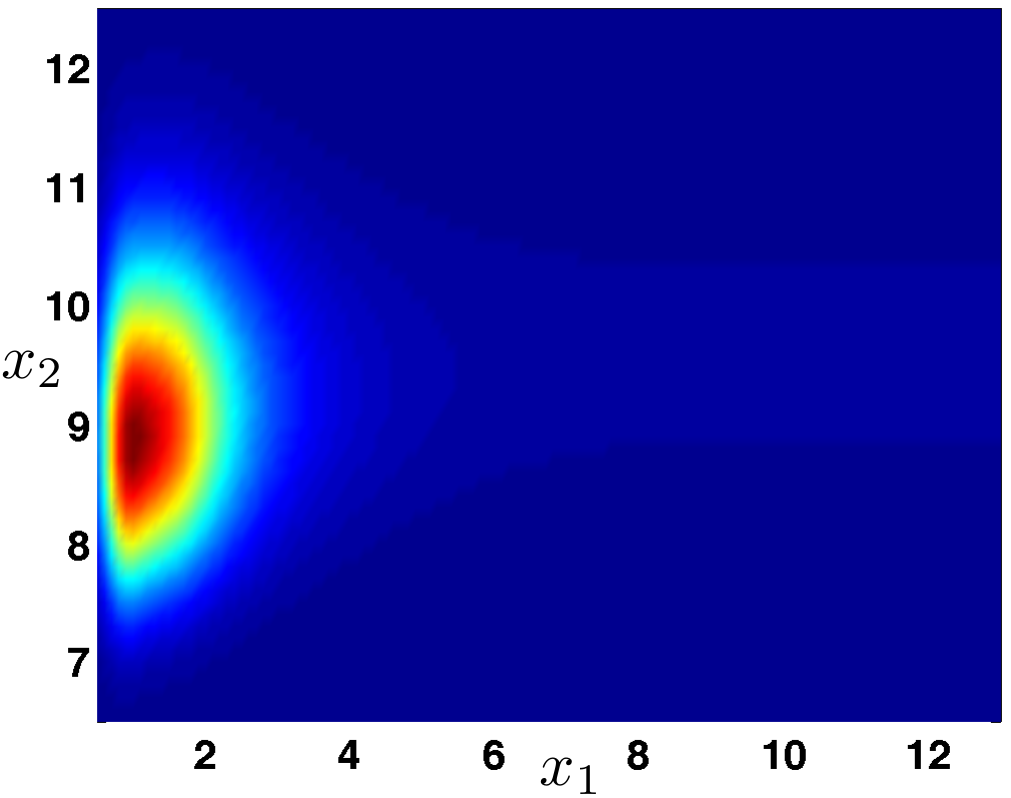}}
}
\vspace{-0.25cm}
\caption{{\bf (a)} MSE (loglog-scale; averaged over $10^3$ independent runs) of GMS (circles) versus the number of candidates $N\in\{1,2,10,20,50,100,250,10^3\}$, but keeping fixed the total number of posterior evaluations $E=NT=1000$, so that $T\in\{1000,500,100,50,20,10,4,1\}$. The MSE values the extreme cases $N=1$, $T=1000$, and $N=1000$, $T=1$, are depicted with dashed lines. In first case, GMS coincides with an adaptive MH scheme (due the adaptation of the proposal, in this example) with a longer chain. The second one represents a static IS scheme (clearly, using the sample proposal than GMS). We can observe the benefit of the dynamic combination of IS estimators obtained by GMS. {\bf (b)} Posterior density $\pi(\x|{\bf  y}, {\bf  Z}, \kappa)$.  
}
\label{FigSIMU2}
\end{center}
\end{figure*}

We generated $P=200$ pairs of data, $\{y_j,{\bf z}_j\}_{j=1}^{P}$, according to the GP model above setting $\delta^*=3$, $\sigma^*=10$, $L=1$, and drawing $z_j\sim\mathcal{U}([0,10])$. We keep fixed these data over the different runs, and the corresponding posterior pdf is given in Figure \ref{FigSIMU2}(b). We computed the ground-truth $\widehat{\x}= [\widehat{\delta}= 3.5200,\widehat{\sigma}= 9.2811]$ using an exhaustive and costly grid approximation, in order to compare the different techniques. 
For both GMS, MTM  and MH schemes, we consider the same adaptive Gaussian proposal pdf $q_t({\bf x}|{\bm \mu}_t,\lambda^2 {\bf I})=\mathcal{N}({\bf x}|{\bm \mu}_t,\lambda^2 {\bf I})$, with $\lambda=5$ and ${\bm \mu}_t$ is adapted considering the arithmetic mean of the outputs after a training period, $t\geq 0.2 T$, in the same fashion of \citep{Haario01} (${\bm \mu}_0=[1,1]^{\top}$).  
First, we test both techniques fixing $T=20$ and varying the number of tries $N$. Then, we set $N=100$ and vary the number of iterations $T$. Figure \ref{FigSIMU} (log-log plot) shows the Mean Square Error (MSE) in the approximation of $\widehat{\x}$ averaged over $10^3$ independent runs. Observe that GMS always outperforms the corresponding MTM scheme. These results confirm the advantage of recycling the auxiliary samples drawn at each iteration during an MTM run.  In Figure \ref{FigSIMU2}(a), we show the MSE obtained by GMS keeping invariant the number of target evaluations $E=NT=10^3$ and varying $N\in\{1,2,10,20,50,100,250,10^3\}$. As a consequence, we have $T\in\{10^3,500,100,50,20,10,4,1\}$. Note that the case  $N=1$, $T=10^3$, corresponds to an adaptive MH (A-MH) method with a longer chain, whereas the case $N=10^3$, $T=1$, corresponds to a static IS scheme (both with the same posterior evaluations $E=NT=10^3$). We observe that the GMS always provides smaller MSE than the static IS approach. Moreover, GMS outperforms A-MH with the exception of two cases where $T\in\{1,4\}$.

\subsection{Localization of a target and tuning of the sensor network}
\begin{figure*}[htbp]
\begin{center}
\centerline{
\subfigure[]{\includegraphics[width=0.40\textwidth]{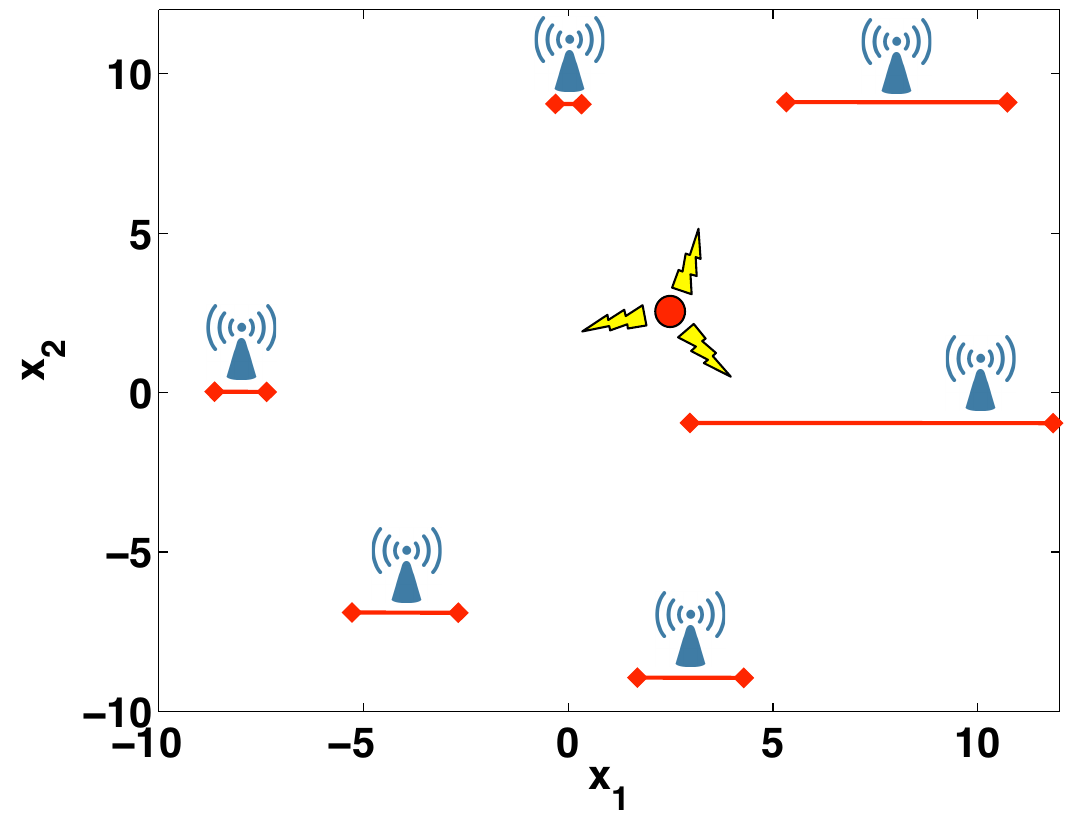}}
\subfigure[]{\includegraphics[width=0.39\textwidth]{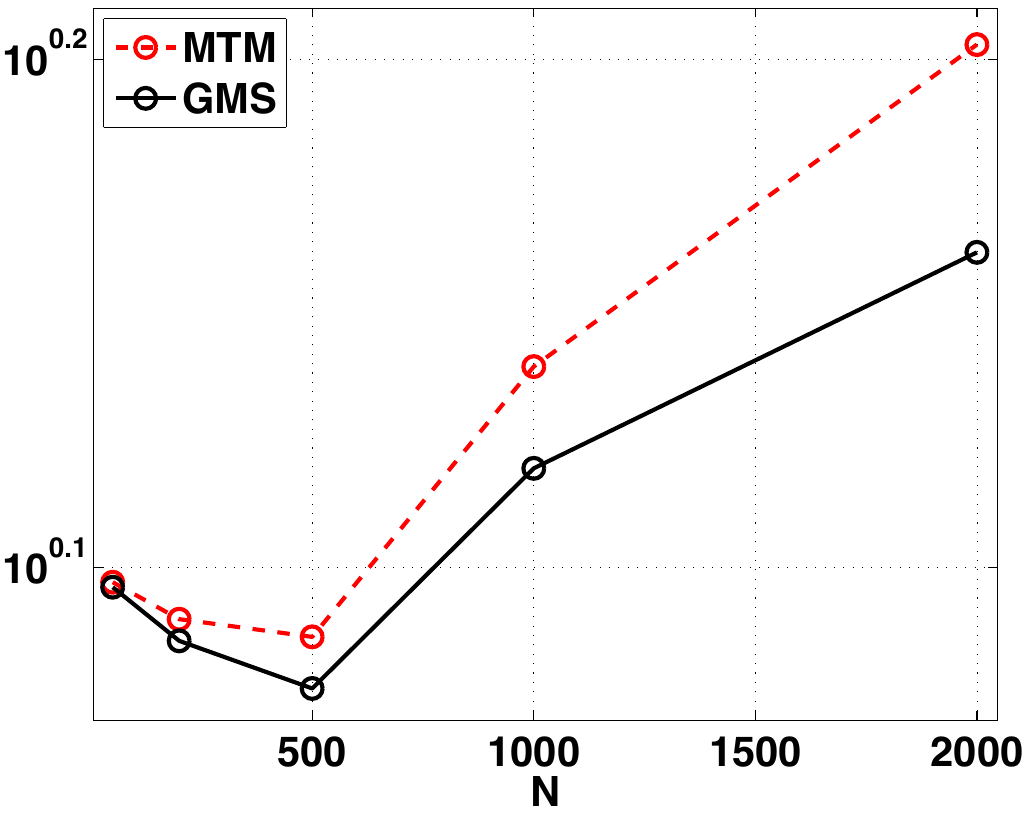}}
}
\vspace{-0.25cm}
\caption{{\bf(a)} Sensor network: the location of the sensors (antennas) and the target (circle) in the numerical example. The solid line represents the different unknown variances of the sensors. {\bf(b)} MSE (log-scale) versus the number of candidates $N\in\{50,200, 500, 1000, 2000\}$ obtained by GMS and the corresponding MTM algorithm, keeping fixed the total number of evaluations $E=NT=10^4$ of the posterior pdf, so that $T\in\{200,50,20,10,5\}$. 
}
\label{FigSIMU_Ex2}
\end{center}
\end{figure*}

We consider the problem of positioning a target in $\mathbb{R}^{2}$ using a range measurements in a wireless sensor network \citep{Ali07,Ihler05}.  
We also assume that the measurements are contaminated by noise with different unknown power, one per each sensor. This situation is common in several practical scenarios. Actually, even if the sensors have the same construction features, the noise perturbation of each the sensor can vary with the time and depends on the location of the sensor. This occurs owing to different causes:  manufacturing defects, obstacles in the reception, different physical environmental conditions (such as humidity and temperature) etc. Moreover, in general, these conditions change along time, hence it is necessary that the central node of the network is able to re-estimate the noise powers jointly with position of the target (and other parameters of the models if required) whenever a new block of observations is processed. More specifically, let us denote the target position with the random vector $\textbf{Z}=[Z_1,Z_2]^{\top}$. The position of the target is then a specific realization ${\bf Z}={\bf z}$. 
The range measurements are obtained from $N_S=6$ sensors located at $\textbf{h}_1=[3, -8]^{\top}$, $\textbf{h}_2=[8,10]^{\top}$, $\textbf{h}_3=[-4,-6]^{\top}$, $\textbf{h}_4=[-8,1]^{\top}$, $\textbf{h}_5=[10,0]^{\top}$ and $\textbf{h}_6=[0,10]^{\top}$ as shown in Figure \ref{FigSIMU_Ex2}(a). The observation models are given by
\begin{gather}
\label{IStemaejemplo}
\begin{split}
Y_{j}=20\log\left(||{\bf z}-{\bf h}_j ||\right)+B_{j}, \quad j=1,\ldots, N_S, \\
\end{split}   
\end{gather}   
where $B_{j}$ are independent Gaussian random variables with pdfs, $\mathcal{N}(b_j;0,\lambda_j^2)$, $j=1,\ldots, N_S$.  We denote 
${\bm \lambda}=[\lambda_1,\ldots,\lambda_{N_S}]$ the vector of standard deviations.
Given the position of the target ${\bf z}^*=[z_1^*=2.5,z_2^*=2.5]^{\top}$ and setting ${\bm \lambda}^*=[\lambda_1^*=1,\lambda_2^*=2,\lambda_3^*=1,\lambda_4^*=0.5,\lambda_5^*=3,\lambda_{6}^*=0.2]$ (since $N_S=6$), we generate $N_O=20$ observations from each sensor according to the model in Eq. \eqref{IStemaejemplo}.
 Then, we finally obtain a measurement matrix ${\bf Y}=[y_{k,1},\ldots, y_{k,N_S}] \in \mathbb{R}^{d_Y}$, where $d_Y=N_ON_S=120$, $k=1,\ldots,N_O$.  
We consider uniform prior $\mathcal{U}(\mathcal{R}_z)$ over the position $[z_1,z_2]^{\top}$  with $\mathcal{R}_z=[-30\times 30]^2$, and a uniform prior over $\lambda_j$, so that ${\bm \lambda}$ has prior $\mathcal{U}(\mathcal{R}_\lambda)$ with $\mathcal{R}_\lambda=[0,20]^{N_S}$. Thus, the posterior pdf is 
\begin{eqnarray}
&&{\bar \pi}({\bf x}|\textbf{Y})={\bar \pi}({\bf z},{\bm \lambda}|\textbf{Y})= \ell(\textbf{y}|z_1,z_2,\lambda_1,\ldots,\lambda_{N_S})\prod_{i=1}^2p(z_i)\prod_{j=1}^{N_S} p(\lambda_j), \\
&=&\left[\prod_{k=1}^{N_O}\prod_{j=1}^{N_S} \frac{1}{\sqrt{2\pi \lambda_j^2}}\exp\left(-\frac{1}{2\lambda_j^2}(y_{k,j}+10\log\left(||{\bf z}-{\bf h}_j ||\right)^2\right) \right]\mathbb{I}_{z}(\mathcal{R}_z) \mathbb{I}_{\lambda}(\mathcal{R}_\lambda), 
\end{eqnarray}
where ${\bf x}=[{\bf z},{\bm \lambda}]^{\top}$ is a vector of parameters of dimension $D=N_S+2=8$ that we desire to infer, and $\mathbb{I}_{c}(\mathcal{R})$ is an indicator variable that is $1$ if $c\in\mathcal{R}$, otherwise is $0$. 

Our goal is to compute the Minimum Mean Square Error (MMSE) estimator, i.e., the expected value of the posterior ${\bar \pi}({\bf x}|\textbf{Y})={\bar \pi}({\bf z},{\bm \lambda}|\textbf{Y})$. Since the MMSE estimator cannot be computed analytically, we apply Monte Carlo methods for approximating it.
We compare GMS, the corresponding MTM scheme,  the Adaptive Multiple Importance Sampling (AMIS) technique \citep{CORNUET12}, and $N$ parallel MH chains with a random walk proposal pdf. For all of them we consider Gaussian proposal densities. For GMS and MTM, we set $q_t({\bf x}|{\bm \mu}_{t},\sigma^2 \mathbb{I})=\mathcal{N}({\bf x}|{\bm \mu}_{t},\sigma^2 \mathbb{I})$ where ${\bm \mu}_{t}$ is adapted considering the empirical mean of the generated samples after a training period, $t\geq 0.2 T$ \citep{Haario01}, ${\bm \mu}_{0}\sim\mathcal{U}([1,5]^{D})$ and $\sigma=1$. For AMIS, we have $q_{t}({\bf x}|{\bm \mu}_{t},{\bf C}_{t})=\mathcal{N}({\bf x}|{\bm \mu}_{t},{\bf C}_{t})$, where ${\bm \mu}_{t}$ is as previously described (with ${\bm \mu}_{0}\sim\mathcal{U}([1,5]^{D})$) and  ${\bf C}_{t}$ is also adapted using the empirical covariance matrix, starting ${\bf C}_0= 4  {\bf I}$. We also test the use of $N$ parallel Metropolis-Hastings (MH) chains (we also consider the case of $N=1$, i.e., a single chain), with a Gaussian random-walk proposal pdf, $q_n({\bm \mu}_{n,t}|{\bm \mu}_{n,t-1},\sigma^2 {\bf I})=\mathcal{N}({\bm \mu}_{n,t}|{\bm \mu}_{n,t-1},\sigma^2 {\bf I})$ with ${\bm \mu}_{n,0}\sim\mathcal{U}([1,5]^{D})$ for all $n$ and $\sigma=1$.

We fix the total number of evaluations of the posterior density as $E=NT=10^4$. Note that, generally, the evaluation of the posterior is the most costly step in MC algorithms (however, AMIS has the additional cost of re-weighting all the samples at each iteration according to the deterministic mixture procedure \citep{Bugallo15,CORNUET12, ElviraMIS15}).  We recall that $T$ denotes the total number of iterations and $N$ the number of samples drawn from each proposal at each iteration.  We consider  ${\bf x}^*=[{\bf z}^*,{\bm \lambda}^*]^{\top}$ as the ground-truth and compute the Mean Square Error (MSE) in the estimation obtained with the different algorithms. The results are averaged over $500$ independent runs and they are provided in Tables \ref{GMS_localization}, \ref{AMIS_localization}, and \ref{MH_localization} and Figure \ref{FigSIMU_Ex2}(b). Note that GMS outperforms AMIS for each a pair $\{N,T\}$ (keeping fixed $E=NT=10^4$), and GMS also provides smaller MSE values than $N$ parallel MH chains (the case $N=1$ corresponds to a unique longer chain). Figure \ref{FigSIMU_Ex2}(b) shows the MSE versus $N$ maintaining $E=NT=10^4$ for GMS and the corresponding MTM method. This figure again confirms the advantage of recycling the samples in a MTM scheme.





\begin{table}[!htb]
\begin{center}
\caption{{\bf Results GMS.}}
\begin{tabular}{|c|c|c|c|c|c|c|c|c|}
\hline 
{\bf MSE}  &  1.30 &  1.24 &    1.22   &  1.21 &    1.22 &    {\bf 1.19} &  1.31  &    {\bf 1.44}   \\ 
\hline
\hline
$N$   &  10 & 20  & 50  & 100 & 200  & 500   & 1000&  2000  \\ 
$T$   &  1000  & 500  & 200  & 100  & 50 & 20  & 10 & 5 \\ 
\hline
$E$  &  \multicolumn{8}{c|}{$NT=10^4$}\\
\hline
{\bf MSE range} &\multicolumn{8}{c|}{ {\bf Min MSE= 1.19}  \quad   --------- \quad    {\bf Max MSE= 1.44} }     \\
\hline
\end{tabular}
\label{GMS_localization}
\end{center}
\end{table}

\begin{table}[!htb]
\begin{center}
\caption{{\bf Results AMIS \citep{CORNUET12}.}}
\begin{tabular}{|c|c|c|c|c|c|c|c|c|}
\hline 
{\bf MSE}   & 1.58 & 1.57 &  1.53   & 1.48 & 1.42  & {\bf 1.29}   &  1.48 &  {\bf 1.71}  \\  
\hline
$N$   &10 & 20 &50 & 100 & 200 & 500 & 1000 & 2000 \\
$T$   & 1000 & 500 & 200 &  100 & 50 & 20 & 10 & 5   \\ 
\hline
$E$  &  \multicolumn{8}{c|}{$NT=10^4$}\\
\hline
{\bf MSE range} &\multicolumn{8}{c|}{ {\bf Min MSE= 1.29}  \quad   --------- \quad    {\bf Max MSE= 1.71} }     \\
\hline
\end{tabular}
\label{AMIS_localization}
\end{center}
\end{table}

\begin{table}[!htb]
\begin{center}
\caption{{\bf Results $N$ parallel MH chains with random-walk proposal pdf.}}
\begin{tabular}{|c|c|c|c|c|c|c|c|c|}
\hline 
{\bf MSE} & 1.42 &  {\bf 1.31} &    1.44 &  2.32 &   2.73 &    {\bf 3.21} &    3.18 &    3.15  \\ 
\hline
\hline
$N$  &1 & 5 & 10 &  50 & 100 & 500 & 1000 &2000 \\
$T$   & $10^4$  & 2000    & 1000 & 200  & 100     &  20   & 10    & 5 \\ 
\hline
$E$  &  \multicolumn{8}{c|}{$NT=10^4$}\\
\hline
{\bf MSE range} &\multicolumn{8}{c|}{ {\bf Min MSE= 1.31}  \quad   --------- \quad    {\bf Max MSE=3.21 } }     \\
\hline
\end{tabular}
\label{MH_localization}
\end{center}
\end{table}

\subsection{Target localization with real data}

In this section, we test the proposed techniques in real data application. More specifically, we consider again a positioning problem in order to localize a target in a bidimensional space using range measurements \citep{Ali07,Patwari03}.

\subsubsection{Setup of the Experiment }
\begin{figure*}[htb]
\centerline{
	\includegraphics[width=7cm]{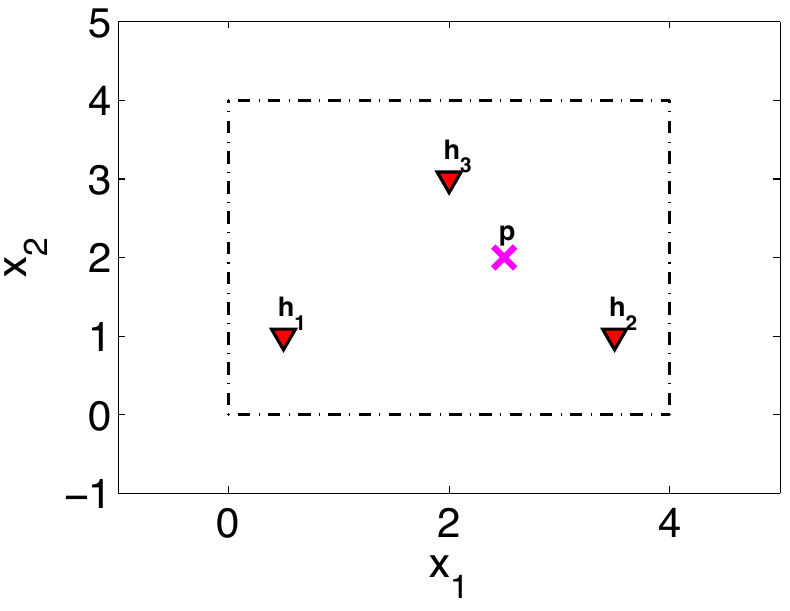}	
  \includegraphics[width=7cm]{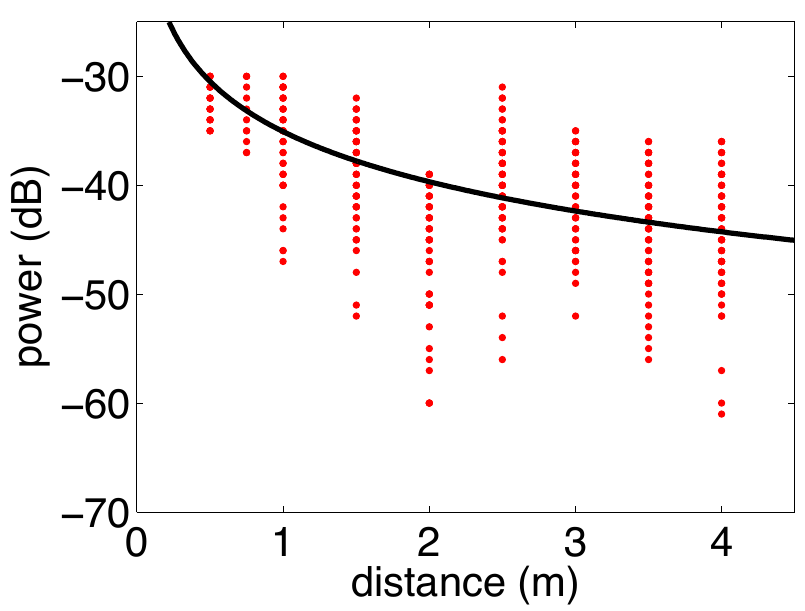}	
}	
\caption{{\bf (a)} Experimental setup: sensors over a rectangular surveillance
area of 4 times 4 meters. The sensors are shown with triangles (denoted by ${\bf h}_i$)  while the target that must be localized is depicted with a cross (denoted by ${\bf p}$). {\bf (b)} The least squares regression to adjust the parameters $l$ and $\gamma$. The points indicate the measurements collected by the sensors at different distances, and the solid curve denotes the function $\hat{l}-10\hat{\gamma}\log\left[\frac{d}{d_{0}}\right]$ with $d_0=0.3$, $\hat{l}=-27.08$ dB and $\hat{\gamma}=1.52$.}  
\label{network}
\end{figure*}

We have designed a network of four nodes. Three of them are placed at fixed positions and play the role of sensors that measure the strength of the radio signals transmitted by the target. The other node plays the role of the target to be localized. All nodes are bluetooth devices (Conceptronic CBT200U2A) with a nominal maximum range of 200 m. 
The deployment of the network is given in Figure \ref{network}(a). We consider a square monitored area of $4 \times 4$ m and place the sensors at fixed positions $\textbf{h}_1=[h_{1,1}=0.5,h_{1,2}=1]$, $\textbf{h}_2=[h_{2,1}=3.5,h_{2,2}=1]$ and $\textbf{h}_3=[h_{3,1}=2,h_{3,2}=3]$, with all coordinates in meters. The target is located at $\textbf{p}=[p_1=2.5,p_2=2]$. 

The measurement equation describes the relationship between the observed radio signal strength $y_i$ obtained by the $i$-th sensor, and the target position $\textbf{p}=[p_1,p_2]$ (see \citep{Rappaport01}), and is given by  
\begin{equation}
\label{modeloobserv}
	y_i=l-10\gamma\log\left[\frac{\sqrt{(p_1-h_{i,1})^2+(p_2-h_{i,2})^2}}{d_{0}}\right]+\theta_{i}  \ \ \ \ \mbox{(dB)},
\end{equation}
where $\gamma$ is a parameter that depends on the physical environment (for instance, in an open space $\gamma\approx 2$), and the constant $l$ is the mean power received by each sensor when the target is located at a reference distance $d_0$. The measurement noise is Gaussian, $N(\theta_i;0,\sigma^2) \propto\exp\left\{-\frac{\theta_{i}^{2}}{2\sigma^2}\right\}$ with $i=1,2,3$. In this experiment, the reference distance has been set to $d_0=0.3$ m. Unlike in the previous numerical example,  here  the parameters $l$, $\gamma$, and $\sigma^2$ have been tuned in advance by least square regression using $200$ measurements with the target placed at known distances from each sensor. As a result, we have obtained  $\hat{l}=-27.08$, $\hat{\gamma}=1.52$, and $\hat{\sigma}=4.41$. Figure \ref{network}(b) shows the measurements obtained at several distances and the fitted curve $\hat{l}-10\hat{\gamma}\log\left[\frac{d}{d_{0}}\right]$, where $d=\sqrt{(p_1-h_{i,1})^2+(p_2-h_{i,2})^2}$. 	

\subsubsection{Posterior density, algorithms and results} 
Assume we collect $M$ independent measurements from each sensor, considering the observation model in Eq. \eqref{modeloobserv}. Let $\textbf{y}=[y_{1,1},\ldots,y_{1,M},y_{2,1},\ldots,y_{2,M},y_{3,1},\ldots,y_{3,M}]$ denote the measurement vector where $y_{i,j}$ is the $j$-th observation of the $i$-th sensor. The the likelihood is 
\begin{gather}
\small
\begin{split}
p(\textbf{y}|\textbf{p})=\prod_{i=1}^{3}\prod_{m=1}^{M}  \mathcal{N}\left(y_{i,m}|\hat{l}-10\hat{\gamma}\log\right(||\textbf{p}-\textbf{h}_i||/d_0\left),\hat{\sigma}^2 \right), \quad {\bf p}\in [0,4] \times  [0,4].	
\end{split}
\end{gather}
We set $p(\textbf{y}|\textbf{p})=0$ if ${\bf p}\notin [0,4] \times  [0,4]$.
We assume a Gaussian prior for each component of ${\bf p}$ with mean $1.5$ and variance $0.5$. Hence, the posterior density is given by
	$p(\textbf{p}|\textbf{y})\propto p(\textbf{y}|\textbf{p})p(\textbf{p})$ where $p(\textbf{p})$ denotes the bidimensional Gaussian prior pdf over the position ${\bf p}$.
Our goal is to approximate the expected value of the posterior $p(\textbf{p}|\textbf{y})$. We collect $M=10$ measurements from each Conceptronic CBT200U2A devices, and compute the ground-truth (i.e., the expected value) with a costly determinist grid. Then, we apply GMS and the corresponding MTM scheme, considering the same bidimensional proposal density $q({\bf p}|{\bm \mu},\lambda^2 \mathbb{I})=\mathcal{N}({\bf p}|{\bm \mu},\lambda^2 \mathbb{I})$ where ${\bm \mu}\sim\mathcal{U}([0,4]\times [0,4])$ (randomly chosen at each independent run) and $\lambda=\sqrt{2}$. We set $N=50$ and $T=100$. We compute the Mean Square Error (MSE) with respect to the ground-truth obtained with GMS and MTM, in $500$ different independent runs. The averaged MSE values are $1.81$ for the GMS method and $2.2$ for the MTM scheme. Therefore, GMS outperforms the corresponding MTM technique also in this experiment.

\subsection{Tracking of biophysical parameters}
\label{LAI}

We consider the challenging problem of estimating biophysical parameters from remote sensing (satellite) observations. In particular, we focus on the estimation of the Leaf Area Index (LAI). 
It is important to track evolution of LAI through time in every spatial position on Earth because LAI plays an important role in vegetation processes such as photosynthesis and transpiration, and is connected to meteorological/climate and ecological land processes \citep{Chen92_LAI}. 
Let us denote LAI  as $x_d\in \mathbb{R}^{+}$ (where $d\in \mathbb{N}^+$  also represents a temporal index) in a specific region at a latitude of $42^{\circ}$ N \citep{Jose}. Since $x_t> 0$,  we consider Gamma prior pdfs over the evolutions of LAI and Gaussian perturbations for the ``in-situ'' received measurements, $y_t$. More specifically, we assume the state-space model (formed by propagation and measurement equations), 
\begin{equation}
\label{StateModel}
\left\{\begin{array}{lll}
g_d(x_d|x_{d-1})&= \mathcal{G}\left(x_d\Big|\frac{x_{d-1}}{b},b\right)&=\frac{1}{c_d} x_d^{ (x_{d-1}-b)/b}\exp\left(-\frac{x_d}{b}\right),  \\
\ell_d(y_d|x_d)&=\mathcal{N}(y_d|x_d,\lambda^2)&=\frac{1}{\sqrt{2\pi\lambda^2}}\exp\left(-\frac{1}{2\lambda^2}(y_d-x_d)^2\right),
\end{array}\right. 
\end{equation}
for $d=2,\ldots,D$, with initial probability $g_1(x_1)=\mathcal{G}(x_1|1,1)$, where $b,\lambda>0$ and $c_d>0$ is a normalizing constant.  Note that the expected value of the Gamma pdf above is $x_{d-1}$ and the variance is $b$. 

\begin{figure}[!htbp]
\centering
\centerline{
\subfigure[$t=2$]{\includegraphics[width=0.33\textwidth]{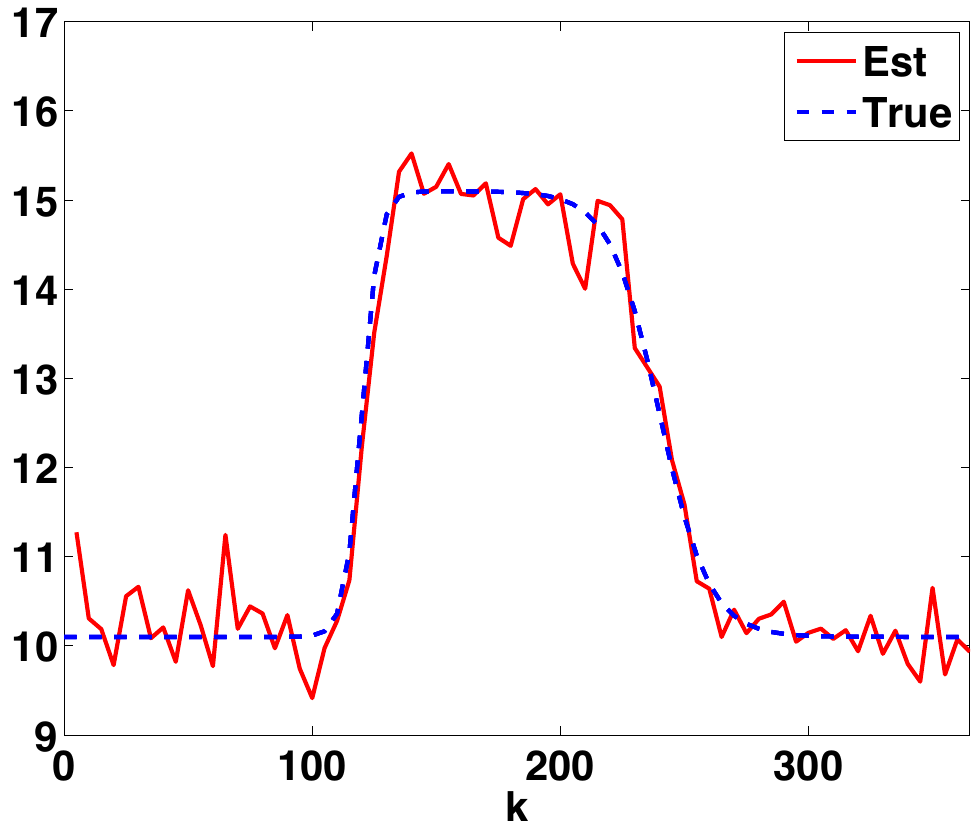} }
\subfigure[$t=10$]{\includegraphics[width=0.33\textwidth]{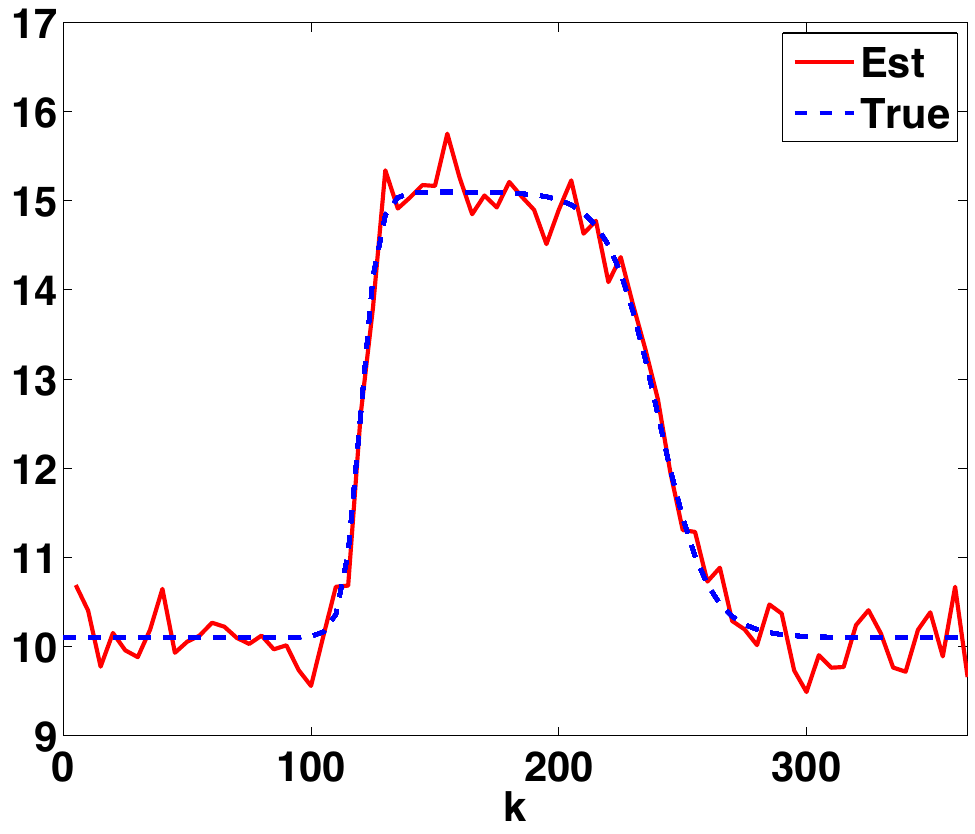} }
\subfigure[$t=100$]{\includegraphics[width=0.33\textwidth]{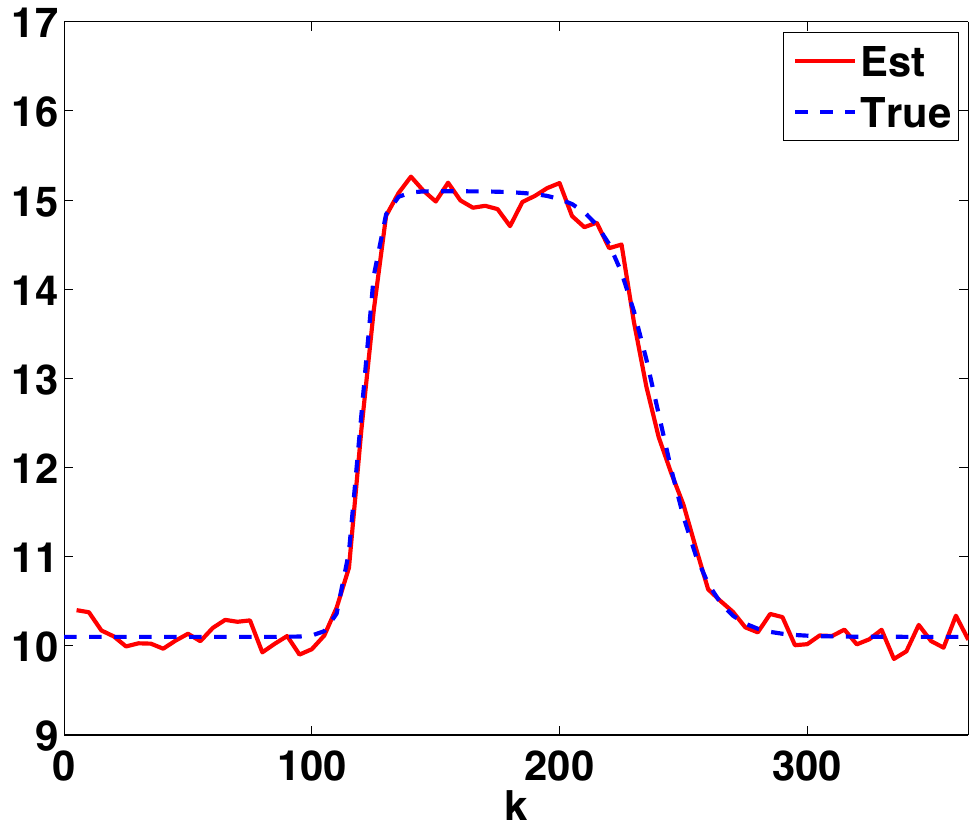} }
}
\caption{Output of DPMH  (with $N=5$, $\lambda=0.1$ and ${\bf b}=[0.01, 0.05, 0.1, 1]^{\top}$) at different iterations {\bf (a)} $t=2$,  {\bf (b)} $t=10$, and {\bf (c)} $t=100$, in one specific run. The true values, $\x^*=x_{1:D}^*$, are shown dashed lines whereas the estimated trajectories by DPMH, $\widehat{\x}_{t}=\widehat{x}_{1:D,t}$, with solid lines.  }
\label{Fig2_Ex3}
\end{figure}

\begin{figure}[htbp]
\centering
\centerline{
\subfigure[]{\includegraphics[width=0.45\textwidth]{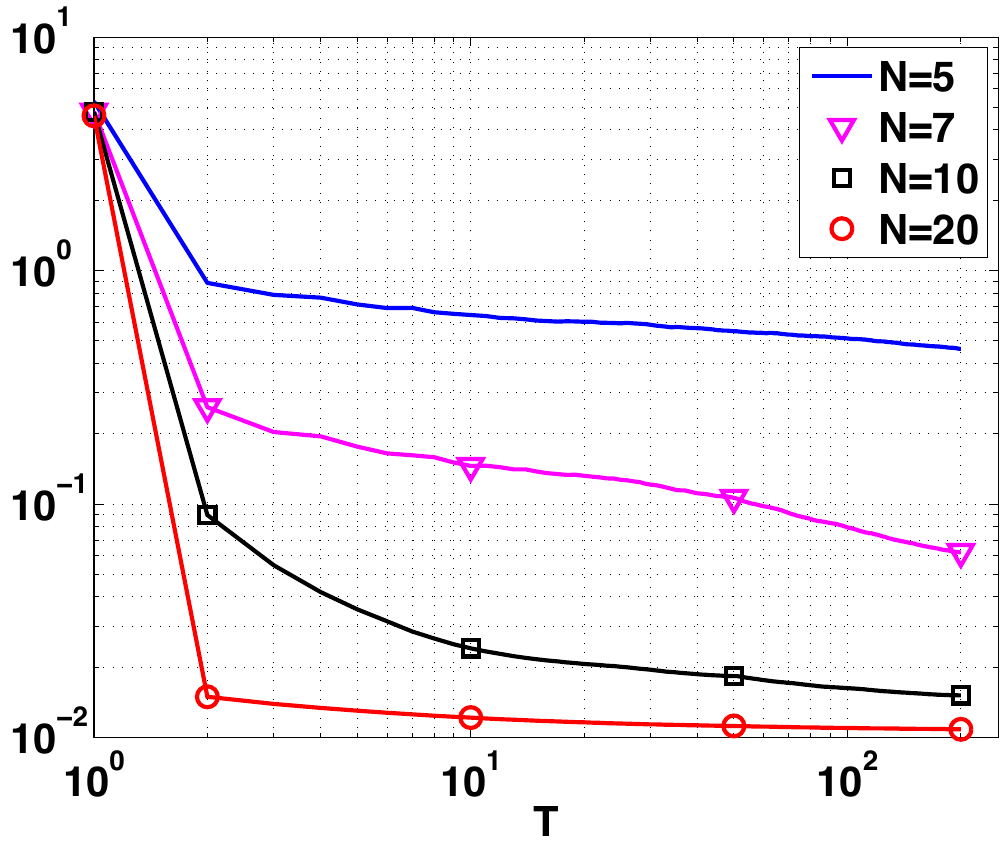} }
\subfigure[]{\includegraphics[width=0.46\textwidth]{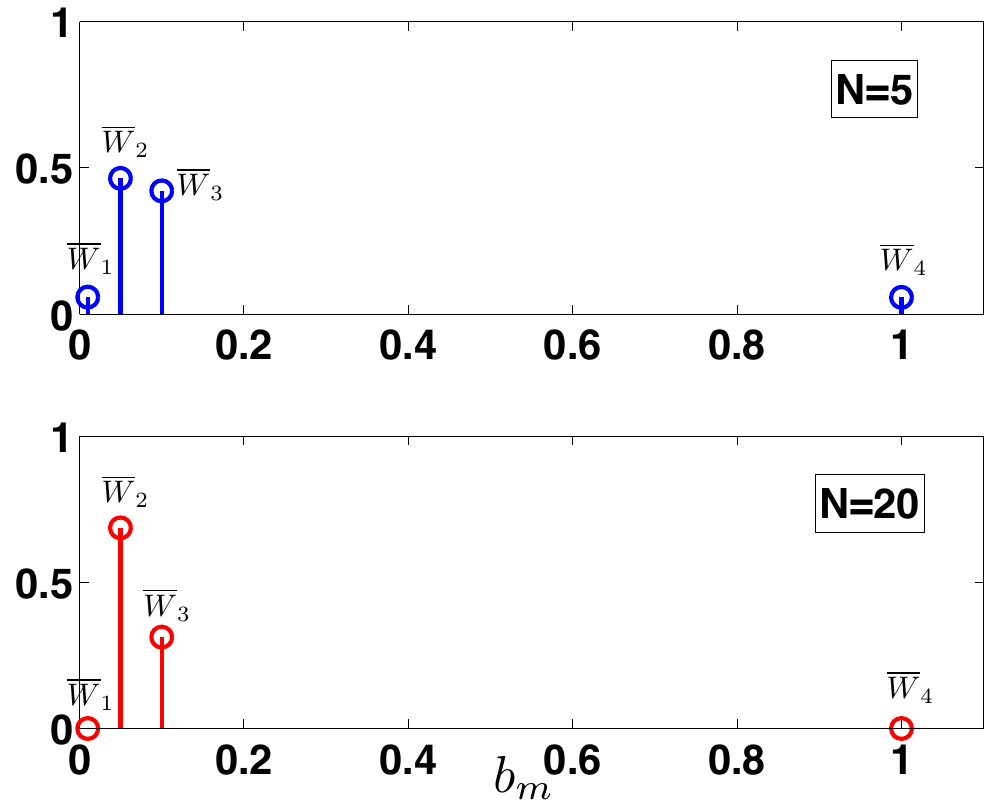} }
}
\caption{{\bf (a)} MSE in estimation of the trajectory (averaged over $2000$ runs) obtained by DPMH as function $T$ and different values of $N\in\{5,7,10,20\}$. As expected, we can see that the performance of DPMH improves as $T$ and $N$ grow. {\bf (b)} Averaged values of the normalized weights $\overline{W}_m=\frac{\widehat{Z}_m}{\sum_{j=1}^M\widehat{Z}_j}$  (with $N=5$ and $N=10$) associated to each filter. DPMH is able to detect the best variances ($b_2$ and $b_3$) of the proposal pdfs among the values $b_1=0.01$, $b_2=0.05$, $b_3=0.1$ and $b_4=1$ (as confirmed by Table \ref{PGMS_DPMH}).
}
\label{Fig1_Ex3}
\end{figure}

\paragraph{First Experiment.} Considering that all the parameters of the model are known, the posterior pdf is   
\begin{eqnarray}
\bar{\pi}(\x|\y)&\propto& \ell(\y|\x)g(\x) =\left[\prod_{d=2}^D \ell_d(y_d|x_d)\right]\left[\left(\prod_{d=2}^D g_d(x_d|x_{d-1})\right) g_1(x_1)\right],
\end{eqnarray}
with $\x=x_{1:D}\in \mathbb{R}^D$. For generating the ground-truth (i.e., the trajectory $\x^*=x_{1:D}^*=[x_1^*,\ldots,x_D^*]$), we simulate the temporal evolution of LAI in one year (i.e., $1 \leq d\leq D=365$) by using a double logistic function (as suggested in the literature \citep{Jose}), i.e.,
\begin{equation}
\label{JoseEq}
x_d=a_1+a_2\left(\frac{1}{1+\exp(a_3(d-a_4))}+\frac{1}{1+\exp(a_5(d-a_6))}+1\right),
\end{equation}
with $a_1=0.1$, $a_2=5$, $a_3=-0.29$, $a_4=120$, $a_5=0.1$ and $a_6=240$ as employed in \citep{Jose}. In Figure \ref{Fig2_Ex3}, the true trajectory $x_{1:D}$ is depicted with dashed lines. 
The observations $\y=y_{2:D}$ are then generated (each run) according to $y_d \sim \ell_d(y_d|x_d)=\frac{1}{\sqrt{2\pi\lambda^2}}\exp\left(-\frac{1}{2\lambda^2}(y_d-x_d)^2\right)$. 
First of all, we test the standard PMH, the particle version of GMS (PGMS), and DPMH  (fixing $\lambda=0.1$).  For DPMH, we use $M=4$ parallel filters with different scale parameters  ${\bf b}=[b_1=0.01, b_2=0.05,b_3=0.1,b_4=1]^{\top}$. Figure \ref{Fig2_Ex3} shows the estimated trajectories $\widehat{\x}_{t}=\widehat{x}_{1:D,t}=\frac{1}{t} \sum_{\tau=1}^{t} \widetilde{\x}_\tau$ (averaged over 2000 runs) obtained by DPMH with $N=5$ at $t\in\{2,10,100\}$, in one specific run. Figure \ref{Fig1_Ex3}(a) depicts the evolution of the MSE obtained by DPMH as a function of $T$ and considering different values of $N\in \{5,7,10,20\}$. The performance of DPMH improves as $T$ and $N$ grow, as expected. DPMH detects the best parameters among the four values in ${\bf b}$, following the weights $\overline{W}_m$  (see Figure \ref{Fig1_Ex3}(b)) and DPMH takes advantage of this ability. Indeed, we compare DPMH with $N=10$, $T=200$, and $M=4$ using ${\bf b}$, with $M=4$ different standard PMH and PGMS algorithms with $N=40$ and $T=200$ (clearly, each one driven by a unique filter, $M=1$) in order to keep the total number of evaluation of the posterior fixed, $E=NMT= 8\cdot 10^3$, each one using a parameter $b_m$, $m=1,\ldots,M$. The results, averaged over 2000 runs, are shown in Table \ref{PGMS_DPMH}. In terms of MSE, DPMH always outperforms the $4$ possible standard PMH methods. PGMS using two parameters, $b_2$ and $b_3$, provides better performance, but DPMH outperforms PGMS averaging the $4$ different MSEs obtained by PGMS.  Moreover, due to the parallelization, in this case DPMH can save $\approx 15 \%$  of the spent computational time.

\begin{figure}[!htbp]
\centering
\centerline{
\includegraphics[width=0.4\textwidth]{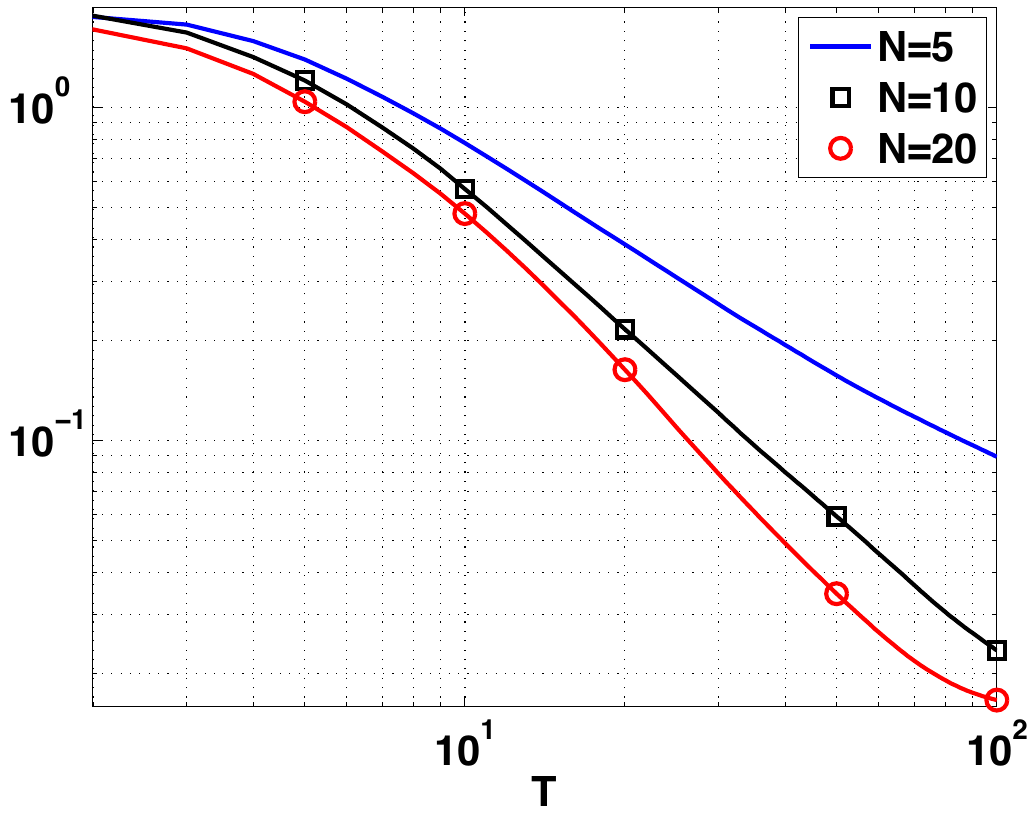} 
}
\caption{ MSE in estimation of $\lambda^*=0.7$  (averaged over $1000$ runs) obtained by DPMMH as function $T$ and different values of $N\in\{5,10,20\}$.
}
\label{Fig3_Ex3}
\end{figure}

 \paragraph{Second Experiment.} Now we consider also the parameter $\lambda$ unknown, so that the complete variable of interest $[\x,\lambda]\in \mathbb{R}^{D+1}$. Then the posterior is $\bar{\pi}(\x,\lambda|\y)\propto \ell(\y|\x,\lambda)g(\x,\lambda)$ according to the model Eq. \eqref{StateModel}, where $g(\x,\lambda)=g(\x) g_\lambda(\lambda)$ and $g_\lambda(\lambda)$ is a uniform pdf in $[0.01,5]$. Then we test the marginal versions of the PMH and DPMH with $q_\lambda(\lambda)=g_\lambda(\lambda)$ (see \ref{MarginalPMH}), for estimating $[\x^*,\lambda^*]$ where  $\x^*=x_{1:D}^*$ is given by Eq. \eqref{JoseEq} and $\lambda^*=0.7$. Figure \ref{Fig3_Ex3} shows the MSE in estimation of $\lambda^*$ (averaged over 1000 runs) obtained by DPMMH as function of $T$ and different number of candidates, $N\in \{5,10,20\}$ (with again $M=4$ and ${\bf b}=[b_1=0.01, b_2=0.05,b_3=0.1,b_4=1]^{\top}$). Table \ref{marginals} compares the standard PMMH and DPMMH for estimating $\lambda^*$ (we set $E=NMT= 4\cdot 10^3$ and $T= 100$). Averaging the results of PMMH, we can observe that DPMMH outperforms the standard PMMH in terms of smaller MSE and smaller computational time.

\begin{table}[!htb]
\begin{center}
\caption{Comparison among PMH, PGMS and DPMH with $E=NMT= 8\cdot 10^3$ and $T= 200$ ($\lambda=0.1$), estimating the trajectory $\x^*=x_{1:D}^*$.}
\begin{tabular}{|l|c|c|c|}
\hline 
\multirow{4}{*}{{\bf Proposal Var}} &  \multicolumn{1}{c|}{{\bf Standard PMH}} &  \multicolumn{1}{c|}{{\bf PGMS}} &   \multicolumn{1}{c|}{{\bf DPMH}}   \\
\cline{2-4}
&  $N=40$  &$N=40$  & $N=10$\\
& $(M=1)$ &  $(M=1)$ & $M=4$\\
\cline{2-4}
 & {\bf MSE} & {\bf MSE}   & {\bf MSE}    \\ 
\hline
$b_1=0.01$   & 0.0422   &  0.0380      & \multirow{4}{*}{0.0108}  \\ 
$b_2=0.05$   & 0.0130 &0.0100    &  \\ 
$b_3=0.1$   &0.0133  & 0.0102   &\\ 
$b_4=1$   &   0.0178 &   0.0140    &\\ 
\hline
{\bf Average} &0.0216  &  0.0181 & 0.0108  \\
\hline
\hline
{\bf Norm. Time} & 1 &  1 &  0.83  \\
\hline
\end{tabular}
\label{PGMS_DPMH}
\end{center}
\end{table}


\begin{table}[!htb]
\begin{center}
\caption{Comparison among PMMH and DMPMH with $E=NMT= 4\cdot 10^3$ and $T= 100$, for estimating $\lambda^*=0.7$.}
\begin{tabular}{|c|c|c|}
\hline 
\multirow{4}{*}{{\bf Proposal Var}} &  \multicolumn{1}{c|}{{\bf PMMH}}  &   \multicolumn{1}{c|}{{\bf DPMMH}}   \\
\cline{2-3}
&  $N=40$   & $N=10$\\
& $(M=1)$  & $M=4$\\
\cline{2-3}
 & {\bf MSE}   & {\bf MSE}    \\ 
\hline
$b_1=0.01$     &  0.0929      & \multirow{4}{*}{0.0234}  \\ 
$b_2=0.05$    & 0.0186   &  \\ 
$b_3=0.1$     &   0.0401 &\\ 
$b_4=1$    &    0.0223   &\\ 
\hline
{\bf Average} & 0.0435 & 0.0234  \\
\hline
\hline
{\bf Norm. Time} & 1  &  0.85  \\
\hline
\end{tabular}
\label{marginals}
\end{center}
\end{table}


\newpage
\section{Conclusions}\label{Conclsect}
In this work, we have described the Group Importance Sampling  (GIS) theory and its application in other Monte Carlo schemes. We have considered the use of GIS in  SIR (a.k.a., particle filtering), showing that GIS is strictly required if the resampling procedure is applied only in a subset of the current population of particles. Moreover we have highlighted that, in the standard SIR method, if GIS is applied there exists two equivalent estimators of the marginal likelihood  (one of them is an estimator of the marginal likelihood {\it only} if the GIS weighting is used), exactly as in Sequential Importance Sampling (SIS). 
We have also shown that the Independent Multiple Try Metropolis (I-MTM) schemes and the Particle Metropolis-Hastings (PMH) algorithm can be interpreted as a classical Metropolis-Hastings (MH) method taking into account the GIS approach. 

Furthermore, two novel methodologies based on GIS have been introduced. One of them (GMS) yields a Markov chain of weighted samples and can be also considered an iterative importance sampler. The second one (DPMH) is a distributed version of the PMH where different parallel particle filters can be jointly employed. These filter cooperate for driving the PMH scheme. Both techniques have been applied successfully in three different numerical experiments (tuning of the hyperparameters for GPs, two localization problems in a wireless sensor network (one with real data), and the tracking of the Leaf Area Index), comparing them with several benchmark methods. Marginal versions of GMS and DPMH have been also discussed and tested in the numerical applications. Three Matlab demos have been also given in order to facilitate the comprehension of the reader. 
As a future line, we plan to design an adaptive DPMH scheme in order to select online the best particle filters among the $M$ run in parallel, and parsimoniously distribute the computational effort.

\section*{Acknowledgements}
This work has been supported by the European Research Council (ERC) through the ERC Consolidator Grant SEDAL ERC-2014-CoG 647423.

\bibliographystyle{unsrtnat}
\bibliography{bibliografia}

\appendix

\section{Proper weighting of a resampled particle}
\label{WeightingAResPar}

Let us consider the particle approximation of $\bar{\pi}$ obtained by the IS approach drawing $N$ particles ${\bf x}_n \sim q(\x)$,  $n=1\ldots,N$, i.e.,
\begin{equation}
\label{PIEQ}
\widehat{\pi}({\bf x}|{\bf x}_{1:N})=\sum_{n=1}^N {\bar w}({\bf x}_n) \delta({\bf x}-{\bf x}_n)=\sum_{n=1}^N \frac{w({\bf x}_n)}{\sum_{i=1}^N w({\bf x}_i)} \delta({\bf x}-{\bf x}_n)= \frac{1}{N \widehat{Z}} \sum_{n=1}^N w({\bf x}_n) \delta({\bf x}-{\bf x}_n).
\end{equation}
where 
\begin{equation}
\widehat{Z} =\widehat{Z}({\bf x}_{1:N})=\frac{1}{N}\sum_{i=1}^N w({\bf x}_i).
\end{equation}
Given the set of particles ${\bf x}_{1:N} \sim \prod_{n=1}^N q(\x_n)$, a resampled particle is generated as $\widetilde{\x}'\sim \widehat{\pi}({\bf x}|{\bf x}_{1:N})$. Let us denote the joint pdf $\widetilde{Q}({\bf x},{\bf x}_{1:N})= \widehat{\pi}({\bf x}|{\bf x}_{1:N}) \left[\prod_{i=1}^N q({\bf x}_i)\right]$. The marginal pdf $\widetilde{q}({\bf x})$ of a resampled particle $\widetilde{\x}'$, integrating out ${\bf x}_{1:N}$ (i.e., $\widetilde{\x}'\sim \widetilde{q}({\bf x})$),  is
\begin{eqnarray}
\widetilde{q}({\bf x})&=&\int_{\mathcal{X}^N}  \widetilde{Q}({\bf x},{\bf x}_{1:N}) d{\bf x}_{1:N} \\
&=& \int_{\mathcal{X}^N}\widehat{\pi}({\bf x}|{\bf x}_{1:N}) \left[\prod_{i=1}^N q({\bf x}_i)\right]  d{\bf x}_{1:N},  \\
&=& \int_{\mathcal{X}^N} \left[\frac{1}{N \widehat{Z}({\bf x}_{1:N})} \sum_{j=1}^N w({\bf x}_j) \delta({\bf x}-{\bf x}_j)\right] \left[\prod_{i=1}^N q({\bf x}_i)\right]  d{\bf x}_{1:N}, \\
&=&   \sum_{j=1}^N \int_{\mathcal{X}^N} \frac{1}{N \widehat{Z}({\bf x}_{1:N})} w({\bf x}_j) \left[\prod_{i=1}^N q({\bf x}_i)\right]   \delta({\bf x}-{\bf x}_j) d{\bf x}_{1:N}, \\ 
&=& \sum_{j=1}^{N} \left({\int_{\mathcal{X}^{N-1}} \frac{w({\bf x})}{N  \widehat{Z}({\bf x}_{1:j-1},{\bf x},{\bf x}_{j+1:N})} \left[q(\x)\prod_{\substack{i=1 \\ i \neq j}}^N q({\bf x}_{i})\right]  d{\bf x}_{\neg j}} \right), \\
&=& \sum_{j=1}^{N} \left({\int_{\mathcal{X}^{N-1}} \frac{w({\bf x})}{N\widehat{Z}} \left[q(\x)\prod_{\substack{i=1 \\ i \neq j}}^N q({\bf x}_{i})\right]  d{\bf x}_{\neg j}} \right), 
\end{eqnarray}
where we have used the integration property of the delta function $\delta({\bf x}-{\bf x}_j)$, i.e, given a generic function $f({\bf v})$, we have $\int_{\mathcal{X}} f({\bf v}) \delta({\bf x}-{\bf v}) d{\bf v}=f({\bf x})$ with ${\bf x} \in \mathcal{X}$ and, in the last equality, we have just used the simplified notation $\widehat{Z} =\widehat{Z}({\bf x}_{1:N})=\widehat{Z}({\bf x}_{1:j-1},{\bf x},{\bf x}_{j+1:N})$. Moreover, since $w({\bf x})=\frac{\pi(\x)}{q(\x)}$,
\begin{eqnarray}
\widetilde{q}({\bf x})&=&\pi(\x) \sum_{j=1}^{N} \left({\int_{\mathcal{X}^{N-1}} \frac{1}{N\widehat{Z}} \left[\prod_{\substack{i=1 \\ i \neq j}}^N q({\bf x}_{i})\right]  d{\bf x}_{\neg j}} \right),  \\
&=&\pi(\x) \cdot N {\int_{\mathcal{X}^{N-1}} \frac{1}{N\widehat{Z}} \left[\prod_{\substack{i=1 \\ i \neq j}}^N q({\bf x}_{i})\right]  d{\bf x}_{\neg j}}, \\
&=&\pi(\x)  {\int_{\mathcal{X}^{N-1}} \frac{1}{\widehat{Z}} \left[\prod_{\substack{i=1 \\ i \neq j}}^N q({\bf x}_{i})\right]  d{\bf x}_{\neg j}}, \quad \quad j\in \{1,\ldots,N\},
\end{eqnarray}
 where we have used that all the $N$ integrals within the sum are equals, due to the symmetry of the integrand function with respect to the $N-1$ integration variables.
Therefore, the standard IS weight of a resampled particle, $\widetilde{\x}'\sim \widetilde{q}({\bf x})$, is
\begin{eqnarray}
w(\widetilde{\x}')=\frac{\pi(\widetilde{\x}')}{\widetilde{q}(\widetilde{\x}')}.
\end{eqnarray}
However, generally $\widetilde{q}({\bf x})$ cannot be evaluated, hence the standard IS weight cannot be computed \citep{Lamberti16,GISssp16,NSMC}, \cite[App. C1]{LAIS17}. An alternative is to use the Liu's definition of proper weighting in Eq. \eqref{eq_liu_1} and look for a weight function $\rho({\bf  \widetilde x})=\rho({\bf  \widetilde x}|{\bf x}_{1:N})$ such that 
\begin{equation}
E_{\widetilde{Q}({\bf x},{\bf x}_{1:N})}[\rho({\bf x}|{\bf x}_{1:N}) h({\bf x})]=c E_{\bar \pi}[h({\bf x})],
\label{eq_liu_2}
\end{equation} 
where $\widetilde{Q}({\bf x},{\bf x}_{1:N})= \widehat{\pi}({\bf x}|{\bf x}_{1:N}) \left[\prod_{i=1}^N q({\bf x}_i)\right]$. Below, we show that a suitable  choice is 
\begin{equation}
\label{SUPER_IMP_DEFINITION}
\rho({\bf  \widetilde x}|{\bf x}_{1:N})=\widehat{Z}({\bf x}_{1:N})=\frac{1}{N}\sum_{i=1}^N w({\bf  x}_i),
 \end{equation}
 since it holds in Eq.  \eqref{eq_liu_2}.

{\PR
 Note that
\begin{eqnarray}
E_{\widetilde{Q}({\bf x},{\bf x}_{1:N})}[\rho({\bf x}|{\bf x}_{1:N}) h({\bf x})]&=&\int_{\mathcal{X}}\int_{\mathcal{X}^N}  \rho({\bf x}|{\bf x}_{1:N}) h({\bf x})\widetilde{Q}({\bf x},{\bf x}_{1:N}) d{\bf x}d{\bf x}_{1:N}, \\
&=&\int_{\mathcal{X}}\int_{\mathcal{X}^N} h({\bf x}) \rho({\bf x}|{\bf x}_{1:N})\widehat{\pi}({\bf x}|{\bf x}_{1:N}) \left[\prod_{i=1}^N q({\bf x}_i)\right]  d{\bf x}d{\bf x}_{1:N}.
\end{eqnarray}
Recalling that $\widehat{\pi}({\bf x}|{\bf x}_{1:N}) =\frac{1}{N \widehat Z} \sum_{j=1}^N w({\bf x}_j) \delta({\bf x}-{\bf x}_j)$,
where ${\widehat Z}={\widehat Z}({\bf x}_{1:N})=\frac{1}{N}\sum_{n=1}^N w({\bf x}_n)$ and $w({\bf x}_n)=\frac{\pi ({\bf x}_n) }{q({\bf x}_n)}$, we can rearrange the expectation above as 
\begin{eqnarray}
E_{\widetilde{Q}({\bf x},{\bf x}_{1:N})}[\rho({\bf x}|{\bf x}_{1:N}) h({\bf x})]&=& \int_{\mathcal{X}} h({\bf x}) \left[  \sum_{j=1}^{N} \left({\int_{\mathcal{X}^{N-1}} \rho({\bf x}|{\bf x}_{1:N})\frac{w({\bf x})}{N\widehat{Z}} \left[q(\x)\prod_{\substack{i=1 \\ i \neq j}}^N q({\bf x}_{i})\right]  d{\bf x}_{\neg j}} \right) \right] d{\bf x},  \\
&=& \int_{\mathcal{X}} h({\bf x})\pi({\bf x})  \left[  \sum_{j=1}^{N} \left({\int_{\mathcal{X}^{N-1}} \rho({\bf x}|{\bf x}_{1:N})\frac{1}{N\widehat{Z}} \left[\prod_{\substack{i=1 \\ i \neq j}}^N q({\bf x}_{i})\right]  d{\bf x}_{\neg j}} \right) \right] d{\bf x}, 
\end{eqnarray}
where  ${\bf x}_{\neg j}=[{\bf x}_1, \ldots, {\bf x}_{j-1},{\bf x}_{j+1}, \ldots, {\bf x}_{N}]$. 
If we choose $\rho({\bf x}|{\bf x}_{1:N})=\widehat{Z}$ and replace it in the expression above, we obtain
\begin{eqnarray}
E_{\widetilde{Q}({\bf x},{\bf x}_{1:N})}[\rho({\bf x}|{\bf x}_{1:N}) h({\bf x})]&=&  \int_{\mathcal{X}} h({\bf x})  \pi({\bf x}) \left[  \sum_{j=1}^{N} \left({\int_{\mathcal{X}^{N-1}} \widehat{Z}\frac{1}{N\widehat{Z}} \left[\prod_{\substack{i=1 \\ i \neq j}}^N q({\bf x}_{i})\right]  d{\bf x}_{\neg j}} \right) \right] d{\bf x},  \\
&=&  \int_{\mathcal{X}} h({\bf x})  \pi({\bf x}) N \frac{1}{N} d{\bf x},  \\
&=& \int_{\mathcal{X}} h({\bf x})  \pi({\bf x}) d{\bf x} \\
&=& c E_{\bar \pi}[h({\bf x})], \label{EstoyLlorado}
\end{eqnarray}
where $c=Z$, that is the normalizing constant of $\pi(\x)$. Note that Eq. \eqref{EstoyLlorado} coincides with \eqref{eq_liu_2}. $\quad\quad \Box$
}

\section{Particle approximation by GIS}
\label{ImportantApp}

Let us consider $S$ samples $\x_{m,n}\sim q_m(\x)$, where $S=\sum_{m=1}^{M}N_m$,  and weight them $w_{m,n}=\frac{\pi(\x_{m,n})}{q_m(\x_{m,n})}$ with $m=1,\ldots,M$ and  $n=1,\ldots,N_m$. Moreover, let us define two types of normalized weights, one within the $m$-th group
\begin{eqnarray}
\bar{w}_{m,n}=\frac{w_{m,n}}{\sum_{k=1}^N w_{m,k}}=\frac{w_{m,n}}{N_m \widehat{Z}_m},
\end{eqnarray}
and the other one considering all the $S$ samples,
\begin{eqnarray}
\bar{r}_{m,n}=\frac{w_{m,n}}{\sum_{j=1}^M\sum_{k=1}^{N_j}w_{j,k}}=\frac{w_{m,n}}{\sum_{j=1}^{M}N_j \widehat{Z}_j}.
\end{eqnarray}
The complete particle approximation of the target distribution is
\begin{eqnarray}
\widehat{\pi}(\x|\x_{1:M,1:N})&=&\frac{1}{\sum_{j=1}^M \sum_{k=1}^{N_j} w_{j,k}}\sum_{m=1}^M \sum_{n=1}^{N_m} w_{m,n} \delta(\x-\x_{m,n}),   \\
&=&\sum_{m=1}^M \sum_{n=1}^{N_m} \bar{r}_{m,n} \delta(\x-\x_{m,n}).
\end{eqnarray}
Note that it can be also rewritten as
\begin{eqnarray}
\widehat{\pi}(\x|\x_{1:M,1:N})&=&\frac{1}{\sum_{j=1}^{M}N_j \widehat{Z}_j}\sum_{m=1}^M N_m \widehat{Z}_m  \sum_{n=1}^N \bar{w}_{m,n} \delta(\x-\x_{m,n}),   \\
&=& \frac{1}{\sum_{j=1}^{M}N_j \widehat{Z}_j}\sum_{m=1}^M N_m \widehat{Z}_m \widehat{\pi}(\x|\x_{m,1:N}), \label{EqAppPartApprox} \\
&=& \sum_{m=1}^M \overline{W}_m \widehat{\pi}(\x|\x_{m,1:N}), 
\end{eqnarray}
where $\widehat{\pi}(\x|\x_{m,1:N})$ are the $m$-th particle approximation and $\overline{W}_m=\frac{N_m \widehat{Z}_m}{\sum_{j=1}^{M}N_j \widehat{Z}_j}$ is the normalized weight of the $m$-th group. If we resample $M$ times $\widetilde{\x}_m\sim \widehat{\pi}(\x|\x_{m,1:N})$ exactly one sample per group, we obtain the particle approximation of Eq. \eqref{PiGroup}, i.e.,
\begin{eqnarray}
\widehat{\pi}(\x|\widetilde{\x}_{1:M})&=&\sum_{m=1}^M  \overline{W}_m \delta(\x-\widetilde{\x}_m).
\end{eqnarray}
Since $\widehat{\pi}(\x|\x_{1:M,1:N})$  is a particle approximation of the target distribution $\bar{\pi}$ (converging to the distribution for $N\rightarrow \infty$), then $\widehat{\pi}(\x|\widetilde{\x}_{1:M})$  is also a particle approximation of $\bar{\pi}$ (converging for $N\rightarrow \infty$ and $M\rightarrow \infty$).
 Therefore, any estimator of the moments of $\bar{\pi}$ obtained using the summary weighted particles as in Eq. \eqref{GroupIS2} is consistent.  

\section{Estimators of the marginal likelihood in SIS and SIR}
\label{SuperAppMargLikeEstimator}
The classical IS estimator of the normalizing constant $Z_d=\int_{\mathbb{R}^{d\times\eta}}  \pi_d(x_{1:d}) d x_{1:d}$ at the $d$-th iteration is
\begin{eqnarray}
\widehat{Z}_d&=&\frac{1}{N} \sum_{n=1}^N w_d^{(n)}=\frac{1}{N} \sum_{n=1}^N w_{d-1}^{(n)}\beta_d^{(n)},  \label{Z_approx2}\\
&=&\frac{1}{N} \sum_{n=1}^N\left[\prod_{j=1}^d \beta_j^{(n)}\right].   \label{Z_approx3}
\end{eqnarray} 
An alternative formulation, denoted as $\overline{Z}_d$, is often used  
\begin{eqnarray}
 \label{EstZ2}
\overline{Z}_d&=& \prod_{j=1}^d\left[\sum_{n=1}^N{\bar w}_{j-1}^{(n)}\beta_j^{(n)}\right]  \\
&=&\prod_{j=1}^d\left[\frac{\sum_{n=1}^N w_{j}^{(n)}}{\sum_{n=1}^N w_{j-1}^{(n)}}\right]=
\widehat{Z}_1\prod_{j=2}^d\left[  \frac{ \widehat{Z}_j}{\widehat{Z}_{j-1}}\right]=\widehat{Z}_{d}.  \label{EstZ2_5}
\end{eqnarray} 
where we have employed ${\bar w}_{j-1}^{(n)}=\frac{w_{j-1}^{(n)}}{\sum_{i=1}^N w_{j-1}^{(i)}}$ and $w_{j}^{(n)}=w_{j-1}^{(n)}\beta_j^{(n)}$ with $w_0^{(n)}=1$ \citep{Doucet01b,Doucet08tut}. Therefore, given Eq. \eqref{EstZ2_5}, in SIS these two estimators  $\widehat{Z}_d$ in Eq. \eqref{Z_approx2}  and $\overline{Z}_d$ in Eq. \eqref{EstZ2} are equivalent approximations of the $d$-th marginal likelihood $Z_d$ \citep{Martino15PF}. Furthermore, note that $\overline{Z}_d$ can be written in a recursive form as
\begin{eqnarray}
 \label{EstZ3}
\overline{Z}_d=\overline{Z}_{d-1}\left[\sum_{n=1}^N{\bar w}_{d-1}^{(n)}\beta_d^{(n)}\right].
\end{eqnarray}

\subsection{Estimators of the marginal likelihood in particle filtering}
\label{SIRsect}

\begin{figure}[htbp]
\begin{center}
\centerline{
 \includegraphics[width=0.7\textwidth]{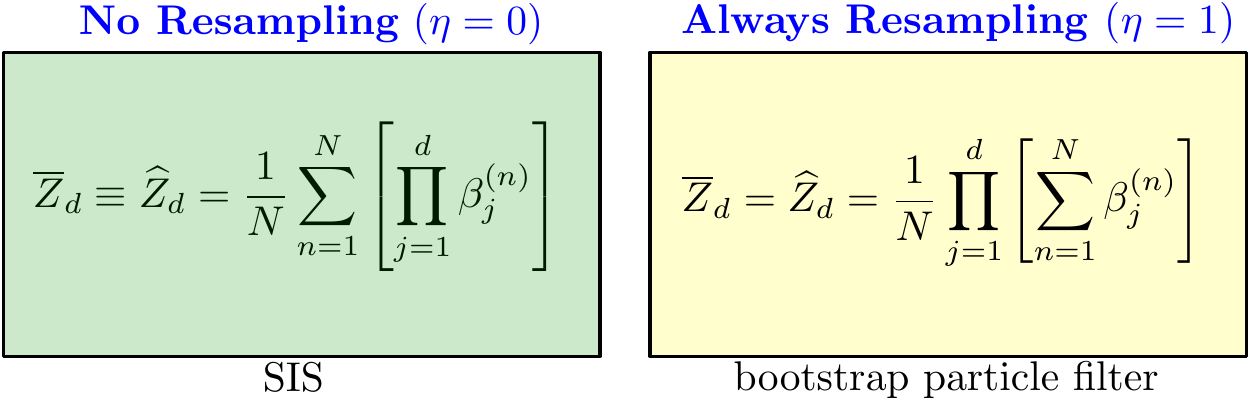}
}
\caption{Expressions of the marginal likelihood estimators $\overline{Z}_d$ and $\widehat{Z}_d$ in two extreme scenarios: without resampling and applying resampling at each iterations.  Note that in the formulations above the operations of sum and product are inverted.  }
\label{Fig1AppMargiLike}
\end{center}
 \end{figure}

Sequential Importance Resampling (SIR) (a.k.a., particle filtering) combines the SIS approach with the application of the resampling procedure corresponding to step \ref{StepResampling2} of Table \ref{alg:SIRpartialRes}. If the GIS weighting is not applied, in SIR only  
\begin{equation}
\overline{Z}_d=\prod_{j=1}^d\left[\sum_{n=1}^N{\bar w}_{j-1}^{(n)}\beta_j^{(n)}\right].
\end{equation}
is a  consistent estimator of $Z_d$. In this case, $\widehat{Z}_d=\frac{1}{N}\sum_{n=1}^N w_{d}^{(n)}$ is not a possible alternative without using GIS. However, considering the proper GIS weighting of the resampled particles (the step \ref{StepProperGIS} of Table \ref{alg:SIRpartialRes}), then $\widehat{Z}_d$ is also a consistent estimator of $Z_d$ and it is equivalent to $\overline{Z}_d$. Below, we analyze three cases considering a resampling applied to the entire set of particles:
\begin{itemize}
\item {\bf No Resampling ($\eta=0$):} this scenario corresponds to SIS where $\widehat{Z}_d$, $\overline{Z}_d$ are equivalent as shown in Eq. \eqref{EstZ2_5}. 
\item {\bf Resampling at each iteration ($\eta=1$):}  using the GIS weighting, $w_{d-1}^{(n)}=\widehat{Z}_{d-1}$ for all $n$ and for all $d$, and replacing in Eq. \eqref{Z_approx2}  we have  
\begin{eqnarray}
\label{ZtotalRes2}
\widehat{Z}_{d}&=&\widehat{Z}_{d-1}\left[\frac{1}{N} \sum_{n=1}^{N} \beta_{d}^{(n)}\right], \\
&=&\frac{1}{N}\prod_{j=1}^d \left[ \sum_{n=1}^{N}  \beta_{j}^{(n)}\right]. \label{ZtotalRes3}
\end{eqnarray}
Since the resampling is applied to the entire set of particles, we have ${\bar w}_{d-1}^{(n)}=\frac{1}{N}$ for all $n$. Replacing it in the expression of $\overline{Z}_{d}$ in \eqref{EstZ3}, we obtain
\begin{equation}
\label{ZtotalRes}
\overline{Z}_{d}=\frac{1}{N}\prod_{j=1}^d \left[  \sum_{n=1}^{N}  \beta_{j}^{(n)}\right],
\end{equation}
that coincides with $\widehat{Z}_{d}$ in Eq. \eqref{ZtotalRes3}.
\item {\bf Adaptive resampling ($0<\eta<1$):} for the sake of simplicity, let us start considering a unique resampling step applied at the $k$-th iteation with $k<d$. We check if both estimators are equal at $d$-th iteration of the recursion. Due to Eq. \eqref{EstZ2_5}, we have $\overline{Z}_{k}\equiv \widehat{Z}_{k}$,\footnote{We consider to compute the estimators before the resampling.} since before the $k$-th iteration no resampling has been applied. 
With the proper weighting $w_{k}^{(n)}=\widehat{Z}_{k}$ for all $n$, at the next iteration we have
\begin{eqnarray}
\widehat{Z}_{k+1}&=&\frac{1}{N} \sum_{n=1}^N w_{k}^{(n)}\beta_{k+1}^{(n)}=\widehat{Z}_{k} \left[\frac{1}{N} \sum_{n=1}^N \beta_{k+1}^{(n)}\right],
\end{eqnarray}
and using Eq. \eqref{EstZ3},  we obtain 
\begin{eqnarray}
\overline{Z}_{k+1}&=&\overline{Z}_{k}\left[\sum_{n=1}^N\frac{1}{N}\beta_{k+1}^{(n)}\right]=\widehat{Z}_{k} \left[\frac{1}{N} \sum_{n=1}^N \beta_{k+1}^{(n)}\right], 
 \end{eqnarray}
so that the estimators are equivalent also at the $(k+1)$-th iteration, $\overline{Z}_{k+1}\equiv \widehat{Z}_{k+1}$. Since we are assuming no resampling steps after the $k$-th iteration and until the $d$-th iteration, we have that $\overline{Z}_{i}\equiv \widehat{Z}_{i}$ for $i=k+2,\ldots,d$ due to we are in a SIS scenario  for $i>k$ (see Eq.  \eqref{EstZ2_5}). This reasoning can be easily extended for different number of resampling steps.  
\end{itemize}
 Figure \ref{Fig1AppMargiLike} summarizes the expressions of the estimators in the extreme cases of $\eta=0$ and $\eta=1$.  
  Note that the operations of sum and product are inverted. See DEMO-1 at \url{https://github.com/lukafree/GIS.git}. 

\section{Particle Marginal Metropolis-Hastings (PMMH) algorithms}
\label{MarginalPMH}
Let us consider ${\bf x}=x_{1:D}=[x_1,x_2,\ldots,x_D]\in \mathcal{X}\subseteq\mathbb{R}^{D\times\xi}$ where $x_d\in \mathbb{R}^{\xi}$ for all $d=1,\ldots,D$ and an additional model parameter ${\bm \theta} \in \mathbb{R}^{d_\theta}$ to be inferred as well. Assuming a prior pdf $g_\theta({\bm \theta})$ over ${\bm \theta}$, and a factorized complete posterior pdf $\bar{\pi}_c({\bf x},{\bm \theta})$ 
\begin{eqnarray}
\bar{\pi}_c({\bf x},{\bm \theta})= \frac{g_\theta({\bm \theta}) \pi({\bf x}|{\bm \theta})}{Z({\bm \theta})} \propto \pi_c({\bf x},{\bm \theta})&=&g_\theta({\bm \theta}) \pi({\bf x}|{\bm \theta}),
\end{eqnarray}
where $\pi({\bf x}|{\bm \theta})=\gamma_1(x_1|{\bm \theta}) \prod_{d=2}^D\gamma_d(x_d|x_{1:d-1},{\bm \theta})$ and $ Z({\bm \theta})=\int_{\mathbb{R}^{d_\theta}}g_\theta({\bm \theta}) \pi({\bf x}|{\bm \theta}) d{\bm \theta}$. Moreover, let us the denote as $\widehat{\pi}(\x|\v_{1:N},{\bm \theta})=\frac{1}{N \widehat{Z}({\bm \theta})}\sum_{n=1}^N w(\v_n|{\bm \theta}) \delta(\x-\v_n)$ a particle approximation of $\pi({\bf x}|{\bm \theta})$ obtained by one run of a particle filter approach, and $\widehat{Z}({\bm \theta})=\frac{1}{N}\sum_{n=1}^N w(\v_n|{\bm \theta})$ is an unbiased estimator of $Z({\bm \theta})$. The Marginal PMH (PMMH) technique is then summarized in Table \ref{alg:PMMH}. PMMH is often used for both smoothing and parameter estimation in state-space models. 
Note that if $q_\theta({\bm \theta}|{\bm \theta}_{t-1})=g_\theta({\bm \theta})$ then the acceptance function becomes
\begin{equation}
\alpha = \min\left[1, \frac{\widehat{Z}({\bm \theta}')}{\widehat{Z}({\bm \theta}_{t-1})}\right].
\end{equation}

\begin{table}[!h]
	\centering
	\caption{{\bf Particle Marginal MH (PMMH) algorithm}}
	    \begin{tabular}{|p{0.95\columnwidth}|}
		\hline
		  \vspace{-0.5cm} 
  \begin{enumerate}
\item Choose $\x_0$, ${\bm \theta}_0$, and obtain a first estimation $\widehat{Z}({\bm \theta}_0)$.
\item For $t=1,\ldots,T:$
\begin{enumerate}
\item Draw ${\bm \theta}'\sim q_\theta({\bm \theta}|{\bm \theta}_{t-1})$ and $\v_j\sim \widehat{\pi}(\x|\v_{1:N},{\bm \theta}')=\frac{1}{N \widehat{Z}({\bm \theta}')}\sum_{n=1}^N w(\v_n|{\bm \theta}') \delta(\x-\v_n)$ (where $\widehat{\pi}$ is obtained with one run of a particle filter).
\item Set  ${\bm \theta}_t={\bm \theta}'$, $\x_t={\bf v}_j$, with probability 
\begin{equation}
\label{AlfaPMMH}
\alpha = \min\left[1, \frac{\widehat{Z}({\bm \theta}') g_\theta({\bm \theta}') q_\theta({\bm \theta}_{t-1}|{\bm \theta}')}{\widehat{Z}({\bm \theta}_{t-1})  g_\theta({\bm \theta}_{t-1}) q_\theta({\bm \theta}|{\bm \theta}_{t-1})}\right].
\end{equation}
Otherwise, set ${\bm \theta}_t={\bm \theta}'$ and $\x_t=\x_{t-1}$.
\end{enumerate}
  \item Return $\{\x_t\}_{t=1}^T$ and $\{{\bm \theta}_t\}_{t=1}^T$.
  \vspace{-0.3cm}
    \end{enumerate} 	\\
	\hline
	\end{tabular}		
	\label{alg:PMMH}
\end{table}

\paragraph{Distributed Particle Marginal  Metropolis-Hastings (DPMMH).} We can easily design a marginal version of DPMH in Section \ref{P-PF-PMH}, drawing ${\bm \theta}'\sim q_\theta({\bm \theta}|{\bm \theta}_{t-1})$ and run $M$ particle filters addressing the target pdf ${\bar \pi}({\bf x}|{\bm \theta}')$.
 The algorithm follows the steps in Table \ref{alg:PMMH} with the difference that $M$ parallel particle filters are used, and
 in this case the acceptance probability is 
\begin{equation}
\label{AlfaPMMH_2}
\alpha = \min\left[1, \frac{\left[\sum_{m=1}^M\widehat{Z}_m({\bm \theta}')\right]g_\theta({\bm \theta}') q_\theta({\bm \theta}_{t-1}|{\bm \theta}')}{\left[\sum_{m=1}^M\widehat{Z}_m({\bm \theta}_{t-1})\right] g_\theta({\bm \theta}_{t-1}) q_\theta({\bm \theta}|{\bm \theta}_{t-1})}\right].
\end{equation}

\section{ GMS as infinite parallel IMTM2 chains}
\label{ConsistencyGMS_2}
In this section, we show how the GMS can be interpreted as the use of infinite number of dependent parallel IMTM2 chains.
We have already seen that we can recover an I-MTM2 chain from the GMS outputs applying one resampling step for each $t$ when $\mathcal{S}_t\neq  \mathcal{S}_{t-1}$, i.e.,       
\begin{gather}
\widetilde{{\bf x}}_{t}= \left\{
\begin{split}
\label{RecChain_APP}
&\widetilde{{\bf v}}_{t} \sim \sum_{n=1}^N \frac{\rho_{n,t}}{\sum_{i=1}^N \rho_{i,t}}  \delta({\bf x}-{\bf x}_{n,t}),   \quad \mbox{ if }  \quad \mathcal{S}_t\neq  \mathcal{S}_{t-1}, \\
&\widetilde{{\bf x}}_{t-1}, \quad\quad\quad\quad\quad\quad\quad\quad\quad\quad\mbox{  } \mbox{ if } \quad \mathcal{S}_t=  \mathcal{S}_{t-1},
\end{split}
\right. 
\end{gather}
for $t=1,\ldots,T$. The sequence $\{\widetilde{{\bf x}}_{t}\}_{t=1}^T$ is a chain obtained by one run of an I-MTM2 technique. Note that 
(a )the sample generation, (b) the acceptance probability function and hence (c) the dynamics of GMS exactly coincide with the corresponding steps of I-MTM2 (or PMH; depending on candidate generation procedure). Hence, the ergodicity of the recovered chain is ensured.
\newline
{\bf Parallel chains from GMS outputs.} 
We can extend the previous consideration for generation $C$ parallel I-MTM2 chains. Indeed, we resample independently $C$ times (instead of only one) within the set of accepted candidates at the $t$-th iteration $\{{\bf x}_{1,t},\ldots,{\bf x}_{N,t}\}$ , i.e.,  
\begin{gather}
\widetilde{{\bf x}}_{t}^{(c)}= \left\{
\begin{split}
\label{RecChain_parallelChains}
&\widetilde{{\bf v}}_{t}^{(c)} \sim \sum_{n=1}^N \frac{\rho_{n,t}}{\sum_{i=1}^N \rho_{i,t}}  \delta({\bf x}-{\bf x}_{n,t}),   \quad \mbox{ if }  \quad \mathcal{S}_t\neq  \mathcal{S}_{t-1}, \\
&\widetilde{{\bf x}}_{t-1}^{(c)}, \quad\quad\quad\quad\quad\quad\quad\quad\quad\quad\mbox{ }\mbox{  }\mbox{  }\mbox{ if } \quad \mathcal{S}_t=  \mathcal{S}_{t-1}, 
\end{split}
\right. 
\end{gather}
for $c=1,\ldots,C$, where the super-index denotes the $c$-th chain (similar procedures have been suggested in \citep{Calderhead14,OMCMC_DSP}).
Clearly, the resulting $C$ parallel chains are not independent, and there is  an evident loss in the performance w.r.t. the case of independent parallel chains (IPCs). However, at each iteration, the number of target evaluations per iteration is only $N$ instead of $NC$, as in the case of IPCs. 
Note that that each chain in ergodic, so that each estimator $\widetilde{I}_T^{(c)}=\frac{1}{T} \sum_{t=1}^T h(\widetilde{{\bf x}}_t^{(c)})$ is consistent for $T\rightarrow \infty$. As a consequence, the arithmetic mean of consistent estimators,
\begin{equation}
\label{AquiApp}
\widetilde{I}_{C,T}=\frac{1}{C} \sum_{c=1}^C\widetilde{I}_T^{(c)}=\frac{1}{CT} \sum_{t=1}^T  \sum_{c=1}^C h(\widetilde{{\bf x}}_t^{(c)}),
\end{equation}
is also consistent, for all values of $C\geq 1$. 
\newline
{\bf GMS as limit case.} 
Let us consider the case $\mathcal{S}_t\neq  \mathcal{S}_{t-1}$ (the other is trivial), at some iteration $t$. In this scenario, the samples of the $C$ parallel I-MTM2 chains, $\widetilde{{\bf x}}_t^{(1)}$,$\widetilde{{\bf x}}_t^{(2)}$,...,$\widetilde{{\bf x}}_t^{(C)}$, are obtained by resampled independently $C$ samples from the set $\{{\bf x}_{1,t},\ldots,{\bf x}_{N,t}\}$ according to the normalized weights  $\bar{\rho}_{n,t}=\frac{\rho_{n,t}}{\sum_{i=1}^N \rho_{i,t}}$, for $n=1,\ldots,N$. Recall that the samples $\widetilde{{\bf x}}_t^{(1)}$,$\widetilde{{\bf x}}_t^{(2)}$,...,$\widetilde{{\bf x}}_t^{(C)}$, will be used in the final estimator $\widetilde{I}_{C,T}$ in Eq. \eqref{AquiApp}.

Let us denote as $\#j$ the number of times that a specific candidate ${\bf x}_{j,t}$ (contained in the set $\{{\bf x }_{n,t}\}_{n=1}^N$) has been selected as state of one of $C$ chains, at the $t$ iteration. As $C\rightarrow \infty$, The fraction  $\frac{\#j}{C}$ approaches exactly the corresponding probability $\bar{\rho}_{j,t}$. Then, for $C\rightarrow \infty$, we have that the estimator in Eq. \eqref{AquiApp} approaches the GMS estimator, i.e.,
\begin{equation}
\label{AquiApp2}
 \widetilde{I}_{T} =\lim_{C\rightarrow \infty} \widetilde{I}_{C,T}= \frac{1}{T} \sum_{t=1}^T \sum_{n=1}^N \bar{\rho}_{n,t} h({\bf x}_{n,t}).
\end{equation}
 Since $ \widetilde{I}_{C,T}$ as $T\rightarrow \infty$ is consistent for all values of $C$, then the GMS estimator $ \widetilde{I}_{T}$ is also consistent (and can be obtained as $C\rightarrow \infty$).

 \end{document}